\documentclass[aps,prl,twocolumn,superscriptaddress]{revtex4-2}
\usepackage{graphics,graphicx, array, afterpage}
\usepackage{qcircuit,amsmath}
\usepackage{grffile}
\usepackage[export]{adjustbox}
\usepackage{lineno}
\usepackage{xcolor}
\usepackage{longtable}
\usepackage{braket}
\usepackage{array}
\usepackage{multirow}
\usepackage{enumerate}
\usepackage{amsmath}
\usepackage{amssymb}
\usepackage{amsthm}
\usepackage{times}
\usepackage{multirow}

\usepackage[colorlinks, citecolor=blue]{hyperref}
\hypersetup{linkcolor=magenta,urlcolor=blue,citecolor=blue,pdfstartview={FitH},urlcolor=blue}

\usepackage[latin9]{inputenc}
\usepackage{color}
\usepackage{tabularx}
\usepackage{algorithm}
\usepackage{algpseudocode}
\usepackage{graphbox}
\usepackage{amsfonts}
\usepackage{dcolumn}
\usepackage{bm}
\usepackage{epstopdf}

\begin{document}

\title{Embedding Quantum Many-Body Scars into Decoherence-Free Subspaces}

\author{He-Ran Wang}
\thanks{These authors contributed equally to this work.}
\affiliation{Institute for Advanced Study, Tsinghua University, Beijing 100084, People's Republic of China}

\author{Dong Yuan}
\thanks{These authors contributed equally to this work.}
\affiliation{Center for Quantum Information, IIIS, Tsinghua University, Beijing 100084, People's Republic of China}

\author{Shun-Yao Zhang}
\thanks{These authors contributed equally to this work.}
\affiliation{Center for Quantum Information, IIIS, Tsinghua University, Beijing 100084, People's Republic of China}

\author{Zhong Wang}
\email{wangzhongemail@tsinghua.edu.cn}
\affiliation{Institute for Advanced Study, Tsinghua University, Beijing 100084, People's Republic of China}

\author{Dong-Ling Deng}
\email{dldeng@tsinghua.edu.cn}
\affiliation{Center for Quantum Information, IIIS, Tsinghua University, Beijing 100084, People's Republic of China}
\affiliation{Hefei National Laboratory, Hefei 230088, People's Republic of China}
\affiliation{Shanghai Qi Zhi Institute, 41st Floor, AI Tower, No. 701 Yunjin Road, Xuhui District, Shanghai 200232, China}

\author{L.-M. Duan}
\affiliation{Center for Quantum Information, IIIS, Tsinghua University, Beijing 100084, People's Republic of China}
\affiliation{Hefei National Laboratory, Hefei 230088, People's Republic of China}
\affiliation{Shanghai Qi Zhi Institute, 41st Floor, AI Tower, No. 701 Yunjin Road, Xuhui District, Shanghai 200232, China}
\affiliation{New Cornerstone Science Laboratory, IIIS, Tsinghua University, Beijing 100084, People's Republic of China}

\begin{abstract}

Quantum many-body scars are non-thermal excited eigenstates of non-integrable Hamiltonians, which could support coherent revival dynamics from special initial states when scars form an equally spaced tower in the energy spectrum. For open quantum systems, engineering many-body scarred dynamics by a controlled coupling to the environment remains largely unexplored. In this paper, we provide a general framework to exactly embed quantum many-body scars into the decoherence-free subspaces of Lindblad master equations. The dissipative scarred dynamics manifest persistent periodic oscillations for \textit{generic} initial states, and can be practically utilized to prepare scar states with potential quantum metrology applications.
We construct the Liouvillian dissipators with the local projectors that annihilate the whole scar towers, and utilize the Hamiltonian part to rotate the undesired states out of the null space of dissipators.
We demonstrate our protocol through several typical models hosting many-body scar towers, and propose an experimental scheme to observe the dissipative scarred dynamics based on digital quantum simulations and resetting ancilla qubits.
\end{abstract}

\maketitle 

\begin{figure}
\hspace*{-0.4\textwidth}
\includegraphics[width=0.4\textwidth]{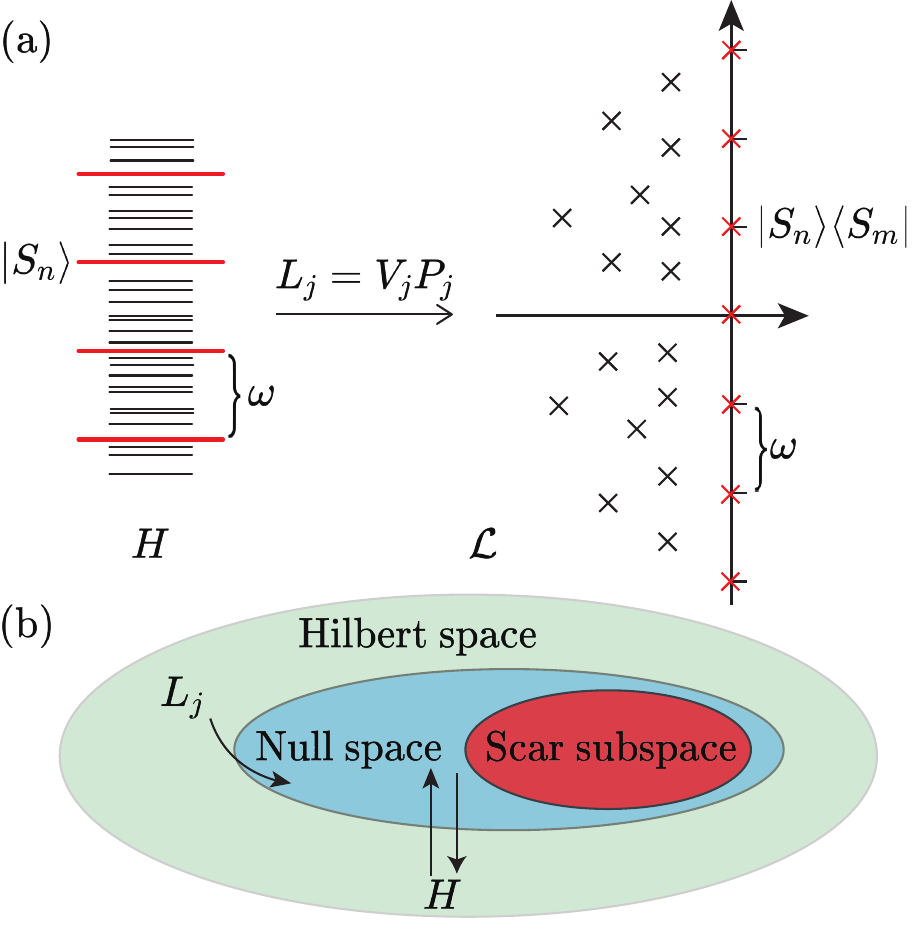} 
\caption{
Schematic illustration of the protocol for embedding quantum many-body scars into decoherence-free subspaces of Lindblad master equations. (a) The equally spaced many-body scar tower $\{\ket{S_n}\}$ (red lines) is embedded onto the imaginary axis of the Liouvillian spectrum as non-decaying eigenmodes in the form of $\{\ket{S_n}\bra{S_m}\}$ (red crosses).  
(b) The dissipators drive the system into their common null space (sometimes equals the scar subspace) and the Hamiltonian part of the Liouvillian rotates the undesired states out of the null space to make them decay away.
}
\label{fig:illustraion}
\end{figure}

Isolated quantum many-body systems typically thermalize under Hamiltonian evolution, during which any local information preserved in the initial states scrambles into the entire system. 
These features of quantum thermalization have been illustrated by the eigenstate thermalization hypothesis (ETH) in the past decades \cite{Deutsch1991Quantum,Srednicki1994Chaos}. 
In recent years, studies of weak ergodicity breaking, namely, a small fraction of ETH-violating eigenstates immersed in a sea of thermal ones, dubbed quantum many-body scars, have attracted considerable attention \cite{Serbyn2021quantum,Moudgalya2022Quantum,Chandran2022Quantum}.  
One of the hallmarks of quantum many-body scars, originally discovered in experiments with Rydberg atoms \cite{Bernien2017Probing,Bluvstein2021Controlling}, is their ability to support long-lived coherent oscillations from initial states that have large overlap with a tower of equally spaced scars in the energy spectrum \cite{Turner2018weak,Turner2018quantum,Moudgalya2018Exact,Moudgalya2018Entanglement,Schecter2019Weak,Choi2019emergent,Iadecola2020Quantum,shibata2020onsager,Chattopadhyay2020quantum}. Despite the fact that such anomalous eigenstates, typically with sub-volume-law entanglement entropy, have been found and carefully analyzed in various Hamiltonians~\cite{Shiraishi2017Systematic,lin2019exact,ok2019topological,bull2019systematic,Surace2020Lattice,hudomal2020quantum,Moudgalya2020eta,Mark2020eta,Scherg2021Observing,Desaules2021Proposal,Banerjee2021Quantum,Langlett2022rainbow,Langlett2021Hilbert,Schindler2022Exact,Desaules2023Weak,Zhang2022Many,Su2023Observation}, the extensions of many-body scars and related coherent revivals into the regime of open quantum systems remain largely unexplored. Here, we add this crucial yet missing block by introducing a general framework to exactly embed quantum many-body scars into the decoherence-free subspaces of Lindblad master equations. 
See Fig.~\ref{fig:illustraion} for a pictorial illustration.

The dynamics of open quantum systems coupled to a Markovian environment are described by the following Lindblad master equation \cite{breuer2002theory}
\begin{equation}\label{eq:lindblad}
    \frac{d \rho}{d t}=-i[H,\rho]+ \gamma \sum_{j} (2L_{j}\rho L_{j}^\dagger-\{L_j^\dagger L_j,\rho\}) \equiv \mathcal{L}(\rho),
\end{equation}
where $\rho$ is the density matrix, $H$ is the Hamiltonian part
governing the unitary dynamics, $\{L_j\}$ are jump operators describing the dissipative quantum channels with strength $\gamma$, and $\mathcal{L}$ is the Liouvillian superoperator. 
In particular, if the evolution dynamics governed by $\mathcal{L}$ are purely unitary within a subspace $W$ and do not suffer from dissipation, $W$ is said to be a decoherence-free subspace of this Lindblad master equation \cite{Lidar1998Decoherence,Bacon1999Robustness}. 
One special case is that all the basis elements $\{ \ket{S_n} \}$ of the subspace $W$ are annihilated by all the dissipators, and $W$ is closed under the action of the Hamiltonian part, i.e., $L_j \ket{S_n} = 0, \forall j, n$ ($\{\ket{S_n}\}$ are therefore ``dark states" of the jump operators), and $H W \subseteq W$.
The decoherence-free subspaces were originally proposed to reduce noises in quantum computation and realize the ``passive" quantum error correction codes \cite{Duan1997Preserving,Zanardi1997Noiseless,Lidar1998Decoherence}. Later works apply similar techniques to realize the dissipative quantum state preparation \cite{Plenio1999Cavity,Diehl2008Quantum,Kraus2008Preparation,Verstraete2009Quantum,Diehl2010Dissipation,Diehl2011Topology}.

In this paper, by designing the dissipators and the Hamiltonian part, we introduce a general protocol to construct local Liouvillians that host \textit{scar-state-only} decoherence-free subspaces. One important consequence reflecting on the Liouvillian spectrum is that all the non-decaying eigenmodes are equally spaced on the imaginary axis, as depicted in Fig.~\ref{fig:illustraion}(a). Hence, unlike their closed-system counterparts, which are highly sensitive to the initial states and vulnerable to instantaneous perturbations, the open-system scarred dynamics manifest persistent periodic oscillations for generic initial states (even mixed states) and exhibit intrinsic tolerance to such disturbances.
We demonstrate our protocol through four typical models hosting many-body scars, with the constructed Liouvillians summarized in Table.~\ref{tab:Liouvillian_summary}. 
We show that our dissipative protocol can be further utilized to prepare each scar state that possesses extensive multipartite entanglement  with potential  applications in quantum enhanced metrology.
In addition, we propose an experimental scheme to observe such dissipative scarred dynamics on current quantum simulators through digital quantum simulations and resetting ancilla qubits.

\begin{table}
\begin{center}
\caption{Summary of the local Hamiltonians and jump operators of the constructed Liouvillians for four typical models. The generic
local operators $\{V_j\}$ are specified in following discussions.}
\label{tab:Liouvillian_summary}
\begin{tabular}{l c c }
\hline
\hline
Model & $H_j$ & $L_j$ \\
\\
\hline
Toy model in~\cite{Choi2019emergent} & $P_j h_j P_j+\Omega\sigma_j^x/2$ & $V_{j,j+1}(1-\vec{\sigma}_j\cdot\vec{\sigma}_{j+1})$  \\
\\
  \multirow{2}{*}{Spin-$1$ $XY$~\cite{Schecter2019Weak}} & $(S_j^x S_{j+1}^x+S_j^y S_{j+1}^y)$ & \multirow{2}{*}{$V_{j,j+1}(S_j^x S_{j+1}^x+S_j^y S_{j+1}^y)$} \\
  & $+h S_j^z+D(S_j^z)^2$ 
\\
\\
 \multirow{2}{*}{AKLT~\cite{Moudgalya2018Exact}} &  \multirow{2}{*}{$T_{j,j+1}^{S=2}$}& $V_{j,j+1}T_{j,j+1}^{S=2,m=-2,-1,0}$, \\
& & $V'_{j-1,j,j+1} T'_{j-1,j,j+1}$
 \\
\\
Domain-wall & $(\sigma^x_{j}-\sigma^z_{j-1}\sigma^x_{j}\sigma^z_{j+1})$ & \multirow{2}{*}{$V_{j,j+1} (\ket{\uparrow\uparrow}\bra{\uparrow\uparrow})_{j,j+1}$} \\
preserving~\cite{Iadecola2020Quantum}  & $+\Delta\sigma_j^z+J\sigma_j^z\sigma_{j+1}^z$ \\
\hline\hline
\end{tabular}
\label{tab}
\end{center}
\end{table}

\textit{Non-Hermitian Shiraishi-Mori embedding.}-- We motivate our protocol from the non-Hermitian generalization of the Shiraishi-Mori embedding method \cite{Shiraishi2017Systematic}, then extend to the Liouvillian formalism. In Ref.~\cite{Shiraishi2017Systematic} Shiraishi and Mori proposed an approach to embed non-thermal eigenstates into the spectrum of non-integrable Hamiltonians. The general Shiraishi-Mori Hamiltonians have the form of $H = \sum_{j} P_j h_j P_j+H'$, where $\{P_j\}$ is a set of local projectors ($P_j^2 = P_j$), $[H',P_j]=0, \forall j$, and $\{h_j\}$ are arbitrary local Hamiltonians. We hereafter refer $j=1,2,\cdots, L$ to the label of sites in a one-dimensional spin chain with periodic boundary condition. 
We denote the common null space annihilated by these local projectors $\{P_j\}$ by $W'$. Since $P_j H \ket{\Psi} = P_j H' \ket{\Psi} = H' P_j \ket{\Psi} = 0$ for $\forall \ket{\Psi} \in W'$, $W'$ is closed under the action of $H$ ($H W' \subseteq W'$), and therefore hosts $\text{dim}(W')$ eigenstates of $H$. For properly chosen $\{P_j\}$ and $H'$, these eigenstates could become many-body scars embedded into the middle of the spectrum of $H$. 
Note that in the present case, the scar subspace $W$ [the red circle in Fig.~\ref{fig:illustraion}(b)] coincides with the common null space $W'$ of the local projectors [the blue circle in Fig.~\ref{fig:illustraion}(b)]. 

Now we consider adding some non-Hermitian terms into the Shiraishi-Mori Hamiltonian
\begin{equation}
    H_{\text{NH}}=H-i\sum_j P_j D_j P_j=\sum_{j} P_j (h_j-iD_j) P_j+H',
\label{eq:nHSMembedding}
\end{equation}
where the local Hermitian operators $\{D_j\}$ are positive definite, such that the imaginary parts of the spectrum of $H_{\text{NH}}$ are upper bounded by zero. Since the non-Hermitian terms still annihilate the embedded scars, their eigenenergies are kept to be purely real. Other thermal eigenstates now acquire negative imaginary parts for their eigenenergies, therefore will decay away once we start the dissipative evolution driven by $H_{\text{NH}}$. Through this simple modification, we build up a relationship between thermalization in the closed systems and decoherence in the open systems for quantum many-body scarred models. We notice that the addition of non-Hermitian terms into many-body scarred Hamiltonians has been carried out in previous works~\cite{Pakrouski2021Group,Chen2023nonHermitian,Omiya2023Quantum} in different frameworks.

\begin{figure*}
\hspace*{-0.99\textwidth}
\includegraphics[width=0.99\textwidth]{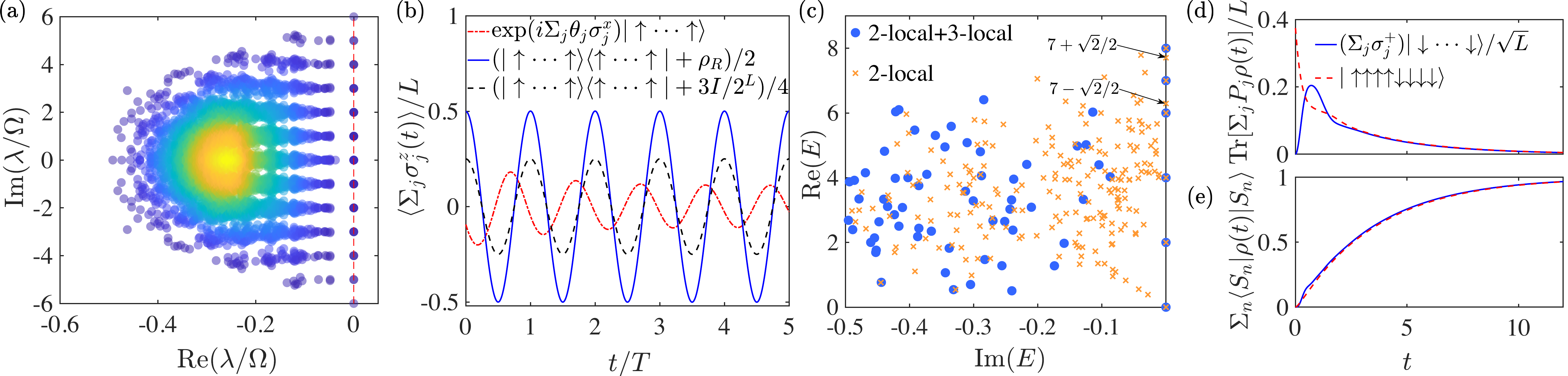} 
\caption{ Numerical results for the Liouvillian spectrum and dissipative scarred dynamics. (a) Liouvillian spectrum of the toy model hosting Dicke states as scars. Scarred eigenmodes are equidistantly embedded on the imaginary axis (the red dotted line) in the form of $\{\ket{S_n}\bra{S_m}\}$. $L=6, \Omega=2\pi, \gamma=1, V_{j,j+1}=\sigma_j^x$. (b) Total spin-$z$ dynamics for the toy model, starting from three different initial states. $\theta_j \in [0, \pi]$ are some random rotation angles. $\rho_R$ is a random physical density matrix. (c) Spectrum of the non-Hermitian AKLT Hamiltonian with or without the three-local projectors. $L=8,\gamma=2$. Liouvillian dynamics of the quantum jump rate (d) and the scar subspace overlap (e) for the domain-wall preserving model, starting from two initial states. $L=8, \Delta=0.5, J=1, \gamma=1, V_{j,j+1}=\sigma_j^x\sigma_{j+1}^x$.}
\label{fig:spectrum_dynamics}
\end{figure*}

However, we emphasize that the description of open quantum dynamics in terms of non-Hermitian Hamiltonians is accurate only for short-time dynamics without quantum jumps, or under post-selection. We thus turn to the Lindblad master equation Eq.~\eqref{eq:lindblad} to describe the full-fledged open quantum scarred dynamics. 
We take the Hamiltonian part of the Liouvillian as the same $H$, and choose the jump operators as $L_j=V_j P_j$, where $\{V_j\}$ are generic local operators. Now the Liouvillian could be written as 
$\mathcal{L}(\rho) = -iH_{\text{eff}}\rho + i \rho H_{\text{eff}}^\dagger + 2\gamma \sum_{j} L_{j} \rho L_{j}^\dagger$
with the effective Hamiltonian $H_{\text{eff}} = H - i \gamma\sum_j L_j^\dagger L_j$ having the same form as the non-Hermitian Shiraishi-Mori embedding Eq.~\eqref{eq:nHSMembedding}. Given that 
$L_j|S_n\rangle=0, \forall j, n$, $H_{\text{eff}}|S_n\rangle = E_n |S_n\rangle$, one can verify that all the $\text{dim}(W)$ scarred eigenstates $\{|S_n\rangle\}$ are embedded into the decoherence-free subspace of the Liouvillian in the form of $\{\ket{S_n}\bra{S_m}\}$ [totally $\text{dim}(W)^2$ basis elements]: $\mathcal{L}(|S_n\rangle\langle S_m|) = -i(E_n-E_m)|S_n\rangle\langle S_m|$. We particularly stress that the non-constant operators $V_j$ in the dissipators $L_j=V_jP_j$ are indispensable. Otherwise, all common eigenstates of $\{P_j\}$ (not necessarily with zero eigenvalues) and $H'$ would enter the decoherence-free subspace, which may include undesired states~\cite{Diehl2008Quantum,Kraus2008Preparation,Verstraete2009Quantum}.

When the Hamiltonian $H$ (not necessarily following the Shiraishi-Mori formalism) exhibits certain restricted spectrum generating algebra in the scarred subspace $W$\cite{Choi2019emergent,Chandran2022Quantum,Moudgalya2022Quantum,SuppMaterials}, i.e., $([H,Q^\dagger]-\omega Q^\dagger)W=0$ for some ladder operator $Q^\dagger$ generating the tower of scar states $\ket{S_n} = (Q^\dagger)^n |S_0\rangle$, 
energy levels of scars are evenly spaced by $\omega$ in the spectrum of $H$.
In our dissipative protocol, inherited from the Hamiltonian, all the non-decaying eigenmodes located on the imaginary axis are uniformly spaced by the same $\omega$. We emphasize that the aforementioned condition imposes less stringent constraints on the Liouvillians than the dynamical symmetry studied in previous literature \cite{buvca2019Non,Booker2020Non,Buvca2020Bethe,Chinzei2020Time,Guarnieri2022Time}, where the entanglement structure of states in the decoherence-free subspace is not the primary focus either (see \cite{dyn_symm_footnote}).

Within the framework of Shiraishi-Mori embedding, we demonstrate two examples as follows. The first toy model \cite{Choi2019emergent} is a one-dimensional spin-1/2 chain with $ H_{\text{toy}} = H' + \sum_j P_j h_j P_j $, where $H'= \Omega (\sum_j \sigma_j^x) / 2$, $P_j = (1-\vec{\sigma}_j\cdot\vec{\sigma}_{j+1}) / 4$ and $h_j=\sum_{\mu,\nu}J_{\mu\nu}\sigma_{j-1}^\mu\sigma_{j+2}^\nu$ is a generic two-spin operator. 
$\sigma_j^\mu\ (\mu=x,y,z)$ are standard Pauli matrices.
Since $\{P_j\}$ project two adjacent spins onto the singlet states, $H_{\text{toy}}$ hosts the $x$-direction Dicke states $\ket{S=L/2,S_x=m}$ as scarred eigenstates with energy spacing $\omega = \Omega$, where $S$ is the total spin and $S_x = \sum_j \sigma_j^x /2$ is the total spin-$x$ polarization, $m=-L/2,-L/2+1,\cdots,L/2$. As for the corresponding Liouvillian, we use the same Hamiltonian $H_{\text{toy}}$, together with $L_j=\sigma_j^x P_j$. By exact diagonalization (ED), we obtain the desired Liouvillian spectrum [Fig.~\ref{fig:spectrum_dynamics}(a)] and persistent coherent oscillations from generic initial states [Fig.~\ref{fig:spectrum_dynamics}(b)]. 
Moreover, when the Liouvillian superoperator respects the strong symmetry $S_x$~\cite{Buca2012Note}, i.e., $[H_{\text{toy}}, S_x]=[ L_j, S_x ]= 0, \forall j$, [the generic forms of local Hamiltonians and dissipators take $h_j=J_1(\sigma_{j-1}^y\sigma_{j+2}^y+\sigma_{j-1}^z\sigma_{j+2}^z)+J_2(\sigma_{j-1}^y\sigma_{j+2}^z-\sigma_{j-1}^z\sigma_{j+2}^y)+J_3\sigma_{j-1}^x\sigma_{j+2}^x$, and $L_j=\sigma_j^x P_j$], 
the value of $S_x$ is preserved during the open scarred dynamics.
In these scenarios, we can effectively prepare any desired $x$-direction Dicke state by starting the Liouvillian evolution from an $x$-direction spin product state in the same symmetry sector~\cite{SuppMaterials}.

The second example is the spin-$1$ $XY$ model \cite{Schecter2019Weak} $H_{XY} = \sum_{j} [ S^x_j S^x_{j+1} + S^y_j S^y_{j+1} + h S^z_j + D (S^z_j)^2 ]$, where there are three 
degrees of freedom on each site ($\ket{-1},\ket{0},\ket{1}$) and 
$S^\mu_j\ (\mu=x,y,z)$ are spin-$1$ operators. 
The $L+1$ scarred eigenstates are generated from the ferromagnetic state $|S_0\rangle=\ket{-1,\cdots,-1}$ by the ladder operator $Q^\dagger=\sum_j (-1)^j(S_j^{+})^2$ with the energy spacing $\omega = 2h$. 
$H_{XY}$ has been shown to be consistent with the Shiraishi-Mori embedding formalism \cite{Schecter2019Weak,mark2020unified}.
The scar-tower states are annihilated by a set of six orthogonal two-local projectors, which commute with the $\sum_j S_j^z$ and $\sum_j (S_j^z)^2$ terms (see \cite{mark2020unified} and \cite{SuppMaterials}). 
Fortunately, the null space of these local projectors coincides with that of the $XY$ interaction term $S^x_j S^x_{j+1} + S^y_j S^y_{j+1}$, so we design the jump operators in a simple form as $L_j = S_j^x (S^x_j S^x_{j+1} + S^y_j S^y_{j+1}) $. 

We remark that the success of our dissipative protocol hinges on finding local projectors annihilating the whole scar towers, which could be achieved by compressing the scar tower into a single matrix product state (MPS) $\ket{S(\beta)}=\exp(\beta Q^\dagger)\ket{S_0} = \sum_{n} \beta^n \ket{S_n} / n!$ and applying standard linear algebra techniques to construct local projectors annihilating the local tensors of $\ket{S(\beta)}$ for any $\beta$ \cite{shibata2020onsager,Chattopadhyay2020quantum,mark2020unified,SuppMaterials}.

\textit{Models beyond Shiraishi-Mori embedding.}--
Our strategy of creating scar-state-only decoherence-free subspace can further apply to many-body scarred models beyond the Shiraishi-Mori embedding formalism. One typical example is the spin-$1$ Affleck-Kennedy-Lieb-Tasaki (AKLT) model $H_{\text{AKLT}}=\sum_j T_{j,j+1}^{S=2}$, where $T_{j,j+1}^{S=2}$ projects two adjacent spin-$1$'s onto a total spin-$2$~\cite{Affleck1987Rigorous}. 
A tower of scarred eigenstates with energy spacing $\omega=2$ is generated from the ground state $|S_0\rangle=|G\rangle$ by the ladder operator $Q^\dagger=\sum_j (-1)^j(S_j^{+})^2 $~\cite{Moudgalya2018Exact,Moudgalya2018Entanglement}. 
Two-local projectors annihilating the scar tower are known as $P_j=T_{j,j+1}^{S=2,m=-2}+T_{j,j+1}^{S=2,m=-1}+T_{j,j+1}^{S=2,m=0}$ \cite{mark2020unified}, where $T_{j,j+1}^{S=2,m}$ projects two spin-$1$'s onto a total spin-$2$ with spin-$z$ polarization equal to $m$. 
The AKLT Hamiltonian can then be decomposed as $H_\text{AKLT}=H' + \sum_j P_j$, where $H'=\sum_j T_{j,j+1}^{S=2,m=1,2}$. 
However, since $[H', P_j]\neq 0$, $H_\text{AKLT}$ goes beyond the Shiraishi-Mori framework in the sense that the null space $W'$ is larger than the desired scar subspace $W$ \cite{Moudgalya2018Exact,mark2020unified}. 
We demonstrate the resulting effect by calculating the spectrum of the non-Hermitian Hamiltonian $H_{\text{NH}}=H_{\text{AKLT}}-i\gamma\sum_j P_j$~\cite{syssize_footnote}. Eigenstates of $H_{\text{NH}}$ with purely real eigenvalues are annihilated by $\{P_j\}$ and are eigenstates of the Hermitian part $H_{\text{AKLT}}$, such that they will be embedded in the decoherence-free subspace of the constructed Liouvillian. 
Apart from the scar states with integer eigenvalues, we observe several undesired irrational eigenvalues on the real axis (orange crosses in Fig. \ref{fig:spectrum_dynamics}(c), see~\cite{SuppMaterials} for detailed discussions), which will contaminate the decoherence-free subspace and ruin the periodic oscillations.

To solve the problem, we introduce the three-local projector $T'_{j-1,j,j+1}=(\ket{T'}\bra{T'})_{j-1,j,j+1}$, obtained by the compressed MPS technique,
\begin{equation}
    \ket{T'}=\frac{1}{\sqrt{2}}(\ket{0,1,1}+\ket{1,1,0})
\end{equation}
to enter the Liouvillian as dissipators.
The three-local projectors also annihilate the whole scar tower and they can effectively kill unwanted states in the decoherence-free subspace. 
As shown by the blue dots in Fig.~\ref{fig:spectrum_dynamics}(c), after adding the three-local projectors, irrational eigenvalues disappear from the real axis, and therefore harmonic scarred oscillations are restored (There still exist a few remaining eigenstates with eigenvalues $L-1$ or $L-2$. See \cite{SuppMaterials} for detailed discussions.)
We remark that the common null space $W'$ of two-local and three-local projectors is still larger than the scar subspace $W$, but the Hamiltonian part $H_{\text{AKLT}}$ of the Liouvillian drives unwanted states out of $W'$ to make them decay away [Fig.~\ref{fig:illustraion}(b)].

To better illustrate the interplay between the dissipators and the Hamiltonian part, we consider the domain-wall preserving model~\cite{Iadecola2020Quantum} $H_{\text{DW}} = H_0 + H_\Delta + H_J$, where $H_0=\sum_j (\sigma_j^x - \sigma_{j-1}^z \sigma_j^x \sigma_{j+1}^z), H_\Delta = \Delta \sum_j\sigma_j^z$, and $H_J=J\sum_j \sigma_j^z\sigma_{j+1}^z$. The ladder operator $Q^\dagger = \sum_j (-1)^j P^0_{j-1} \sigma_j^+ P^0_{j+1}$ [$P^0_j=(1-\sigma_j^z)/2$] generates the scar tower from the reference state $\ket{S_0}=\ket{\downarrow\downarrow\cdots \downarrow}$ with energy spacing $\omega = 2\Delta-4J$. 
The scar-tower states are subject to the emergent Rydberg-blockade constraints that are absent in $H_{\text{DW}}$: Two neighboring spins can not both be in the up states. For the constructed Liouvillian, we therefore take $P_j = (\ket{\uparrow\uparrow}\bra{\uparrow\uparrow})_{j,j+1}, V_{j,j+1}=\sigma_j^x\sigma_{j+1}^x $, such that $L_j = V_{j,j+1}P_{j}=\sigma^-_j \sigma^-_{j+1}$ (In \cite{SuppMaterials} we show that $\{P_j\}$ are the only two-local projectors annihilating the scar tower).
We emphasize that $[H_0, P_j]\neq 0$, and the null space $W'$ of $\{P_j\}$ is \textit{exponentially} large with respect to $L$ \cite{Turner2018weak,Turner2018quantum}, while the dimension of the scar subspace $W$ is only $L/2+2$ \cite{SuppMaterials}. The Hamiltonian part of the Liouvillian, $H_{\text{DW}}$, thus plays an indispensable role in creating a scar-state-only decoherence-free subspace, which we demonstrate through the following Liouvillian dynamics.
We use the quantum jump rate $\textnormal{Tr}[\sum_j P_j\rho(t)]/L$ to characterize whether a state has reached the null space (zero value implies the state is within $W'$). As shown in Fig.~\ref{fig:spectrum_dynamics}(d), for an initial state in $W'$ but out of $W$ (blue solid line), the quantum jump rate increases up from zero, then decays back to zero, indicating that the state is driven out of $W'$ by the Hamiltonian part and converges to the scar subspace due to dissipation of other eigenmodes. As a comparison, an initial state out of the null space is driven into the scar subspace directly (red dashed line). Meanwhile, we compute the dynamics of the scar subspace overlap for these two initial states, which approaches one monotonically [Fig.~\ref{fig:spectrum_dynamics}(e)]. More numerical results are displayed in~\cite{SuppMaterials}.

\begin{figure}
\hspace*{-0.48\textwidth}
\includegraphics[width=0.48\textwidth]{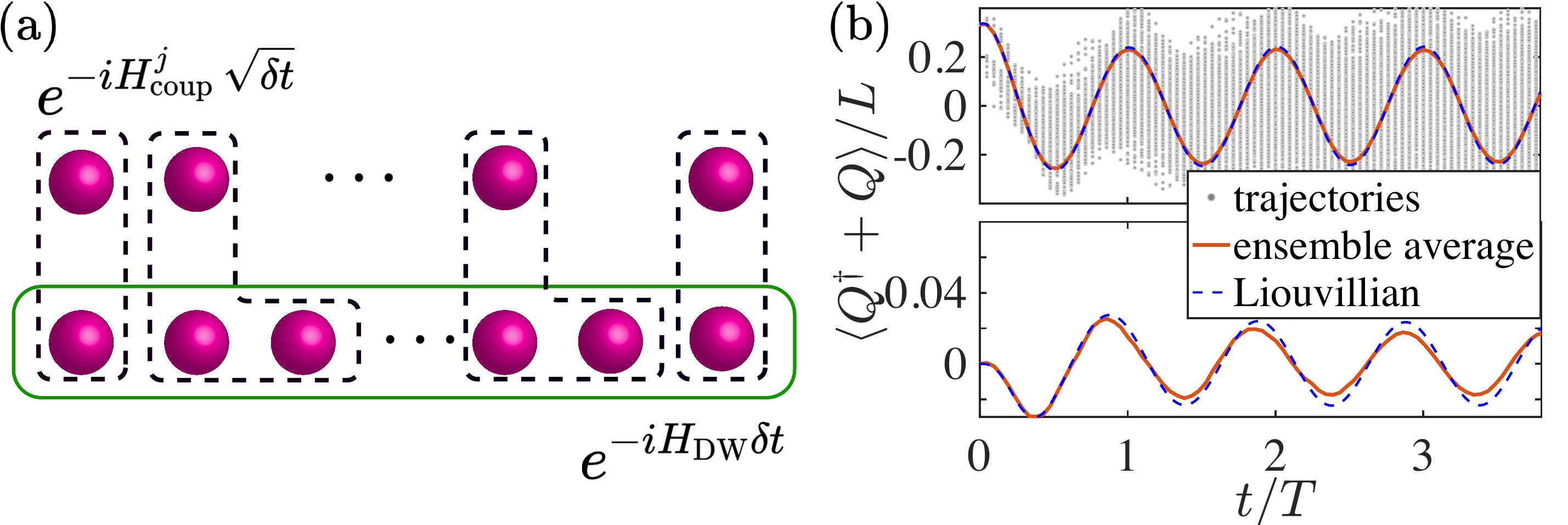} 
\caption{(a) An illustration of the experimental scheme to implement the dissipative scarred dynamics. (b) Observable dynamics simulated by the quantum trajectory method with the initial state $\exp(i \Sigma_j \theta_j \sigma^x_j) \left(\prod_{j=2}^{L-1} [ 1 + (-1)^j P^0_{j-1} \sigma_j^{+} P^0_{j+1} ]|\downarrow \cdots \downarrow\rangle\right)$~\cite{Iadecola2020Quantum}, $\theta_j \in [0,0.2 \pi]$ (top panel) and $|\uparrow \uparrow \uparrow \uparrow \downarrow\downarrow\downarrow \downarrow\rangle$ (bottom panel, trajectories omitted due to the plot range). For both panels we take $1000$ trajectories. $\delta t = 0.1$, $L_{j=1,L}= \sigma_{j}^-$, other $L_{j\neq 1,L} = \sigma_j^- \sigma_{j+1}^-$, $L=8, \Delta=0.5, J=1, \gamma=1$. 
}
\label{fig:Experiment}
\end{figure}

\textit{Experimental realization.}-- 
The dissipative scarred dynamics can be readily implemented~\cite{Lloyd2001Engineering,Bacon2001Universal} using currently available quantum simulation technologies, as we demonstrate with the domain-wall preserving model below.
Consider a one-dimensional qubit chain coupled to another array of ancilla qubits [Fig.~\ref{fig:Experiment}(a)]. We digitally simulate the Liouvillian evolution through three steps, similar to the formalism of quantum collision models~\cite{Ciccarello2022Quantum,Guarnieri2022Time,Cattaneo2022Brief,Gillman2023Using} : 
Suppose at time $t$ the entire system has a quantum state in the decoupled form: $\ket{\psi(t)}\otimes\ket{\downarrow\cdots\downarrow}$, with all the ancilla qubits set to $\ket{\downarrow}$. (1) We apply the unitary operator $\exp(-i H_{\text{DW}} \delta t)$ (could be Trotterized to local gates) on $\ket{\psi(t)}$, which plays the role of Hermitian Hamiltonian evolution; 
(2) We then apply the local unitary gates $\prod_j \exp(-i H_\text{coup}^j \sqrt{\delta t} )$ with 
\begin{equation}
    H_\text{coup}^j = \sqrt{2\gamma} (L_j \tau_j^+ + L_j^\dagger \tau_j^-),
\end{equation}
which couple the system and ancilla ($\tau^{\pm}_j$) qubits to create probabilistic quantum jumps induced by $\{L_j\}$; (3) Finally we reset all the ancilla qubits back to $\ket{\downarrow}$ via measurements or optical pumping~\cite{Shankar2013Autonomously,Han2021Experimental,Cai2021High,Google2023Stable,Barreiro2011Open,Lin2013Dissipative,Schindler2013quantum,Weimer2010Rydberg}. We rigorously prove that the above protocol faithfully reproduces the many-body Liouvillian dynamics up to error of order $O(\delta t^2)$ \cite{SuppMaterials}. Moreover, we numerically simulate the three-step dynamical process by the quantum trajectory method~\cite{Daley2014Quantum}. As shown in Fig.~\ref{fig:Experiment}(b), with a moderate $\delta t$, the observable dynamics of $Q^\dagger + Q = \sum_j (-1)^j P^0_{j-1} \sigma_j^x P^0_{j+1}$ (requiring only two measurement settings) computed by the ensemble average of trajectories agree well with the exact Liouvillian evolution. We particularly choose two initial states that are easy to prepare on experimental platforms -- The first one mimics an imperfectly prepared bond-dimension-two MPS, and the second one is a product state.

\textit{Conclusions.--} 
In summary, our protocol utilizes the synergy between the dissipators and the Hamiltonian part of the Liouvillian to create a scar-state-only decoherence-free subspace. We systematically obtain local projectors annihilating the whole scar towers by the compressed MPS technique. Meanwhile, maximizing the power of the Hamiltonian part is crucial to keep the designed dissipators as local as possible. 
On the one hand, our framework introduces many-body scarred dynamics into the open quantum system regime. An intriguing advantage compared to the closed-system counterpart is that, the dissipative scarred dynamics is independent of the initial states and naturally tolerate instantaneous perturbations. The constructed decoherence-free subspaces can be utilized to prepare scar states with extensive multipartite entanglement~\cite{Dooley2021Robust,Desaules2022Extensive,Dooley2023Entanglement} by engineered short-range dissipation, which makes them promising candidates for quantum enhanced metrology~\cite{Pezze2018Quantum}.
On the other hand, our work introduces as well new principles and techniques to construct local Liouvillians hosting decoherence-free subspaces with special entanglement structures and equally spaced non-decaying eigenmodes. These scar-state-only decoherence-free subspaces support non-stationary coherent many-body dynamics under dissipation, which have profound connections with certain dissipative kinetically constrained models~\cite{Olmos2012Facilitated,Macieszczak2016Towards} and open up an avenue towards the realization of dissipative time crystals~\cite{Maskara2021discrete,gong2018discrete,kessler2021observation,buvca2019Non,Booker2020Non}.~In the current work the non-decaying scarred eigenmodes of Liouvillians are inherited from the original Hamiltonians. It is also interesting to consider the intrinsic scarred eigenmodes in open quantum systems, which could possibly be distinguished by relatively small operator entanglement~\cite{Zanardi2000Entangling,Zanardi2001Entanglement,Prosen2007Operator,Pizorn2009Operator,Zhou2017Operator}.

We acknowledge helpful discussions with Berislav Bu\v{c}a, Alexey Gorshkov, Fernando Iemini, Francisco Machado, Lei Ying and Yukai Wu, communications with Juan P. Garrahan, and previous collaborations with Thomas Iadecola and Shenglong Xu.  
This work was supported by the National Natural Science Foundation of China (Grants No. 12125405, No. 12075128 and T2225008),  Shanghai Qi Zhi Institute,
the Innovation Program for Quantum Science and Technology (No. 2021ZD0302203, No. 2021ZD0301601, and No. 2021ZD0302502), National Key R$\&$D Program of China (No. 2023YFA1406702), the
Tsinghua University Initiative Scientific Research Program, Tsinghua University Dushi Program, and the Ministry of Education of China.

\bibliographystyle{apsrev4-1-title}
\bibliography{DengQAIGroup,QMBS,Heran}

\begin{thebibliography}{99}%
\makeatletter
\providecommand \@ifxundefined [1]{%
 \@ifx{#1\undefined}
}%
\providecommand \@ifnum [1]{%
 \ifnum #1\expandafter \@firstoftwo
 \else \expandafter \@secondoftwo
 \fi
}%
\providecommand \@ifx [1]{%
 \ifx #1\expandafter \@firstoftwo
 \else \expandafter \@secondoftwo
 \fi
}%
\providecommand \natexlab [1]{#1}%
\providecommand \enquote  [1]{``#1''}%
\providecommand \bibnamefont  [1]{#1}%
\providecommand \bibfnamefont [1]{#1}%
\providecommand \citenamefont [1]{#1}%
\providecommand \href@noop [0]{\@secondoftwo}%
\providecommand \href [0]{\begingroup \@sanitize@url \@href}%
\providecommand \@href[1]{\@@startlink{#1}\@@href}%
\providecommand \@@href[1]{\endgroup#1\@@endlink}%
\providecommand \@sanitize@url [0]{\catcode `\\12\catcode `\$12\catcode
  `\&12\catcode `\#12\catcode `\^12\catcode `\_12\catcode `\%12\relax}%
\providecommand \@@startlink[1]{}%
\providecommand \@@endlink[0]{}%
\providecommand \url  [0]{\begingroup\@sanitize@url \@url }%
\providecommand \@url [1]{\endgroup\@href {#1}{\urlprefix }}%
\providecommand \urlprefix  [0]{URL }%
\providecommand \Eprint [0]{\href }%
\providecommand \doibase [0]{http://dx.doi.org/}%
\providecommand \selectlanguage [0]{\@gobble}%
\providecommand \bibinfo  [0]{\@secondoftwo}%
\providecommand \bibfield  [0]{\@secondoftwo}%
\providecommand \translation [1]{[#1]}%
\providecommand \BibitemOpen [0]{}%
\providecommand \bibitemStop [0]{}%
\providecommand \bibitemNoStop [0]{.\EOS\space}%
\providecommand \EOS [0]{\spacefactor3000\relax}%
\providecommand \BibitemShut  [1]{\csname bibitem#1\endcsname}%
\let\auto@bib@innerbib\@empty
\bibitem [{\citenamefont {Deutsch}(1991)}]{Deutsch1991Quantum}%
  \BibitemOpen
  \bibfield  {author} {\bibinfo {author} {\bibfnamefont {J.~M.}\ \bibnamefont
  {Deutsch}},\ }\bibfield  {title} {\enquote {\bibinfo {title} {Quantum
  statistical mechanics in a closed system},}\ }\href
  {https://link.aps.org/doi/10.1103/PhysRevA.43.2046} {\bibfield  {journal}
  {\bibinfo  {journal} {Phys. Rev. A}\ }\textbf {\bibinfo {volume} {43}},\
  \bibinfo {pages} {2046} (\bibinfo {year} {1991})}\BibitemShut {NoStop}%
\bibitem [{\citenamefont {Srednicki}(1994)}]{Srednicki1994Chaos}%
  \BibitemOpen
  \bibfield  {author} {\bibinfo {author} {\bibfnamefont {M.}~\bibnamefont
  {Srednicki}},\ }\bibfield  {title} {\enquote {\bibinfo {title} {Chaos and
  quantum thermalization},}\ }\href {\doibase 10.1103/PhysRevE.50.888}
  {\bibfield  {journal} {\bibinfo  {journal} {Phys. Rev. E}\ }\textbf {\bibinfo
  {volume} {50}},\ \bibinfo {pages} {888} (\bibinfo {year} {1994})}\BibitemShut
  {NoStop}%
\bibitem [{\citenamefont {Serbyn}\ \emph {et~al.}(2021)\citenamefont {Serbyn},
  \citenamefont {Abanin},\ and\ \citenamefont {Papi{\'c}}}]{Serbyn2021quantum}%
  \BibitemOpen
  \bibfield  {author} {\bibinfo {author} {\bibfnamefont {M.}~\bibnamefont
  {Serbyn}}, \bibinfo {author} {\bibfnamefont {D.~A.}\ \bibnamefont {Abanin}},
  \ and\ \bibinfo {author} {\bibfnamefont {Z.}~\bibnamefont {Papi{\'c}}},\
  }\bibfield  {title} {\enquote {\bibinfo {title} {Quantum many-body scars and
  weak breaking of ergodicity},}\ }\href
  {https://www.nature.com/articles/s41567-021-01230-2} {\bibfield  {journal}
  {\bibinfo  {journal} {Nat. Phys.}\ }\textbf {\bibinfo {volume} {17}},\
  \bibinfo {pages} {675} (\bibinfo {year} {2021})}\BibitemShut {NoStop}%
\bibitem [{\citenamefont {Moudgalya}\ \emph {et~al.}(2022)\citenamefont
  {Moudgalya}, \citenamefont {Bernevig},\ and\ \citenamefont
  {Regnault}}]{Moudgalya2022Quantum}%
  \BibitemOpen
  \bibfield  {author} {\bibinfo {author} {\bibfnamefont {S.}~\bibnamefont
  {Moudgalya}}, \bibinfo {author} {\bibfnamefont {B.~A.}\ \bibnamefont
  {Bernevig}}, \ and\ \bibinfo {author} {\bibfnamefont {N.}~\bibnamefont
  {Regnault}},\ }\bibfield  {title} {\enquote {\bibinfo {title} {Quantum
  many-body scars and hilbert space fragmentation: A review of exact
  results},}\ }\href
  {https://iopscience.iop.org/article/10.1088/1361-6633/ac73a0} {\bibfield
  {journal} {\bibinfo  {journal} {Rep. Prog. Phys.}\ } (\bibinfo {year}
  {2022})}\BibitemShut {NoStop}%
\bibitem [{\citenamefont {Chandran}\ \emph {et~al.}(2022)\citenamefont
  {Chandran}, \citenamefont {Iadecola}, \citenamefont {Khemani},\ and\
  \citenamefont {Moessner}}]{Chandran2022Quantum}%
  \BibitemOpen
  \bibfield  {author} {\bibinfo {author} {\bibfnamefont {A.}~\bibnamefont
  {Chandran}}, \bibinfo {author} {\bibfnamefont {T.}~\bibnamefont {Iadecola}},
  \bibinfo {author} {\bibfnamefont {V.}~\bibnamefont {Khemani}}, \ and\
  \bibinfo {author} {\bibfnamefont {R.}~\bibnamefont {Moessner}},\ }\bibfield
  {title} {\enquote {\bibinfo {title} {Quantum many-body scars: A quasiparticle
  perspective},}\ }\href
  {https://www.annualreviews.org/doi/abs/10.1146/annurev-conmatphys-031620-101617}
  {\bibfield  {journal} {\bibinfo  {journal} {Annu. Rev. Condens. Matter
  Phys.}\ }\textbf {\bibinfo {volume} {14}} (\bibinfo {year}
  {2022})}\BibitemShut {NoStop}%
\bibitem [{\citenamefont {Bernien}\ \emph {et~al.}(2017)\citenamefont
  {Bernien}, \citenamefont {Schwartz}, \citenamefont {Keesling}, \citenamefont
  {Levine}, \citenamefont {Omran}, \citenamefont {Pichler}, \citenamefont
  {Choi}, \citenamefont {Zibrov}, \citenamefont {Endres}, \citenamefont
  {Greiner} \emph {et~al.}}]{Bernien2017Probing}%
  \BibitemOpen
  \bibfield  {author} {\bibinfo {author} {\bibfnamefont {H.}~\bibnamefont
  {Bernien}}, \bibinfo {author} {\bibfnamefont {S.}~\bibnamefont {Schwartz}},
  \bibinfo {author} {\bibfnamefont {A.}~\bibnamefont {Keesling}}, \bibinfo
  {author} {\bibfnamefont {H.}~\bibnamefont {Levine}}, \bibinfo {author}
  {\bibfnamefont {A.}~\bibnamefont {Omran}}, \bibinfo {author} {\bibfnamefont
  {H.}~\bibnamefont {Pichler}}, \bibinfo {author} {\bibfnamefont
  {S.}~\bibnamefont {Choi}}, \bibinfo {author} {\bibfnamefont {A.~S.}\
  \bibnamefont {Zibrov}}, \bibinfo {author} {\bibfnamefont {M.}~\bibnamefont
  {Endres}}, \bibinfo {author} {\bibfnamefont {M.}~\bibnamefont {Greiner}},
  \emph {et~al.},\ }\bibfield  {title} {\enquote {\bibinfo {title} {Probing
  many-body dynamics on a 51-atom quantum simulator},}\ }\href
  {https://www.nature.com/articles/nature24622} {\bibfield  {journal} {\bibinfo
   {journal} {Nature}\ }\textbf {\bibinfo {volume} {551}},\ \bibinfo {pages}
  {579} (\bibinfo {year} {2017})}\BibitemShut {NoStop}%
\bibitem [{\citenamefont {Bluvstein}\ \emph {et~al.}(2021)\citenamefont
  {Bluvstein}, \citenamefont {Omran}, \citenamefont {Levine}, \citenamefont
  {Keesling}, \citenamefont {Semeghini}, \citenamefont {Ebadi}, \citenamefont
  {Wang}, \citenamefont {Michailidis}, \citenamefont {Maskara}, \citenamefont
  {Ho} \emph {et~al.}}]{Bluvstein2021Controlling}%
  \BibitemOpen
  \bibfield  {author} {\bibinfo {author} {\bibfnamefont {D.}~\bibnamefont
  {Bluvstein}}, \bibinfo {author} {\bibfnamefont {A.}~\bibnamefont {Omran}},
  \bibinfo {author} {\bibfnamefont {H.}~\bibnamefont {Levine}}, \bibinfo
  {author} {\bibfnamefont {A.}~\bibnamefont {Keesling}}, \bibinfo {author}
  {\bibfnamefont {G.}~\bibnamefont {Semeghini}}, \bibinfo {author}
  {\bibfnamefont {S.}~\bibnamefont {Ebadi}}, \bibinfo {author} {\bibfnamefont
  {T.~T.}\ \bibnamefont {Wang}}, \bibinfo {author} {\bibfnamefont {A.~A.}\
  \bibnamefont {Michailidis}}, \bibinfo {author} {\bibfnamefont
  {N.}~\bibnamefont {Maskara}}, \bibinfo {author} {\bibfnamefont {W.~W.}\
  \bibnamefont {Ho}},  \emph {et~al.},\ }\bibfield  {title} {\enquote {\bibinfo
  {title} {Controlling quantum many-body dynamics in driven rydberg atom
  arrays},}\ }\href {https://www.science.org/doi/10.1126/science.abg2530}
  {\bibfield  {journal} {\bibinfo  {journal} {Science}\ }\textbf {\bibinfo
  {volume} {371}},\ \bibinfo {pages} {1355} (\bibinfo {year}
  {2021})}\BibitemShut {NoStop}%
\bibitem [{\citenamefont {Turner}\ \emph
  {et~al.}(2018{\natexlab{a}})\citenamefont {Turner}, \citenamefont
  {Michailidis}, \citenamefont {Abanin}, \citenamefont {Serbyn},\ and\
  \citenamefont {Papi{\'c}}}]{Turner2018weak}%
  \BibitemOpen
  \bibfield  {author} {\bibinfo {author} {\bibfnamefont {C.~J.}\ \bibnamefont
  {Turner}}, \bibinfo {author} {\bibfnamefont {A.~A.}\ \bibnamefont
  {Michailidis}}, \bibinfo {author} {\bibfnamefont {D.~A.}\ \bibnamefont
  {Abanin}}, \bibinfo {author} {\bibfnamefont {M.}~\bibnamefont {Serbyn}}, \
  and\ \bibinfo {author} {\bibfnamefont {Z.}~\bibnamefont {Papi{\'c}}},\
  }\bibfield  {title} {\enquote {\bibinfo {title} {Weak ergodicity breaking
  from quantum many-body scars},}\ }\href
  {https://www.nature.com/articles/s41567-018-0137-5} {\bibfield  {journal}
  {\bibinfo  {journal} {Nat. Phys.}\ }\textbf {\bibinfo {volume} {14}},\
  \bibinfo {pages} {745} (\bibinfo {year} {2018}{\natexlab{a}})}\BibitemShut
  {NoStop}%
\bibitem [{\citenamefont {Turner}\ \emph
  {et~al.}(2018{\natexlab{b}})\citenamefont {Turner}, \citenamefont
  {Michailidis}, \citenamefont {Abanin}, \citenamefont {Serbyn},\ and\
  \citenamefont {Papi\ifmmode~\acute{c}\else \'{c}\fi{}}}]{Turner2018quantum}%
  \BibitemOpen
  \bibfield  {author} {\bibinfo {author} {\bibfnamefont {C.~J.}\ \bibnamefont
  {Turner}}, \bibinfo {author} {\bibfnamefont {A.~A.}\ \bibnamefont
  {Michailidis}}, \bibinfo {author} {\bibfnamefont {D.~A.}\ \bibnamefont
  {Abanin}}, \bibinfo {author} {\bibfnamefont {M.}~\bibnamefont {Serbyn}}, \
  and\ \bibinfo {author} {\bibfnamefont {Z.}~\bibnamefont
  {Papi\ifmmode~\acute{c}\else \'{c}\fi{}}},\ }\bibfield  {title} {\enquote
  {\bibinfo {title} {Quantum scarred eigenstates in a rydberg atom chain:
  Entanglement, breakdown of thermalization, and stability to perturbations},}\
  }\href {\doibase 10.1103/PhysRevB.98.155134} {\bibfield  {journal} {\bibinfo
  {journal} {Phys. Rev. B}\ }\textbf {\bibinfo {volume} {98}},\ \bibinfo
  {pages} {155134} (\bibinfo {year} {2018}{\natexlab{b}})}\BibitemShut
  {NoStop}%
\bibitem [{\citenamefont {Moudgalya}\ \emph
  {et~al.}(2018{\natexlab{a}})\citenamefont {Moudgalya}, \citenamefont
  {Rachel}, \citenamefont {Bernevig},\ and\ \citenamefont
  {Regnault}}]{Moudgalya2018Exact}%
  \BibitemOpen
  \bibfield  {author} {\bibinfo {author} {\bibfnamefont {S.}~\bibnamefont
  {Moudgalya}}, \bibinfo {author} {\bibfnamefont {S.}~\bibnamefont {Rachel}},
  \bibinfo {author} {\bibfnamefont {B.~A.}\ \bibnamefont {Bernevig}}, \ and\
  \bibinfo {author} {\bibfnamefont {N.}~\bibnamefont {Regnault}},\ }\bibfield
  {title} {\enquote {\bibinfo {title} {Exact excited states of nonintegrable
  models},}\ }\href {\doibase 10.1103/PhysRevB.98.235155} {\bibfield  {journal}
  {\bibinfo  {journal} {Phys. Rev. B}\ }\textbf {\bibinfo {volume} {98}},\
  \bibinfo {pages} {235155} (\bibinfo {year} {2018}{\natexlab{a}})}\BibitemShut
  {NoStop}%
\bibitem [{\citenamefont {Moudgalya}\ \emph
  {et~al.}(2018{\natexlab{b}})\citenamefont {Moudgalya}, \citenamefont
  {Regnault},\ and\ \citenamefont {Bernevig}}]{Moudgalya2018Entanglement}%
  \BibitemOpen
  \bibfield  {author} {\bibinfo {author} {\bibfnamefont {S.}~\bibnamefont
  {Moudgalya}}, \bibinfo {author} {\bibfnamefont {N.}~\bibnamefont {Regnault}},
  \ and\ \bibinfo {author} {\bibfnamefont {B.~A.}\ \bibnamefont {Bernevig}},\
  }\bibfield  {title} {\enquote {\bibinfo {title} {Entanglement of exact
  excited states of affleck-kennedy-lieb-tasaki models: Exact results,
  many-body scars, and violation of the strong eigenstate thermalization
  hypothesis},}\ }\href {\doibase 10.1103/PhysRevB.98.235156} {\bibfield
  {journal} {\bibinfo  {journal} {Phys. Rev. B}\ }\textbf {\bibinfo {volume}
  {98}},\ \bibinfo {pages} {235156} (\bibinfo {year}
  {2018}{\natexlab{b}})}\BibitemShut {NoStop}%
\bibitem [{\citenamefont {Schecter}\ and\ \citenamefont
  {Iadecola}(2019)}]{Schecter2019Weak}%
  \BibitemOpen
  \bibfield  {author} {\bibinfo {author} {\bibfnamefont {M.}~\bibnamefont
  {Schecter}}\ and\ \bibinfo {author} {\bibfnamefont {T.}~\bibnamefont
  {Iadecola}},\ }\bibfield  {title} {\enquote {\bibinfo {title} {Weak
  ergodicity breaking and quantum many-body scars in spin-1 $xy$ magnets},}\
  }\href {\doibase 10.1103/PhysRevLett.123.147201} {\bibfield  {journal}
  {\bibinfo  {journal} {Phys. Rev. Lett.}\ }\textbf {\bibinfo {volume} {123}},\
  \bibinfo {pages} {147201} (\bibinfo {year} {2019})}\BibitemShut {NoStop}%
\bibitem [{\citenamefont {Choi}\ \emph {et~al.}(2019)\citenamefont {Choi},
  \citenamefont {Turner}, \citenamefont {Pichler}, \citenamefont {Ho},
  \citenamefont {Michailidis}, \citenamefont {Papi\ifmmode~\acute{c}\else
  \'{c}\fi{}}, \citenamefont {Serbyn}, \citenamefont {Lukin},\ and\
  \citenamefont {Abanin}}]{Choi2019emergent}%
  \BibitemOpen
  \bibfield  {author} {\bibinfo {author} {\bibfnamefont {S.}~\bibnamefont
  {Choi}}, \bibinfo {author} {\bibfnamefont {C.~J.}\ \bibnamefont {Turner}},
  \bibinfo {author} {\bibfnamefont {H.}~\bibnamefont {Pichler}}, \bibinfo
  {author} {\bibfnamefont {W.~W.}\ \bibnamefont {Ho}}, \bibinfo {author}
  {\bibfnamefont {A.~A.}\ \bibnamefont {Michailidis}}, \bibinfo {author}
  {\bibfnamefont {Z.}~\bibnamefont {Papi\ifmmode~\acute{c}\else \'{c}\fi{}}},
  \bibinfo {author} {\bibfnamefont {M.}~\bibnamefont {Serbyn}}, \bibinfo
  {author} {\bibfnamefont {M.~D.}\ \bibnamefont {Lukin}}, \ and\ \bibinfo
  {author} {\bibfnamefont {D.~A.}\ \bibnamefont {Abanin}},\ }\bibfield  {title}
  {\enquote {\bibinfo {title} {Emergent su(2) dynamics and perfect quantum
  many-body scars},}\ }\href {\doibase 10.1103/PhysRevLett.122.220603}
  {\bibfield  {journal} {\bibinfo  {journal} {Phys. Rev. Lett.}\ }\textbf
  {\bibinfo {volume} {122}},\ \bibinfo {pages} {220603} (\bibinfo {year}
  {2019})}\BibitemShut {NoStop}%
\bibitem [{\citenamefont {Iadecola}\ and\ \citenamefont
  {Schecter}(2020)}]{Iadecola2020Quantum}%
  \BibitemOpen
  \bibfield  {author} {\bibinfo {author} {\bibfnamefont {T.}~\bibnamefont
  {Iadecola}}\ and\ \bibinfo {author} {\bibfnamefont {M.}~\bibnamefont
  {Schecter}},\ }\bibfield  {title} {\enquote {\bibinfo {title} {Quantum
  many-body scar states with emergent kinetic constraints and
  finite-entanglement revivals},}\ }\href
  {https://link.aps.org/doi/10.1103/PhysRevB.101.024306} {\bibfield  {journal}
  {\bibinfo  {journal} {Phys. Rev. B}\ }\textbf {\bibinfo {volume} {101}},\
  \bibinfo {pages} {024306} (\bibinfo {year} {2020})}\BibitemShut {NoStop}%
\bibitem [{\citenamefont {Shibata}\ \emph {et~al.}(2020)\citenamefont
  {Shibata}, \citenamefont {Yoshioka},\ and\ \citenamefont
  {Katsura}}]{shibata2020onsager}%
  \BibitemOpen
  \bibfield  {author} {\bibinfo {author} {\bibfnamefont {N.}~\bibnamefont
  {Shibata}}, \bibinfo {author} {\bibfnamefont {N.}~\bibnamefont {Yoshioka}}, \
  and\ \bibinfo {author} {\bibfnamefont {H.}~\bibnamefont {Katsura}},\
  }\bibfield  {title} {\enquote {\bibinfo {title} {Onsager's scars in
  disordered spin chains},}\ }\href {\doibase 10.1103/PhysRevLett.124.180604}
  {\bibfield  {journal} {\bibinfo  {journal} {Phys. Rev. Lett.}\ }\textbf
  {\bibinfo {volume} {124}},\ \bibinfo {pages} {180604} (\bibinfo {year}
  {2020})}\BibitemShut {NoStop}%
\bibitem [{\citenamefont {Chattopadhyay}\ \emph {et~al.}(2020)\citenamefont
  {Chattopadhyay}, \citenamefont {Pichler}, \citenamefont {Lukin},\ and\
  \citenamefont {Ho}}]{Chattopadhyay2020quantum}%
  \BibitemOpen
  \bibfield  {author} {\bibinfo {author} {\bibfnamefont {S.}~\bibnamefont
  {Chattopadhyay}}, \bibinfo {author} {\bibfnamefont {H.}~\bibnamefont
  {Pichler}}, \bibinfo {author} {\bibfnamefont {M.~D.}\ \bibnamefont {Lukin}},
  \ and\ \bibinfo {author} {\bibfnamefont {W.~W.}\ \bibnamefont {Ho}},\
  }\bibfield  {title} {\enquote {\bibinfo {title} {Quantum many-body scars from
  virtual entangled pairs},}\ }\href {\doibase 10.1103/PhysRevB.101.174308}
  {\bibfield  {journal} {\bibinfo  {journal} {Phys. Rev. B}\ }\textbf {\bibinfo
  {volume} {101}},\ \bibinfo {pages} {174308} (\bibinfo {year}
  {2020})}\BibitemShut {NoStop}%
\bibitem [{\citenamefont {Shiraishi}\ and\ \citenamefont
  {Mori}(2017)}]{Shiraishi2017Systematic}%
  \BibitemOpen
  \bibfield  {author} {\bibinfo {author} {\bibfnamefont {N.}~\bibnamefont
  {Shiraishi}}\ and\ \bibinfo {author} {\bibfnamefont {T.}~\bibnamefont
  {Mori}},\ }\bibfield  {title} {\enquote {\bibinfo {title} {Systematic
  construction of counterexamples to the eigenstate thermalization
  hypothesis},}\ }\href {\doibase 10.1103/PhysRevLett.119.030601} {\bibfield
  {journal} {\bibinfo  {journal} {Phys. Rev. Lett.}\ }\textbf {\bibinfo
  {volume} {119}},\ \bibinfo {pages} {030601} (\bibinfo {year}
  {2017})}\BibitemShut {NoStop}%
\bibitem [{\citenamefont {Lin}\ and\ \citenamefont
  {Motrunich}(2019)}]{lin2019exact}%
  \BibitemOpen
  \bibfield  {author} {\bibinfo {author} {\bibfnamefont {C.-J.}\ \bibnamefont
  {Lin}}\ and\ \bibinfo {author} {\bibfnamefont {O.~I.}\ \bibnamefont
  {Motrunich}},\ }\bibfield  {title} {\enquote {\bibinfo {title} {Exact quantum
  many-body scar states in the rydberg-blockaded atom chain},}\ }\href
  {https://journals.aps.org/prl/abstract/10.1103/PhysRevLett.122.173401}
  {\bibfield  {journal} {\bibinfo  {journal} {Phys. Rev. Lett.}\ }\textbf
  {\bibinfo {volume} {122}},\ \bibinfo {pages} {173401} (\bibinfo {year}
  {2019})}\BibitemShut {NoStop}%
\bibitem [{\citenamefont {Ok}\ \emph {et~al.}(2019)\citenamefont {Ok},
  \citenamefont {Choo}, \citenamefont {Mudry}, \citenamefont {Castelnovo},
  \citenamefont {Chamon},\ and\ \citenamefont {Neupert}}]{ok2019topological}%
  \BibitemOpen
  \bibfield  {author} {\bibinfo {author} {\bibfnamefont {S.}~\bibnamefont
  {Ok}}, \bibinfo {author} {\bibfnamefont {K.}~\bibnamefont {Choo}}, \bibinfo
  {author} {\bibfnamefont {C.}~\bibnamefont {Mudry}}, \bibinfo {author}
  {\bibfnamefont {C.}~\bibnamefont {Castelnovo}}, \bibinfo {author}
  {\bibfnamefont {C.}~\bibnamefont {Chamon}}, \ and\ \bibinfo {author}
  {\bibfnamefont {T.}~\bibnamefont {Neupert}},\ }\bibfield  {title} {\enquote
  {\bibinfo {title} {Topological many-body scar states in dimensions one, two,
  and three},}\ }\href
  {https://journals.aps.org/prresearch/abstract/10.1103/PhysRevResearch.1.033144}
  {\bibfield  {journal} {\bibinfo  {journal} {Phys. Rev. Research}\ }\textbf
  {\bibinfo {volume} {1}},\ \bibinfo {pages} {033144} (\bibinfo {year}
  {2019})}\BibitemShut {NoStop}%
\bibitem [{\citenamefont {Bull}\ \emph {et~al.}(2019)\citenamefont {Bull},
  \citenamefont {Martin},\ and\ \citenamefont
  {Papi{\'c}}}]{bull2019systematic}%
  \BibitemOpen
  \bibfield  {author} {\bibinfo {author} {\bibfnamefont {K.}~\bibnamefont
  {Bull}}, \bibinfo {author} {\bibfnamefont {I.}~\bibnamefont {Martin}}, \ and\
  \bibinfo {author} {\bibfnamefont {Z.}~\bibnamefont {Papi{\'c}}},\ }\bibfield
  {title} {\enquote {\bibinfo {title} {Systematic construction of scarred
  many-body dynamics in 1d lattice models},}\ }\href
  {https://journals.aps.org/prl/abstract/10.1103/PhysRevLett.123.030601}
  {\bibfield  {journal} {\bibinfo  {journal} {Phys. Rev. Lett.}\ }\textbf
  {\bibinfo {volume} {123}},\ \bibinfo {pages} {030601} (\bibinfo {year}
  {2019})}\BibitemShut {NoStop}%
\bibitem [{\citenamefont {Surace}\ \emph {et~al.}(2020)\citenamefont {Surace},
  \citenamefont {Mazza}, \citenamefont {Giudici}, \citenamefont {Lerose},
  \citenamefont {Gambassi},\ and\ \citenamefont
  {Dalmonte}}]{Surace2020Lattice}%
  \BibitemOpen
  \bibfield  {author} {\bibinfo {author} {\bibfnamefont {F.~M.}\ \bibnamefont
  {Surace}}, \bibinfo {author} {\bibfnamefont {P.~P.}\ \bibnamefont {Mazza}},
  \bibinfo {author} {\bibfnamefont {G.}~\bibnamefont {Giudici}}, \bibinfo
  {author} {\bibfnamefont {A.}~\bibnamefont {Lerose}}, \bibinfo {author}
  {\bibfnamefont {A.}~\bibnamefont {Gambassi}}, \ and\ \bibinfo {author}
  {\bibfnamefont {M.}~\bibnamefont {Dalmonte}},\ }\bibfield  {title} {\enquote
  {\bibinfo {title} {Lattice gauge theories and string dynamics in rydberg atom
  quantum simulators},}\ }\href {\doibase 10.1103/PhysRevX.10.021041}
  {\bibfield  {journal} {\bibinfo  {journal} {Phys. Rev. X}\ }\textbf {\bibinfo
  {volume} {10}},\ \bibinfo {pages} {021041} (\bibinfo {year}
  {2020})}\BibitemShut {NoStop}%
\bibitem [{\citenamefont {Hudomal}\ \emph {et~al.}(2020)\citenamefont
  {Hudomal}, \citenamefont {Vasi{\'c}}, \citenamefont {Regnault},\ and\
  \citenamefont {Papi{\'c}}}]{hudomal2020quantum}%
  \BibitemOpen
  \bibfield  {author} {\bibinfo {author} {\bibfnamefont {A.}~\bibnamefont
  {Hudomal}}, \bibinfo {author} {\bibfnamefont {I.}~\bibnamefont {Vasi{\'c}}},
  \bibinfo {author} {\bibfnamefont {N.}~\bibnamefont {Regnault}}, \ and\
  \bibinfo {author} {\bibfnamefont {Z.}~\bibnamefont {Papi{\'c}}},\ }\bibfield
  {title} {\enquote {\bibinfo {title} {Quantum scars of bosons with correlated
  hopping},}\ }\href {https://www.nature.com/articles/s42005-020-0364-9}
  {\bibfield  {journal} {\bibinfo  {journal} {Commun. Phys.}\ }\textbf
  {\bibinfo {volume} {3}},\ \bibinfo {pages} {1} (\bibinfo {year}
  {2020})}\BibitemShut {NoStop}%
\bibitem [{\citenamefont {Moudgalya}\ \emph {et~al.}(2020)\citenamefont
  {Moudgalya}, \citenamefont {Regnault},\ and\ \citenamefont
  {Bernevig}}]{Moudgalya2020eta}%
  \BibitemOpen
  \bibfield  {author} {\bibinfo {author} {\bibfnamefont {S.}~\bibnamefont
  {Moudgalya}}, \bibinfo {author} {\bibfnamefont {N.}~\bibnamefont {Regnault}},
  \ and\ \bibinfo {author} {\bibfnamefont {B.~A.}\ \bibnamefont {Bernevig}},\
  }\bibfield  {title} {\enquote {\bibinfo {title} {$\ensuremath{\eta}$-pairing
  in hubbard models: From spectrum generating algebras to quantum many-body
  scars},}\ }\href {\doibase 10.1103/PhysRevB.102.085140} {\bibfield  {journal}
  {\bibinfo  {journal} {Phys. Rev. B}\ }\textbf {\bibinfo {volume} {102}},\
  \bibinfo {pages} {085140} (\bibinfo {year} {2020})}\BibitemShut {NoStop}%
\bibitem [{\citenamefont {Mark}\ and\ \citenamefont
  {Motrunich}(2020)}]{Mark2020eta}%
  \BibitemOpen
  \bibfield  {author} {\bibinfo {author} {\bibfnamefont {D.~K.}\ \bibnamefont
  {Mark}}\ and\ \bibinfo {author} {\bibfnamefont {O.~I.}\ \bibnamefont
  {Motrunich}},\ }\bibfield  {title} {\enquote {\bibinfo {title}
  {$\ensuremath{\eta}$-pairing states as true scars in an extended hubbard
  model},}\ }\href {\doibase 10.1103/PhysRevB.102.075132} {\bibfield  {journal}
  {\bibinfo  {journal} {Phys. Rev. B}\ }\textbf {\bibinfo {volume} {102}},\
  \bibinfo {pages} {075132} (\bibinfo {year} {2020})}\BibitemShut {NoStop}%
\bibitem [{\citenamefont {Scherg}\ \emph {et~al.}(2021)\citenamefont {Scherg},
  \citenamefont {Kohlert}, \citenamefont {Sala}, \citenamefont {Pollmann},
  \citenamefont {Madhusudhana}, \citenamefont {Bloch},\ and\ \citenamefont
  {Aidelsburger}}]{Scherg2021Observing}%
  \BibitemOpen
  \bibfield  {author} {\bibinfo {author} {\bibfnamefont {S.}~\bibnamefont
  {Scherg}}, \bibinfo {author} {\bibfnamefont {T.}~\bibnamefont {Kohlert}},
  \bibinfo {author} {\bibfnamefont {P.}~\bibnamefont {Sala}}, \bibinfo {author}
  {\bibfnamefont {F.}~\bibnamefont {Pollmann}}, \bibinfo {author}
  {\bibfnamefont {B.~H.}\ \bibnamefont {Madhusudhana}}, \bibinfo {author}
  {\bibfnamefont {I.}~\bibnamefont {Bloch}}, \ and\ \bibinfo {author}
  {\bibfnamefont {M.}~\bibnamefont {Aidelsburger}},\ }\bibfield  {title}
  {\enquote {\bibinfo {title} {Observing non-ergodicity due to kinetic
  constraints in tilted fermi-hubbard chains},}\ }\href
  {https://www.nature.com/articles/s41467-021-24726-0} {\bibfield  {journal}
  {\bibinfo  {journal} {Nat. Commun.}\ }\textbf {\bibinfo {volume} {12}},\
  \bibinfo {pages} {1} (\bibinfo {year} {2021})}\BibitemShut {NoStop}%
\bibitem [{\citenamefont {Desaules}\ \emph {et~al.}(2021)\citenamefont
  {Desaules}, \citenamefont {Hudomal}, \citenamefont {Turner},\ and\
  \citenamefont {Papi\ifmmode~\acute{c}\else
  \'{c}\fi{}}}]{Desaules2021Proposal}%
  \BibitemOpen
  \bibfield  {author} {\bibinfo {author} {\bibfnamefont {J.-Y.}\ \bibnamefont
  {Desaules}}, \bibinfo {author} {\bibfnamefont {A.}~\bibnamefont {Hudomal}},
  \bibinfo {author} {\bibfnamefont {C.~J.}\ \bibnamefont {Turner}}, \ and\
  \bibinfo {author} {\bibfnamefont {Z.}~\bibnamefont
  {Papi\ifmmode~\acute{c}\else \'{c}\fi{}}},\ }\bibfield  {title} {\enquote
  {\bibinfo {title} {Proposal for realizing quantum scars in the tilted 1d
  fermi-hubbard model},}\ }\href {\doibase 10.1103/PhysRevLett.126.210601}
  {\bibfield  {journal} {\bibinfo  {journal} {Phys. Rev. Lett.}\ }\textbf
  {\bibinfo {volume} {126}},\ \bibinfo {pages} {210601} (\bibinfo {year}
  {2021})}\BibitemShut {NoStop}%
\bibitem [{\citenamefont {Banerjee}\ and\ \citenamefont
  {Sen}(2021)}]{Banerjee2021Quantum}%
  \BibitemOpen
  \bibfield  {author} {\bibinfo {author} {\bibfnamefont {D.}~\bibnamefont
  {Banerjee}}\ and\ \bibinfo {author} {\bibfnamefont {A.}~\bibnamefont {Sen}},\
  }\bibfield  {title} {\enquote {\bibinfo {title} {Quantum scars from zero
  modes in an abelian lattice gauge theory on ladders},}\ }\href
  {https://link.aps.org/doi/10.1103/PhysRevLett.126.220601} {\bibfield
  {journal} {\bibinfo  {journal} {Phys. Rev. Lett.}\ }\textbf {\bibinfo
  {volume} {126}},\ \bibinfo {pages} {220601} (\bibinfo {year}
  {2021})}\BibitemShut {NoStop}%
\bibitem [{\citenamefont {Langlett}\ \emph {et~al.}(2022)\citenamefont
  {Langlett}, \citenamefont {Yang}, \citenamefont {Wildeboer}, \citenamefont
  {Gorshkov}, \citenamefont {Iadecola},\ and\ \citenamefont
  {Xu}}]{Langlett2022rainbow}%
  \BibitemOpen
  \bibfield  {author} {\bibinfo {author} {\bibfnamefont {C.~M.}\ \bibnamefont
  {Langlett}}, \bibinfo {author} {\bibfnamefont {Z.-C.}\ \bibnamefont {Yang}},
  \bibinfo {author} {\bibfnamefont {J.}~\bibnamefont {Wildeboer}}, \bibinfo
  {author} {\bibfnamefont {A.~V.}\ \bibnamefont {Gorshkov}}, \bibinfo {author}
  {\bibfnamefont {T.}~\bibnamefont {Iadecola}}, \ and\ \bibinfo {author}
  {\bibfnamefont {S.}~\bibnamefont {Xu}},\ }\bibfield  {title} {\enquote
  {\bibinfo {title} {Rainbow scars: From area to volume law},}\ }\href
  {\doibase 10.1103/PhysRevB.105.L060301} {\bibfield  {journal} {\bibinfo
  {journal} {Phys. Rev. B}\ }\textbf {\bibinfo {volume} {105}},\ \bibinfo
  {pages} {L060301} (\bibinfo {year} {2022})}\BibitemShut {NoStop}%
\bibitem [{\citenamefont {Langlett}\ and\ \citenamefont
  {Xu}(2021)}]{Langlett2021Hilbert}%
  \BibitemOpen
  \bibfield  {author} {\bibinfo {author} {\bibfnamefont {C.~M.}\ \bibnamefont
  {Langlett}}\ and\ \bibinfo {author} {\bibfnamefont {S.}~\bibnamefont {Xu}},\
  }\bibfield  {title} {\enquote {\bibinfo {title} {Hilbert space fragmentation
  and exact scars of generalized fredkin spin chains},}\ }\href
  {https://link.aps.org/doi/10.1103/PhysRevB.103.L220304} {\bibfield  {journal}
  {\bibinfo  {journal} {Phys. Rev. B}\ }\textbf {\bibinfo {volume} {103}},\
  \bibinfo {pages} {L220304} (\bibinfo {year} {2021})}\BibitemShut {NoStop}%
\bibitem [{\citenamefont {Schindler}\ \emph {et~al.}(2022)\citenamefont
  {Schindler}, \citenamefont {Regnault},\ and\ \citenamefont
  {Bernevig}}]{Schindler2022Exact}%
  \BibitemOpen
  \bibfield  {author} {\bibinfo {author} {\bibfnamefont {F.}~\bibnamefont
  {Schindler}}, \bibinfo {author} {\bibfnamefont {N.}~\bibnamefont {Regnault}},
  \ and\ \bibinfo {author} {\bibfnamefont {B.~A.}\ \bibnamefont {Bernevig}},\
  }\bibfield  {title} {\enquote {\bibinfo {title} {Exact quantum scars in the
  chiral nonlinear luttinger liquid},}\ }\href
  {https://link.aps.org/doi/10.1103/PhysRevB.105.035146} {\bibfield  {journal}
  {\bibinfo  {journal} {Phys. Rev. B}\ }\textbf {\bibinfo {volume} {105}},\
  \bibinfo {pages} {035146} (\bibinfo {year} {2022})}\BibitemShut {NoStop}%
\bibitem [{\citenamefont {Desaules}\ \emph {et~al.}(2023)\citenamefont
  {Desaules}, \citenamefont {Banerjee}, \citenamefont {Hudomal}, \citenamefont
  {Papi\ifmmode~\acute{c}\else \'{c}\fi{}}, \citenamefont {Sen},\ and\
  \citenamefont {Halimeh}}]{Desaules2023Weak}%
  \BibitemOpen
  \bibfield  {author} {\bibinfo {author} {\bibfnamefont {J.-Y.}\ \bibnamefont
  {Desaules}}, \bibinfo {author} {\bibfnamefont {D.}~\bibnamefont {Banerjee}},
  \bibinfo {author} {\bibfnamefont {A.}~\bibnamefont {Hudomal}}, \bibinfo
  {author} {\bibfnamefont {Z.}~\bibnamefont {Papi\ifmmode~\acute{c}\else
  \'{c}\fi{}}}, \bibinfo {author} {\bibfnamefont {A.}~\bibnamefont {Sen}}, \
  and\ \bibinfo {author} {\bibfnamefont {J.~C.}\ \bibnamefont {Halimeh}},\
  }\bibfield  {title} {\enquote {\bibinfo {title} {Weak ergodicity breaking in
  the schwinger model},}\ }\href {\doibase 10.1103/PhysRevB.107.L201105}
  {\bibfield  {journal} {\bibinfo  {journal} {Phys. Rev. B}\ }\textbf {\bibinfo
  {volume} {107}},\ \bibinfo {pages} {L201105} (\bibinfo {year}
  {2023})}\BibitemShut {NoStop}%
\bibitem [{\citenamefont {Zhang}\ \emph {et~al.}(2022)\citenamefont {Zhang},
  \citenamefont {Dong}, \citenamefont {Gao}, \citenamefont {Zhao},
  \citenamefont {Hao}, \citenamefont {Desaules}, \citenamefont {Guo},
  \citenamefont {Chen}, \citenamefont {Deng}, \citenamefont {Liu} \emph
  {et~al.}}]{Zhang2022Many}%
  \BibitemOpen
  \bibfield  {author} {\bibinfo {author} {\bibfnamefont {P.}~\bibnamefont
  {Zhang}}, \bibinfo {author} {\bibfnamefont {H.}~\bibnamefont {Dong}},
  \bibinfo {author} {\bibfnamefont {Y.}~\bibnamefont {Gao}}, \bibinfo {author}
  {\bibfnamefont {L.}~\bibnamefont {Zhao}}, \bibinfo {author} {\bibfnamefont
  {J.}~\bibnamefont {Hao}}, \bibinfo {author} {\bibfnamefont {J.-Y.}\
  \bibnamefont {Desaules}}, \bibinfo {author} {\bibfnamefont {Q.}~\bibnamefont
  {Guo}}, \bibinfo {author} {\bibfnamefont {J.}~\bibnamefont {Chen}}, \bibinfo
  {author} {\bibfnamefont {J.}~\bibnamefont {Deng}}, \bibinfo {author}
  {\bibfnamefont {B.}~\bibnamefont {Liu}},  \emph {et~al.},\ }\bibfield
  {title} {\enquote {\bibinfo {title} {Many-body hilbert space scarring on a
  superconducting processor},}\ }\href
  {https://www.nature.com/articles/s41567-022-01784-9} {\bibfield  {journal}
  {\bibinfo  {journal} {Nat. Phys.}\ }\textbf {\bibinfo {volume} {19}},\
  \bibinfo {pages} {120} (\bibinfo {year} {2022})}\BibitemShut {NoStop}%
\bibitem [{\citenamefont {Su}\ \emph {et~al.}(2023)\citenamefont {Su},
  \citenamefont {Sun}, \citenamefont {Hudomal}, \citenamefont {Desaules},
  \citenamefont {Zhou}, \citenamefont {Yang}, \citenamefont {Halimeh},
  \citenamefont {Yuan}, \citenamefont {Papi\ifmmode~\acute{c}\else
  \'{c}\fi{}},\ and\ \citenamefont {Pan}}]{Su2023Observation}%
  \BibitemOpen
  \bibfield  {author} {\bibinfo {author} {\bibfnamefont {G.-X.}\ \bibnamefont
  {Su}}, \bibinfo {author} {\bibfnamefont {H.}~\bibnamefont {Sun}}, \bibinfo
  {author} {\bibfnamefont {A.}~\bibnamefont {Hudomal}}, \bibinfo {author}
  {\bibfnamefont {J.-Y.}\ \bibnamefont {Desaules}}, \bibinfo {author}
  {\bibfnamefont {Z.-Y.}\ \bibnamefont {Zhou}}, \bibinfo {author}
  {\bibfnamefont {B.}~\bibnamefont {Yang}}, \bibinfo {author} {\bibfnamefont
  {J.~C.}\ \bibnamefont {Halimeh}}, \bibinfo {author} {\bibfnamefont {Z.-S.}\
  \bibnamefont {Yuan}}, \bibinfo {author} {\bibfnamefont {Z.}~\bibnamefont
  {Papi\ifmmode~\acute{c}\else \'{c}\fi{}}}, \ and\ \bibinfo {author}
  {\bibfnamefont {J.-W.}\ \bibnamefont {Pan}},\ }\bibfield  {title} {\enquote
  {\bibinfo {title} {Observation of many-body scarring in a bose-hubbard
  quantum simulator},}\ }\href {\doibase 10.1103/PhysRevResearch.5.023010}
  {\bibfield  {journal} {\bibinfo  {journal} {Phys. Rev. Research}\ }\textbf
  {\bibinfo {volume} {5}},\ \bibinfo {pages} {023010} (\bibinfo {year}
  {2023})}\BibitemShut {NoStop}%
\bibitem [{\citenamefont {Breuer}\ and\ \citenamefont
  {Petruccione}(2002)}]{breuer2002theory}%
  \BibitemOpen
  \bibfield  {author} {\bibinfo {author} {\bibfnamefont {H.-P.}\ \bibnamefont
  {Breuer}}\ and\ \bibinfo {author} {\bibfnamefont {F.}~\bibnamefont
  {Petruccione}},\ }\href@noop {} {\emph {\bibinfo {title} {The theory of open
  quantum systems}}}\ (\bibinfo  {publisher} {Oxford University Press on
  Demand},\ \bibinfo {year} {2002})\BibitemShut {NoStop}%
\bibitem [{\citenamefont {Lidar}\ \emph {et~al.}(1998)\citenamefont {Lidar},
  \citenamefont {Chuang},\ and\ \citenamefont {Whaley}}]{Lidar1998Decoherence}%
  \BibitemOpen
  \bibfield  {author} {\bibinfo {author} {\bibfnamefont {D.~A.}\ \bibnamefont
  {Lidar}}, \bibinfo {author} {\bibfnamefont {I.~L.}\ \bibnamefont {Chuang}}, \
  and\ \bibinfo {author} {\bibfnamefont {K.~B.}\ \bibnamefont {Whaley}},\
  }\bibfield  {title} {\enquote {\bibinfo {title} {Decoherence-free subspaces
  for quantum computation},}\ }\href {\doibase 10.1103/PhysRevLett.81.2594}
  {\bibfield  {journal} {\bibinfo  {journal} {Phys. Rev. Lett.}\ }\textbf
  {\bibinfo {volume} {81}},\ \bibinfo {pages} {2594} (\bibinfo {year}
  {1998})}\BibitemShut {NoStop}%
\bibitem [{\citenamefont {Bacon}\ \emph {et~al.}(1999)\citenamefont {Bacon},
  \citenamefont {Lidar},\ and\ \citenamefont {Whaley}}]{Bacon1999Robustness}%
  \BibitemOpen
  \bibfield  {author} {\bibinfo {author} {\bibfnamefont {D.}~\bibnamefont
  {Bacon}}, \bibinfo {author} {\bibfnamefont {D.~A.}\ \bibnamefont {Lidar}}, \
  and\ \bibinfo {author} {\bibfnamefont {K.~B.}\ \bibnamefont {Whaley}},\
  }\bibfield  {title} {\enquote {\bibinfo {title} {Robustness of
  decoherence-free subspaces for quantum computation},}\ }\href {\doibase
  https://doi.org/10.1103/PhysRevA.60.1944} {\bibfield  {journal} {\bibinfo
  {journal} {Phys. Rev. A}\ }\textbf {\bibinfo {volume} {60}},\ \bibinfo
  {pages} {1944} (\bibinfo {year} {1999})}\BibitemShut {NoStop}%
\bibitem [{\citenamefont {Duan}\ and\ \citenamefont
  {Guo}(1997)}]{Duan1997Preserving}%
  \BibitemOpen
  \bibfield  {author} {\bibinfo {author} {\bibfnamefont {L.-M.}\ \bibnamefont
  {Duan}}\ and\ \bibinfo {author} {\bibfnamefont {G.-C.}\ \bibnamefont {Guo}},\
  }\bibfield  {title} {\enquote {\bibinfo {title} {Preserving coherence in
  quantum computation by pairing quantum bits},}\ }\href {\doibase
  https://doi.org/10.1103/PhysRevLett.79.1953} {\bibfield  {journal} {\bibinfo
  {journal} {Phys. Rev. Lett.}\ }\textbf {\bibinfo {volume} {79}},\ \bibinfo
  {pages} {1953} (\bibinfo {year} {1997})}\BibitemShut {NoStop}%
\bibitem [{\citenamefont {Zanardi}\ and\ \citenamefont
  {Rasetti}(1997)}]{Zanardi1997Noiseless}%
  \BibitemOpen
  \bibfield  {author} {\bibinfo {author} {\bibfnamefont {P.}~\bibnamefont
  {Zanardi}}\ and\ \bibinfo {author} {\bibfnamefont {M.}~\bibnamefont
  {Rasetti}},\ }\bibfield  {title} {\enquote {\bibinfo {title} {Noiseless
  quantum codes},}\ }\href {\doibase 10.1103/PhysRevLett.79.3306} {\bibfield
  {journal} {\bibinfo  {journal} {Phys. Rev. Lett.}\ }\textbf {\bibinfo
  {volume} {79}},\ \bibinfo {pages} {3306} (\bibinfo {year}
  {1997})}\BibitemShut {NoStop}%
\bibitem [{\citenamefont {Plenio}\ \emph {et~al.}(1999)\citenamefont {Plenio},
  \citenamefont {Huelga}, \citenamefont {Beige},\ and\ \citenamefont
  {Knight}}]{Plenio1999Cavity}%
  \BibitemOpen
  \bibfield  {author} {\bibinfo {author} {\bibfnamefont {M.~B.}\ \bibnamefont
  {Plenio}}, \bibinfo {author} {\bibfnamefont {S.~F.}\ \bibnamefont {Huelga}},
  \bibinfo {author} {\bibfnamefont {A.}~\bibnamefont {Beige}}, \ and\ \bibinfo
  {author} {\bibfnamefont {P.~L.}\ \bibnamefont {Knight}},\ }\bibfield  {title}
  {\enquote {\bibinfo {title} {Cavity-loss-induced generation of entangled
  atoms},}\ }\href {\doibase 10.1103/PhysRevA.59.2468} {\bibfield  {journal}
  {\bibinfo  {journal} {Phys. Rev. A}\ }\textbf {\bibinfo {volume} {59}},\
  \bibinfo {pages} {2468} (\bibinfo {year} {1999})}\BibitemShut {NoStop}%
\bibitem [{\citenamefont {Diehl}\ \emph {et~al.}(2008)\citenamefont {Diehl},
  \citenamefont {Micheli}, \citenamefont {Kantian}, \citenamefont {Kraus},
  \citenamefont {B{\"u}chler},\ and\ \citenamefont
  {Zoller}}]{Diehl2008Quantum}%
  \BibitemOpen
  \bibfield  {author} {\bibinfo {author} {\bibfnamefont {S.}~\bibnamefont
  {Diehl}}, \bibinfo {author} {\bibfnamefont {A.}~\bibnamefont {Micheli}},
  \bibinfo {author} {\bibfnamefont {A.}~\bibnamefont {Kantian}}, \bibinfo
  {author} {\bibfnamefont {B.}~\bibnamefont {Kraus}}, \bibinfo {author}
  {\bibfnamefont {H.}~\bibnamefont {B{\"u}chler}}, \ and\ \bibinfo {author}
  {\bibfnamefont {P.}~\bibnamefont {Zoller}},\ }\bibfield  {title} {\enquote
  {\bibinfo {title} {Quantum states and phases in driven open quantum systems
  with cold atoms},}\ }\href {https://www.nature.com/articles/nphys1073}
  {\bibfield  {journal} {\bibinfo  {journal} {Nat. Phys.}\ }\textbf {\bibinfo
  {volume} {4}},\ \bibinfo {pages} {878} (\bibinfo {year} {2008})}\BibitemShut
  {NoStop}%
\bibitem [{\citenamefont {Kraus}\ \emph {et~al.}(2008)\citenamefont {Kraus},
  \citenamefont {B\"uchler}, \citenamefont {Diehl}, \citenamefont {Kantian},
  \citenamefont {Micheli},\ and\ \citenamefont
  {Zoller}}]{Kraus2008Preparation}%
  \BibitemOpen
  \bibfield  {author} {\bibinfo {author} {\bibfnamefont {B.}~\bibnamefont
  {Kraus}}, \bibinfo {author} {\bibfnamefont {H.~P.}\ \bibnamefont
  {B\"uchler}}, \bibinfo {author} {\bibfnamefont {S.}~\bibnamefont {Diehl}},
  \bibinfo {author} {\bibfnamefont {A.}~\bibnamefont {Kantian}}, \bibinfo
  {author} {\bibfnamefont {A.}~\bibnamefont {Micheli}}, \ and\ \bibinfo
  {author} {\bibfnamefont {P.}~\bibnamefont {Zoller}},\ }\bibfield  {title}
  {\enquote {\bibinfo {title} {Preparation of entangled states by quantum
  markov processes},}\ }\href {\doibase 10.1103/PhysRevA.78.042307} {\bibfield
  {journal} {\bibinfo  {journal} {Phys. Rev. A}\ }\textbf {\bibinfo {volume}
  {78}},\ \bibinfo {pages} {042307} (\bibinfo {year} {2008})}\BibitemShut
  {NoStop}%
\bibitem [{\citenamefont {Verstraete}\ \emph {et~al.}(2009)\citenamefont
  {Verstraete}, \citenamefont {Wolf},\ and\ \citenamefont
  {Cirac}}]{Verstraete2009Quantum}%
  \BibitemOpen
  \bibfield  {author} {\bibinfo {author} {\bibfnamefont {F.}~\bibnamefont
  {Verstraete}}, \bibinfo {author} {\bibfnamefont {M.~M.}\ \bibnamefont
  {Wolf}}, \ and\ \bibinfo {author} {\bibfnamefont {J.~I.}\ \bibnamefont
  {Cirac}},\ }\bibfield  {title} {\enquote {\bibinfo {title} {Quantum
  computation and quantum-state engineering driven by dissipation},}\ }\href
  {https://www.nature.com/articles/nphys1342/} {\bibfield  {journal} {\bibinfo
  {journal} {Nat. Phys.}\ }\textbf {\bibinfo {volume} {5}},\ \bibinfo {pages}
  {633} (\bibinfo {year} {2009})}\BibitemShut {NoStop}%
\bibitem [{\citenamefont {Diehl}\ \emph {et~al.}(2010)\citenamefont {Diehl},
  \citenamefont {Yi}, \citenamefont {Daley},\ and\ \citenamefont
  {Zoller}}]{Diehl2010Dissipation}%
  \BibitemOpen
  \bibfield  {author} {\bibinfo {author} {\bibfnamefont {S.}~\bibnamefont
  {Diehl}}, \bibinfo {author} {\bibfnamefont {W.}~\bibnamefont {Yi}}, \bibinfo
  {author} {\bibfnamefont {A.~J.}\ \bibnamefont {Daley}}, \ and\ \bibinfo
  {author} {\bibfnamefont {P.}~\bibnamefont {Zoller}},\ }\bibfield  {title}
  {\enquote {\bibinfo {title} {Dissipation-induced $d$-wave pairing of
  fermionic atoms in an optical lattice},}\ }\href {\doibase
  https://doi.org/10.1103/PhysRevLett.105.227001} {\bibfield  {journal}
  {\bibinfo  {journal} {Phys. Rev. Lett.}\ }\textbf {\bibinfo {volume} {105}},\
  \bibinfo {pages} {227001} (\bibinfo {year} {2010})}\BibitemShut {NoStop}%
\bibitem [{\citenamefont {Diehl}\ \emph {et~al.}(2011)\citenamefont {Diehl},
  \citenamefont {Rico}, \citenamefont {Baranov},\ and\ \citenamefont
  {Zoller}}]{Diehl2011Topology}%
  \BibitemOpen
  \bibfield  {author} {\bibinfo {author} {\bibfnamefont {S.}~\bibnamefont
  {Diehl}}, \bibinfo {author} {\bibfnamefont {E.}~\bibnamefont {Rico}},
  \bibinfo {author} {\bibfnamefont {M.~A.}\ \bibnamefont {Baranov}}, \ and\
  \bibinfo {author} {\bibfnamefont {P.}~\bibnamefont {Zoller}},\ }\bibfield
  {title} {\enquote {\bibinfo {title} {Topology by dissipation in atomic
  quantum wires},}\ }\href {\doibase 10.1038/nphys2106} {\bibfield  {journal}
  {\bibinfo  {journal} {Nat. Phys.}\ }\textbf {\bibinfo {volume} {7}},\
  \bibinfo {pages} {971} (\bibinfo {year} {2011})}\BibitemShut {NoStop}%
\bibitem [{\citenamefont {Pakrouski}\ \emph {et~al.}(2021)\citenamefont
  {Pakrouski}, \citenamefont {Pallegar}, \citenamefont {Popov},\ and\
  \citenamefont {Klebanov}}]{Pakrouski2021Group}%
  \BibitemOpen
  \bibfield  {author} {\bibinfo {author} {\bibfnamefont {K.}~\bibnamefont
  {Pakrouski}}, \bibinfo {author} {\bibfnamefont {P.~N.}\ \bibnamefont
  {Pallegar}}, \bibinfo {author} {\bibfnamefont {F.~K.}\ \bibnamefont {Popov}},
  \ and\ \bibinfo {author} {\bibfnamefont {I.~R.}\ \bibnamefont {Klebanov}},\
  }\bibfield  {title} {\enquote {\bibinfo {title} {Group theoretic approach to
  many-body scar states in fermionic lattice models},}\ }\href {\doibase
  https://doi.org/10.1103/PhysRevResearch.3.043156} {\bibfield  {journal}
  {\bibinfo  {journal} {Phys. Rev. Research}\ }\textbf {\bibinfo {volume}
  {3}},\ \bibinfo {pages} {043156} (\bibinfo {year} {2021})}\BibitemShut
  {NoStop}%
\bibitem [{\citenamefont {Chen}\ \emph {et~al.}(2023)\citenamefont {Chen},
  \citenamefont {Chen},\ and\ \citenamefont {Zhu}}]{Chen2023nonHermitian}%
  \BibitemOpen
  \bibfield  {author} {\bibinfo {author} {\bibfnamefont {Q.}~\bibnamefont
  {Chen}}, \bibinfo {author} {\bibfnamefont {S.~A.}\ \bibnamefont {Chen}}, \
  and\ \bibinfo {author} {\bibfnamefont {Z.}~\bibnamefont {Zhu}},\ }\bibfield
  {title} {\enquote {\bibinfo {title} {Weak ergodicity breaking in
  non-hermitian many-body systems},}\ }\href {\doibase
  https://doi.org/10.21468/SciPostPhys.15.2.052} {\bibfield  {journal}
  {\bibinfo  {journal} {SciPost Phys.}\ }\textbf {\bibinfo {volume} {15}},\
  \bibinfo {pages} {052} (\bibinfo {year} {2023})}\BibitemShut {NoStop}%
\bibitem [{\citenamefont {Omiya}\ and\ \citenamefont
  {M\"uller}(2023)}]{Omiya2023Quantum}%
  \BibitemOpen
  \bibfield  {author} {\bibinfo {author} {\bibfnamefont {K.}~\bibnamefont
  {Omiya}}\ and\ \bibinfo {author} {\bibfnamefont {M.}~\bibnamefont
  {M\"uller}},\ }\bibfield  {title} {\enquote {\bibinfo {title} {Quantum
  many-body scars in bipartite rydberg arrays originating from hidden projector
  embedding},}\ }\href {\doibase 10.1103/PhysRevA.107.023318} {\bibfield
  {journal} {\bibinfo  {journal} {Phys. Rev. A}\ }\textbf {\bibinfo {volume}
  {107}},\ \bibinfo {pages} {023318} (\bibinfo {year} {2023})}\BibitemShut
  {NoStop}%
\bibitem [{Sup()}]{SuppMaterials}%
  \BibitemOpen
  \href@noop {} {}\bibinfo {note} {See the Supplementary Materials at [URL will
  be inserted by publisher] for details about the compressed matrix product
  state technique to obtain the local projectors annihilating the whole scar
  towers, the special states in the AKLT model, more numerical results, more
  discussions about the experimental realizations, and the dissipative
  preparation of scar states with quantum metrology applications, which further
  includes
  Ref.~\cite{Perez-Garcia2007Matrix,Shen2023Construction,Google2023Nonabelian,Mesoscopic2009Muller,Evered2023High,Mcewen2021Removing,Miao2022Overcoming,Braunstein1994Statistical,Pezze2009Entanglement,Hyllus2012Fisher,Toth2012Multipartite,Iadecola2019Quantum,Kitagawa1993Squeezed}.}\BibitemShut
  {Stop}%
\bibitem [{\citenamefont {Bu{\v{c}}a}\ \emph {et~al.}(2019)\citenamefont
  {Bu{\v{c}}a}, \citenamefont {Tindall},\ and\ \citenamefont
  {Jaksch}}]{buvca2019Non}%
  \BibitemOpen
  \bibfield  {author} {\bibinfo {author} {\bibfnamefont {B.}~\bibnamefont
  {Bu{\v{c}}a}}, \bibinfo {author} {\bibfnamefont {J.}~\bibnamefont {Tindall}},
  \ and\ \bibinfo {author} {\bibfnamefont {D.}~\bibnamefont {Jaksch}},\
  }\bibfield  {title} {\enquote {\bibinfo {title} {Non-stationary coherent
  quantum many-body dynamics through dissipation},}\ }\href {\doibase
  https://doi.org/10.1038/s41467-019-09757-y} {\bibfield  {journal} {\bibinfo
  {journal} {Nat. Commun.}\ }\textbf {\bibinfo {volume} {10}},\ \bibinfo
  {pages} {1730} (\bibinfo {year} {2019})}\BibitemShut {NoStop}%
\bibitem [{\citenamefont {Booker}\ \emph {et~al.}(2020)\citenamefont {Booker},
  \citenamefont {Bu{\v{c}}a},\ and\ \citenamefont {Jaksch}}]{Booker2020Non}%
  \BibitemOpen
  \bibfield  {author} {\bibinfo {author} {\bibfnamefont {C.}~\bibnamefont
  {Booker}}, \bibinfo {author} {\bibfnamefont {B.}~\bibnamefont {Bu{\v{c}}a}},
  \ and\ \bibinfo {author} {\bibfnamefont {D.}~\bibnamefont {Jaksch}},\
  }\bibfield  {title} {\enquote {\bibinfo {title} {Non-stationarity and
  dissipative time crystals: spectral properties and finite-size effects},}\
  }\href {https://iopscience.iop.org/article/10.1088/1367-2630/ababc4/meta}
  {\bibfield  {journal} {\bibinfo  {journal} {New J. Phys.}\ }\textbf {\bibinfo
  {volume} {22}},\ \bibinfo {pages} {085007} (\bibinfo {year}
  {2020})}\BibitemShut {NoStop}%
\bibitem [{\citenamefont {Bu{\v{c}}a}\ \emph {et~al.}(2020)\citenamefont
  {Bu{\v{c}}a}, \citenamefont {Booker}, \citenamefont {Medenjak},\ and\
  \citenamefont {Jaksch}}]{Buvca2020Bethe}%
  \BibitemOpen
  \bibfield  {author} {\bibinfo {author} {\bibfnamefont {B.}~\bibnamefont
  {Bu{\v{c}}a}}, \bibinfo {author} {\bibfnamefont {C.}~\bibnamefont {Booker}},
  \bibinfo {author} {\bibfnamefont {M.}~\bibnamefont {Medenjak}}, \ and\
  \bibinfo {author} {\bibfnamefont {D.}~\bibnamefont {Jaksch}},\ }\bibfield
  {title} {\enquote {\bibinfo {title} {Bethe ansatz approach for dissipation:
  exact solutions of quantum many-body dynamics under loss},}\ }\href {\doibase
  https://doi.org/10.1088/1367-2630/abd124} {\bibfield  {journal} {\bibinfo
  {journal} {New J. Phys.}\ }\textbf {\bibinfo {volume} {22}},\ \bibinfo
  {pages} {123040} (\bibinfo {year} {2020})}\BibitemShut {NoStop}%
\bibitem [{\citenamefont {Chinzei}\ and\ \citenamefont
  {Ikeda}(2020)}]{Chinzei2020Time}%
  \BibitemOpen
  \bibfield  {author} {\bibinfo {author} {\bibfnamefont {K.}~\bibnamefont
  {Chinzei}}\ and\ \bibinfo {author} {\bibfnamefont {T.~N.}\ \bibnamefont
  {Ikeda}},\ }\bibfield  {title} {\enquote {\bibinfo {title} {Time crystals
  protected by floquet dynamical symmetry in hubbard models},}\ }\href
  {\doibase 10.1103/PhysRevLett.125.060601} {\bibfield  {journal} {\bibinfo
  {journal} {Phys. Rev. Lett.}\ }\textbf {\bibinfo {volume} {125}},\ \bibinfo
  {pages} {060601} (\bibinfo {year} {2020})}\BibitemShut {NoStop}%
\bibitem [{\citenamefont {Guarnieri}\ \emph {et~al.}(2022)\citenamefont
  {Guarnieri}, \citenamefont {Mitchison}, \citenamefont {Purkayastha},
  \citenamefont {Jaksch}, \citenamefont {Bu\ifmmode~\check{c}\else
  \v{c}\fi{}a},\ and\ \citenamefont {Goold}}]{Guarnieri2022Time}%
  \BibitemOpen
  \bibfield  {author} {\bibinfo {author} {\bibfnamefont {G.}~\bibnamefont
  {Guarnieri}}, \bibinfo {author} {\bibfnamefont {M.~T.}\ \bibnamefont
  {Mitchison}}, \bibinfo {author} {\bibfnamefont {A.}~\bibnamefont
  {Purkayastha}}, \bibinfo {author} {\bibfnamefont {D.}~\bibnamefont {Jaksch}},
  \bibinfo {author} {\bibfnamefont {B.}~\bibnamefont {Bu\ifmmode~\check{c}\else
  \v{c}\fi{}a}}, \ and\ \bibinfo {author} {\bibfnamefont {J.}~\bibnamefont
  {Goold}},\ }\bibfield  {title} {\enquote {\bibinfo {title} {Time periodicity
  from randomness in quantum systems},}\ }\href {\doibase
  https://doi.org/10.1103/PhysRevA.106.022209} {\bibfield  {journal} {\bibinfo
  {journal} {Phys. Rev. A}\ }\textbf {\bibinfo {volume} {106}},\ \bibinfo
  {pages} {022209} (\bibinfo {year} {2022})}\BibitemShut {NoStop}%
\bibitem [{dyn()}]{dyn_symm_footnote}%
  \BibitemOpen
  \href@noop {} {}\bibinfo {note} {Despite the fact that both conditions lead
  to equally spaced non-decaying eigenmodes and non-stationary periodic
  dynamics, the dynamical symmetry necessitates $[H,Q^\dagger]= \omega
  Q^\dagger$ to be valid in the entire Hilbert space, and $[L_j,
  Q^\dagger]=[L_j^\dagger, Q^\dagger] = 0 $ for all the dissipators. In
  contrast, our protocol only requires the first condition to hold in the scar
  subspace $W$ and does not need the second one. Furthermore, $[H,Q^\dagger] =
  \omega Q^\dagger$ poses too strong symmetry in the Hamiltonian to qualify the
  tower of states $\{\ket{S_n}\}$ as genuine scar states
  \cite{Moudgalya2020eta,Mark2020eta,Moudgalya2022Quantum,Chandran2022Quantum}.}\BibitemShut
  {Stop}%
\bibitem [{\citenamefont {Bu{\v{c}}a}\ and\ \citenamefont
  {Prosen}(2012)}]{Buca2012Note}%
  \BibitemOpen
  \bibfield  {author} {\bibinfo {author} {\bibfnamefont {B.}~\bibnamefont
  {Bu{\v{c}}a}}\ and\ \bibinfo {author} {\bibfnamefont {T.}~\bibnamefont
  {Prosen}},\ }\bibfield  {title} {\enquote {\bibinfo {title} {A note on
  symmetry reductions of the lindblad equation: transport in constrained open
  spin chains},}\ }\href {\doibase 10.1088/1367-2630/14/7/073007} {\bibfield
  {journal} {\bibinfo  {journal} {New J. Phys.}\ }\textbf {\bibinfo {volume}
  {14}},\ \bibinfo {pages} {073007} (\bibinfo {year} {2012})}\BibitemShut
  {NoStop}%
\bibitem [{\citenamefont {Mark}\ \emph {et~al.}(2020)\citenamefont {Mark},
  \citenamefont {Lin},\ and\ \citenamefont {Motrunich}}]{mark2020unified}%
  \BibitemOpen
  \bibfield  {author} {\bibinfo {author} {\bibfnamefont {D.~K.}\ \bibnamefont
  {Mark}}, \bibinfo {author} {\bibfnamefont {C.-J.}\ \bibnamefont {Lin}}, \
  and\ \bibinfo {author} {\bibfnamefont {O.~I.}\ \bibnamefont {Motrunich}},\
  }\bibfield  {title} {\enquote {\bibinfo {title} {Unified structure for exact
  towers of scar states in the affleck-kennedy-lieb-tasaki and other models},}\
  }\href {\doibase 10.1103/PhysRevB.101.195131} {\bibfield  {journal} {\bibinfo
   {journal} {Phys. Rev. B}\ }\textbf {\bibinfo {volume} {101}},\ \bibinfo
  {pages} {195131} (\bibinfo {year} {2020})}\BibitemShut {NoStop}%
\bibitem [{\citenamefont {Affleck}\ \emph {et~al.}(1987)\citenamefont
  {Affleck}, \citenamefont {Kennedy}, \citenamefont {Lieb},\ and\ \citenamefont
  {Tasaki}}]{Affleck1987Rigorous}%
  \BibitemOpen
  \bibfield  {author} {\bibinfo {author} {\bibfnamefont {I.}~\bibnamefont
  {Affleck}}, \bibinfo {author} {\bibfnamefont {T.}~\bibnamefont {Kennedy}},
  \bibinfo {author} {\bibfnamefont {E.~H.}\ \bibnamefont {Lieb}}, \ and\
  \bibinfo {author} {\bibfnamefont {H.}~\bibnamefont {Tasaki}},\ }\bibfield
  {title} {\enquote {\bibinfo {title} {Rigorous results on valence-bond ground
  states in antiferromagnets},}\ }\href {\doibase 10.1103/PhysRevLett.59.799}
  {\bibfield  {journal} {\bibinfo  {journal} {Phys. Rev. Lett.}\ }\textbf
  {\bibinfo {volume} {59}},\ \bibinfo {pages} {799} (\bibinfo {year}
  {1987})}\BibitemShut {NoStop}%
\bibitem [{sys()}]{syssize_footnote}%
  \BibitemOpen
  \href@noop {} {}\bibinfo {note} {We compute the spectrum of the non-Hermitian
  AKLT Hamiltonian instead of the whole Liouvillian, in order to access to
  larger system sizes ($L\ge 8$) to observe the irrational real eigenvalues
  (see \cite{SuppMaterials} for details).}\BibitemShut {Stop}%
\bibitem [{\citenamefont {Lloyd}\ and\ \citenamefont
  {Viola}(2001)}]{Lloyd2001Engineering}%
  \BibitemOpen
  \bibfield  {author} {\bibinfo {author} {\bibfnamefont {S.}~\bibnamefont
  {Lloyd}}\ and\ \bibinfo {author} {\bibfnamefont {L.}~\bibnamefont {Viola}},\
  }\bibfield  {title} {\enquote {\bibinfo {title} {Engineering quantum
  dynamics},}\ }\href {\doibase 10.1103/PhysRevA.65.010101} {\bibfield
  {journal} {\bibinfo  {journal} {Phys. Rev. A}\ }\textbf {\bibinfo {volume}
  {65}},\ \bibinfo {pages} {010101} (\bibinfo {year} {2001})}\BibitemShut
  {NoStop}%
\bibitem [{\citenamefont {Bacon}\ \emph {et~al.}(2001)\citenamefont {Bacon},
  \citenamefont {Childs}, \citenamefont {Chuang}, \citenamefont {Kempe},
  \citenamefont {Leung},\ and\ \citenamefont {Zhou}}]{Bacon2001Universal}%
  \BibitemOpen
  \bibfield  {author} {\bibinfo {author} {\bibfnamefont {D.}~\bibnamefont
  {Bacon}}, \bibinfo {author} {\bibfnamefont {A.~M.}\ \bibnamefont {Childs}},
  \bibinfo {author} {\bibfnamefont {I.~L.}\ \bibnamefont {Chuang}}, \bibinfo
  {author} {\bibfnamefont {J.}~\bibnamefont {Kempe}}, \bibinfo {author}
  {\bibfnamefont {D.~W.}\ \bibnamefont {Leung}}, \ and\ \bibinfo {author}
  {\bibfnamefont {X.}~\bibnamefont {Zhou}},\ }\bibfield  {title} {\enquote
  {\bibinfo {title} {Universal simulation of markovian quantum dynamics},}\
  }\href {\doibase 10.1103/PhysRevA.64.062302} {\bibfield  {journal} {\bibinfo
  {journal} {Phys. Rev. A}\ }\textbf {\bibinfo {volume} {64}},\ \bibinfo
  {pages} {062302} (\bibinfo {year} {2001})}\BibitemShut {NoStop}%
\bibitem [{\citenamefont {Ciccarello}\ \emph {et~al.}(2022)\citenamefont
  {Ciccarello}, \citenamefont {Lorenzo}, \citenamefont {Giovannetti},\ and\
  \citenamefont {Palma}}]{Ciccarello2022Quantum}%
  \BibitemOpen
  \bibfield  {author} {\bibinfo {author} {\bibfnamefont {F.}~\bibnamefont
  {Ciccarello}}, \bibinfo {author} {\bibfnamefont {S.}~\bibnamefont {Lorenzo}},
  \bibinfo {author} {\bibfnamefont {V.}~\bibnamefont {Giovannetti}}, \ and\
  \bibinfo {author} {\bibfnamefont {G.~M.}\ \bibnamefont {Palma}},\ }\bibfield
  {title} {\enquote {\bibinfo {title} {Quantum collision models: Open system
  dynamics from repeated interactions},}\ }\href
  {https://www.sciencedirect.com/science/article/abs/pii/S0370157322000035}
  {\bibfield  {journal} {\bibinfo  {journal} {Phys. Rep.}\ }\textbf {\bibinfo
  {volume} {954}},\ \bibinfo {pages} {1} (\bibinfo {year} {2022})}\BibitemShut
  {NoStop}%
\bibitem [{\citenamefont {Cattaneo}\ \emph {et~al.}(2022)\citenamefont
  {Cattaneo}, \citenamefont {Giorgi}, \citenamefont {Zambrini},\ and\
  \citenamefont {Maniscalco}}]{Cattaneo2022Brief}%
  \BibitemOpen
  \bibfield  {author} {\bibinfo {author} {\bibfnamefont {M.}~\bibnamefont
  {Cattaneo}}, \bibinfo {author} {\bibfnamefont {G.~L.}\ \bibnamefont
  {Giorgi}}, \bibinfo {author} {\bibfnamefont {R.}~\bibnamefont {Zambrini}}, \
  and\ \bibinfo {author} {\bibfnamefont {S.}~\bibnamefont {Maniscalco}},\
  }\bibfield  {title} {\enquote {\bibinfo {title} {A brief journey through
  collision models for multipartite open quantum dynamics},}\ }\href
  {https://www.worldscientific.com/doi/10.1142/S1230161222500159} {\bibfield
  {journal} {\bibinfo  {journal} {Open Syst. Inf. Dyn.}\ }\textbf {\bibinfo
  {volume} {29}},\ \bibinfo {pages} {2250015} (\bibinfo {year}
  {2022})}\BibitemShut {NoStop}%
\bibitem [{\citenamefont {Gillman}\ \emph {et~al.}(2023)\citenamefont
  {Gillman}, \citenamefont {Carollo},\ and\ \citenamefont
  {Lesanovsky}}]{Gillman2023Using}%
  \BibitemOpen
  \bibfield  {author} {\bibinfo {author} {\bibfnamefont {E.}~\bibnamefont
  {Gillman}}, \bibinfo {author} {\bibfnamefont {F.}~\bibnamefont {Carollo}}, \
  and\ \bibinfo {author} {\bibfnamefont {I.}~\bibnamefont {Lesanovsky}},\
  }\bibfield  {title} {\enquote {\bibinfo {title} {Using $(1+1)d$ quantum
  cellular automata for exploring collective effects in large-scale quantum
  neural networks},}\ }\href
  {https://link.aps.org/doi/10.1103/PhysRevE.107.L022102} {\bibfield  {journal}
  {\bibinfo  {journal} {Phys. Rev. E}\ }\textbf {\bibinfo {volume} {107}},\
  \bibinfo {pages} {L022102} (\bibinfo {year} {2023})}\BibitemShut {NoStop}%
\bibitem [{\citenamefont {Shankar}\ \emph {et~al.}(2013)\citenamefont
  {Shankar}, \citenamefont {Hatridge}, \citenamefont {Leghtas}, \citenamefont
  {Sliwa}, \citenamefont {Narla}, \citenamefont {Vool}, \citenamefont {Girvin},
  \citenamefont {Frunzio}, \citenamefont {Mirrahimi},\ and\ \citenamefont
  {Devoret}}]{Shankar2013Autonomously}%
  \BibitemOpen
  \bibfield  {author} {\bibinfo {author} {\bibfnamefont {S.}~\bibnamefont
  {Shankar}}, \bibinfo {author} {\bibfnamefont {M.}~\bibnamefont {Hatridge}},
  \bibinfo {author} {\bibfnamefont {Z.}~\bibnamefont {Leghtas}}, \bibinfo
  {author} {\bibfnamefont {K.}~\bibnamefont {Sliwa}}, \bibinfo {author}
  {\bibfnamefont {A.}~\bibnamefont {Narla}}, \bibinfo {author} {\bibfnamefont
  {U.}~\bibnamefont {Vool}}, \bibinfo {author} {\bibfnamefont {S.~M.}\
  \bibnamefont {Girvin}}, \bibinfo {author} {\bibfnamefont {L.}~\bibnamefont
  {Frunzio}}, \bibinfo {author} {\bibfnamefont {M.}~\bibnamefont {Mirrahimi}},
  \ and\ \bibinfo {author} {\bibfnamefont {M.~H.}\ \bibnamefont {Devoret}},\
  }\bibfield  {title} {\enquote {\bibinfo {title} {Autonomously stabilized
  entanglement between two superconducting quantum bits},}\ }\href {\doibase
  https://doi.org/10.1038/nature12802} {\bibfield  {journal} {\bibinfo
  {journal} {Nature}\ }\textbf {\bibinfo {volume} {504}},\ \bibinfo {pages}
  {419} (\bibinfo {year} {2013})}\BibitemShut {NoStop}%
\bibitem [{\citenamefont {Han}\ \emph {et~al.}(2021)\citenamefont {Han},
  \citenamefont {Cai}, \citenamefont {Hu}, \citenamefont {Mu}, \citenamefont
  {Ma}, \citenamefont {Xu}, \citenamefont {Wang}, \citenamefont {Wang},
  \citenamefont {Song}, \citenamefont {Zou},\ and\ \citenamefont
  {Sun}}]{Han2021Experimental}%
  \BibitemOpen
  \bibfield  {author} {\bibinfo {author} {\bibfnamefont {J.}~\bibnamefont
  {Han}}, \bibinfo {author} {\bibfnamefont {W.}~\bibnamefont {Cai}}, \bibinfo
  {author} {\bibfnamefont {L.}~\bibnamefont {Hu}}, \bibinfo {author}
  {\bibfnamefont {X.}~\bibnamefont {Mu}}, \bibinfo {author} {\bibfnamefont
  {Y.}~\bibnamefont {Ma}}, \bibinfo {author} {\bibfnamefont {Y.}~\bibnamefont
  {Xu}}, \bibinfo {author} {\bibfnamefont {W.}~\bibnamefont {Wang}}, \bibinfo
  {author} {\bibfnamefont {H.}~\bibnamefont {Wang}}, \bibinfo {author}
  {\bibfnamefont {Y.~P.}\ \bibnamefont {Song}}, \bibinfo {author}
  {\bibfnamefont {C.-L.}\ \bibnamefont {Zou}}, \ and\ \bibinfo {author}
  {\bibfnamefont {L.}~\bibnamefont {Sun}},\ }\bibfield  {title} {\enquote
  {\bibinfo {title} {Experimental simulation of open quantum system dynamics
  via trotterization},}\ }\href {\doibase
  https://doi.org/10.1103/PhysRevLett.127.020504} {\bibfield  {journal}
  {\bibinfo  {journal} {Phys. Rev. Lett.}\ }\textbf {\bibinfo {volume} {127}},\
  \bibinfo {pages} {020504} (\bibinfo {year} {2021})}\BibitemShut {NoStop}%
\bibitem [{\citenamefont {Cai}\ \emph {et~al.}(2021)\citenamefont {Cai},
  \citenamefont {Han}, \citenamefont {Hu}, \citenamefont {Ma}, \citenamefont
  {Mu}, \citenamefont {Wang}, \citenamefont {Xu}, \citenamefont {Hua},
  \citenamefont {Wang}, \citenamefont {Song}, \citenamefont {Zhang},
  \citenamefont {Zou},\ and\ \citenamefont {Sun}}]{Cai2021High}%
  \BibitemOpen
  \bibfield  {author} {\bibinfo {author} {\bibfnamefont {W.}~\bibnamefont
  {Cai}}, \bibinfo {author} {\bibfnamefont {J.}~\bibnamefont {Han}}, \bibinfo
  {author} {\bibfnamefont {L.}~\bibnamefont {Hu}}, \bibinfo {author}
  {\bibfnamefont {Y.}~\bibnamefont {Ma}}, \bibinfo {author} {\bibfnamefont
  {X.}~\bibnamefont {Mu}}, \bibinfo {author} {\bibfnamefont {W.}~\bibnamefont
  {Wang}}, \bibinfo {author} {\bibfnamefont {Y.}~\bibnamefont {Xu}}, \bibinfo
  {author} {\bibfnamefont {Z.}~\bibnamefont {Hua}}, \bibinfo {author}
  {\bibfnamefont {H.}~\bibnamefont {Wang}}, \bibinfo {author} {\bibfnamefont
  {Y.~P.}\ \bibnamefont {Song}}, \bibinfo {author} {\bibfnamefont {J.-N.}\
  \bibnamefont {Zhang}}, \bibinfo {author} {\bibfnamefont {C.-L.}\ \bibnamefont
  {Zou}}, \ and\ \bibinfo {author} {\bibfnamefont {L.}~\bibnamefont {Sun}},\
  }\bibfield  {title} {\enquote {\bibinfo {title} {High-efficiency arbitrary
  quantum operation on a high-dimensional quantum system},}\ }\href {\doibase
  https://doi.org/10.1103/PhysRevLett.127.090504} {\bibfield  {journal}
  {\bibinfo  {journal} {Phys. Rev. Lett.}\ }\textbf {\bibinfo {volume} {127}},\
  \bibinfo {pages} {090504} (\bibinfo {year} {2021})}\BibitemShut {NoStop}%
\bibitem [{\citenamefont {\text{Google Quantum AI and
  Collaborators}}(2024)}]{Google2023Stable}%
  \BibitemOpen
  \bibfield  {author} {\bibinfo {author} {\bibnamefont {\text{Google Quantum AI
  and Collaborators}}},\ }\bibfield  {title} {\enquote {\bibinfo {title}
  {Stable quantum-correlated many-body states via engineered dissipation},}\
  }\href {\doibase 10.1126/science.adh9932} {\bibfield  {journal} {\bibinfo
  {journal} {Science}\ }\textbf {\bibinfo {volume} {383}},\ \bibinfo {pages}
  {1332} (\bibinfo {year} {2024})}\BibitemShut {NoStop}%
\bibitem [{\citenamefont {Barreiro}\ \emph {et~al.}(2011)\citenamefont
  {Barreiro}, \citenamefont {M{\"u}ller}, \citenamefont {Schindler},
  \citenamefont {Nigg}, \citenamefont {Monz}, \citenamefont {Chwalla},
  \citenamefont {Hennrich}, \citenamefont {Roos}, \citenamefont {Zoller},\ and\
  \citenamefont {Blatt}}]{Barreiro2011Open}%
  \BibitemOpen
  \bibfield  {author} {\bibinfo {author} {\bibfnamefont {J.~T.}\ \bibnamefont
  {Barreiro}}, \bibinfo {author} {\bibfnamefont {M.}~\bibnamefont
  {M{\"u}ller}}, \bibinfo {author} {\bibfnamefont {P.}~\bibnamefont
  {Schindler}}, \bibinfo {author} {\bibfnamefont {D.}~\bibnamefont {Nigg}},
  \bibinfo {author} {\bibfnamefont {T.}~\bibnamefont {Monz}}, \bibinfo {author}
  {\bibfnamefont {M.}~\bibnamefont {Chwalla}}, \bibinfo {author} {\bibfnamefont
  {M.}~\bibnamefont {Hennrich}}, \bibinfo {author} {\bibfnamefont {C.~F.}\
  \bibnamefont {Roos}}, \bibinfo {author} {\bibfnamefont {P.}~\bibnamefont
  {Zoller}}, \ and\ \bibinfo {author} {\bibfnamefont {R.}~\bibnamefont
  {Blatt}},\ }\bibfield  {title} {\enquote {\bibinfo {title} {An open-system
  quantum simulator with trapped ions},}\ }\href
  {https://www.nature.com/articles/nature09801} {\bibfield  {journal} {\bibinfo
   {journal} {Nature}\ }\textbf {\bibinfo {volume} {470}},\ \bibinfo {pages}
  {486} (\bibinfo {year} {2011})}\BibitemShut {NoStop}%
\bibitem [{\citenamefont {Lin}\ \emph {et~al.}(2013)\citenamefont {Lin},
  \citenamefont {Gaebler}, \citenamefont {Reiter}, \citenamefont {Tan},
  \citenamefont {Bowler}, \citenamefont {S{\o}rensen}, \citenamefont
  {Leibfried},\ and\ \citenamefont {Wineland}}]{Lin2013Dissipative}%
  \BibitemOpen
  \bibfield  {author} {\bibinfo {author} {\bibfnamefont {Y.}~\bibnamefont
  {Lin}}, \bibinfo {author} {\bibfnamefont {J.}~\bibnamefont {Gaebler}},
  \bibinfo {author} {\bibfnamefont {F.}~\bibnamefont {Reiter}}, \bibinfo
  {author} {\bibfnamefont {T.~R.}\ \bibnamefont {Tan}}, \bibinfo {author}
  {\bibfnamefont {R.}~\bibnamefont {Bowler}}, \bibinfo {author} {\bibfnamefont
  {A.}~\bibnamefont {S{\o}rensen}}, \bibinfo {author} {\bibfnamefont
  {D.}~\bibnamefont {Leibfried}}, \ and\ \bibinfo {author} {\bibfnamefont
  {D.~J.}\ \bibnamefont {Wineland}},\ }\bibfield  {title} {\enquote {\bibinfo
  {title} {Dissipative production of a maximally entangled steady state of two
  quantum bits},}\ }\href {\doibase https://doi.org/10.1038/nature12801}
  {\bibfield  {journal} {\bibinfo  {journal} {Nature}\ }\textbf {\bibinfo
  {volume} {504}},\ \bibinfo {pages} {415} (\bibinfo {year}
  {2013})}\BibitemShut {NoStop}%
\bibitem [{\citenamefont {Schindler}\ \emph {et~al.}(2013)\citenamefont
  {Schindler}, \citenamefont {M{\"u}ller}, \citenamefont {Nigg}, \citenamefont
  {Barreiro}, \citenamefont {Martinez}, \citenamefont {Hennrich}, \citenamefont
  {Monz}, \citenamefont {Diehl}, \citenamefont {Zoller},\ and\ \citenamefont
  {Blatt}}]{Schindler2013quantum}%
  \BibitemOpen
  \bibfield  {author} {\bibinfo {author} {\bibfnamefont {P.}~\bibnamefont
  {Schindler}}, \bibinfo {author} {\bibfnamefont {M.}~\bibnamefont
  {M{\"u}ller}}, \bibinfo {author} {\bibfnamefont {D.}~\bibnamefont {Nigg}},
  \bibinfo {author} {\bibfnamefont {J.~T.}\ \bibnamefont {Barreiro}}, \bibinfo
  {author} {\bibfnamefont {E.~A.}\ \bibnamefont {Martinez}}, \bibinfo {author}
  {\bibfnamefont {M.}~\bibnamefont {Hennrich}}, \bibinfo {author}
  {\bibfnamefont {T.}~\bibnamefont {Monz}}, \bibinfo {author} {\bibfnamefont
  {S.}~\bibnamefont {Diehl}}, \bibinfo {author} {\bibfnamefont
  {P.}~\bibnamefont {Zoller}}, \ and\ \bibinfo {author} {\bibfnamefont
  {R.}~\bibnamefont {Blatt}},\ }\bibfield  {title} {\enquote {\bibinfo {title}
  {Quantum simulation of dynamical maps with trapped ions},}\ }\href
  {https://www.nature.com/articles/nphys2630} {\bibfield  {journal} {\bibinfo
  {journal} {Nat. Phys.}\ }\textbf {\bibinfo {volume} {9}},\ \bibinfo {pages}
  {361} (\bibinfo {year} {2013})}\BibitemShut {NoStop}%
\bibitem [{\citenamefont {Weimer}\ \emph {et~al.}(2010)\citenamefont {Weimer},
  \citenamefont {M{\"u}ller}, \citenamefont {Lesanovsky}, \citenamefont
  {Zoller},\ and\ \citenamefont {B{\"u}chler}}]{Weimer2010Rydberg}%
  \BibitemOpen
  \bibfield  {author} {\bibinfo {author} {\bibfnamefont {H.}~\bibnamefont
  {Weimer}}, \bibinfo {author} {\bibfnamefont {M.}~\bibnamefont {M{\"u}ller}},
  \bibinfo {author} {\bibfnamefont {I.}~\bibnamefont {Lesanovsky}}, \bibinfo
  {author} {\bibfnamefont {P.}~\bibnamefont {Zoller}}, \ and\ \bibinfo {author}
  {\bibfnamefont {H.~P.}\ \bibnamefont {B{\"u}chler}},\ }\bibfield  {title}
  {\enquote {\bibinfo {title} {A rydberg quantum simulator},}\ }\href {\doibase
  https://doi.org/10.1038/nphys1614} {\bibfield  {journal} {\bibinfo  {journal}
  {Nat. Phys.}\ }\textbf {\bibinfo {volume} {6}},\ \bibinfo {pages} {382}
  (\bibinfo {year} {2010})}\BibitemShut {NoStop}%
\bibitem [{\citenamefont {Daley}(2014)}]{Daley2014Quantum}%
  \BibitemOpen
  \bibfield  {author} {\bibinfo {author} {\bibfnamefont {A.~J.}\ \bibnamefont
  {Daley}},\ }\bibfield  {title} {\enquote {\bibinfo {title} {Quantum
  trajectories and open many-body quantum systems},}\ }\href
  {https://www.tandfonline.com/doi/abs/10.1080/00018732.2014.933502} {\bibfield
   {journal} {\bibinfo  {journal} {Adv. Phys.}\ }\textbf {\bibinfo {volume}
  {63}},\ \bibinfo {pages} {77} (\bibinfo {year} {2014})}\BibitemShut {NoStop}%
\bibitem [{\citenamefont {Dooley}(2021)}]{Dooley2021Robust}%
  \BibitemOpen
  \bibfield  {author} {\bibinfo {author} {\bibfnamefont {S.}~\bibnamefont
  {Dooley}},\ }\bibfield  {title} {\enquote {\bibinfo {title} {Robust quantum
  sensing in strongly interacting systems with many-body scars},}\ }\href
  {\doibase https://doi.org/10.1103/PRXQuantum.2.020330} {\bibfield  {journal}
  {\bibinfo  {journal} {PRX Quantum}\ }\textbf {\bibinfo {volume} {2}},\
  \bibinfo {pages} {020330} (\bibinfo {year} {2021})}\BibitemShut {NoStop}%
\bibitem [{\citenamefont {Desaules}\ \emph {et~al.}(2022)\citenamefont
  {Desaules}, \citenamefont {Pietracaprina}, \citenamefont
  {Papi\ifmmode~\acute{c}\else \'{c}\fi{}}, \citenamefont {Goold},\ and\
  \citenamefont {Pappalardi}}]{Desaules2022Extensive}%
  \BibitemOpen
  \bibfield  {author} {\bibinfo {author} {\bibfnamefont {J.-Y.}\ \bibnamefont
  {Desaules}}, \bibinfo {author} {\bibfnamefont {F.}~\bibnamefont
  {Pietracaprina}}, \bibinfo {author} {\bibfnamefont {Z.}~\bibnamefont
  {Papi\ifmmode~\acute{c}\else \'{c}\fi{}}}, \bibinfo {author} {\bibfnamefont
  {J.}~\bibnamefont {Goold}}, \ and\ \bibinfo {author} {\bibfnamefont
  {S.}~\bibnamefont {Pappalardi}},\ }\bibfield  {title} {\enquote {\bibinfo
  {title} {Extensive multipartite entanglement from su(2) quantum many-body
  scars},}\ }\href {\doibase 10.1103/PhysRevLett.129.020601} {\bibfield
  {journal} {\bibinfo  {journal} {Phys. Rev. Lett.}\ }\textbf {\bibinfo
  {volume} {129}},\ \bibinfo {pages} {020601} (\bibinfo {year}
  {2022})}\BibitemShut {NoStop}%
\bibitem [{\citenamefont {Dooley}\ \emph {et~al.}(2023)\citenamefont {Dooley},
  \citenamefont {Pappalardi},\ and\ \citenamefont
  {Goold}}]{Dooley2023Entanglement}%
  \BibitemOpen
  \bibfield  {author} {\bibinfo {author} {\bibfnamefont {S.}~\bibnamefont
  {Dooley}}, \bibinfo {author} {\bibfnamefont {S.}~\bibnamefont {Pappalardi}},
  \ and\ \bibinfo {author} {\bibfnamefont {J.}~\bibnamefont {Goold}},\
  }\bibfield  {title} {\enquote {\bibinfo {title} {Entanglement enhanced
  metrology with quantum many-body scars},}\ }\href {\doibase
  https://doi.org/10.1103/PhysRevB.107.035123} {\bibfield  {journal} {\bibinfo
  {journal} {Phys. Rev. B}\ }\textbf {\bibinfo {volume} {107}},\ \bibinfo
  {pages} {035123} (\bibinfo {year} {2023})}\BibitemShut {NoStop}%
\bibitem [{\citenamefont {Pezz\`e}\ \emph {et~al.}(2018)\citenamefont
  {Pezz\`e}, \citenamefont {Smerzi}, \citenamefont {Oberthaler}, \citenamefont
  {Schmied},\ and\ \citenamefont {Treutlein}}]{Pezze2018Quantum}%
  \BibitemOpen
  \bibfield  {author} {\bibinfo {author} {\bibfnamefont {L.}~\bibnamefont
  {Pezz\`e}}, \bibinfo {author} {\bibfnamefont {A.}~\bibnamefont {Smerzi}},
  \bibinfo {author} {\bibfnamefont {M.~K.}\ \bibnamefont {Oberthaler}},
  \bibinfo {author} {\bibfnamefont {R.}~\bibnamefont {Schmied}}, \ and\
  \bibinfo {author} {\bibfnamefont {P.}~\bibnamefont {Treutlein}},\ }\bibfield
  {title} {\enquote {\bibinfo {title} {Quantum metrology with nonclassical
  states of atomic ensembles},}\ }\href {\doibase
  https://doi.org/10.1103/RevModPhys.90.035005} {\bibfield  {journal} {\bibinfo
   {journal} {Rev. Mod. Phys.}\ }\textbf {\bibinfo {volume} {90}},\ \bibinfo
  {pages} {035005} (\bibinfo {year} {2018})}\BibitemShut {NoStop}%
\bibitem [{\citenamefont {Olmos}\ \emph {et~al.}(2012)\citenamefont {Olmos},
  \citenamefont {Lesanovsky},\ and\ \citenamefont
  {Garrahan}}]{Olmos2012Facilitated}%
  \BibitemOpen
  \bibfield  {author} {\bibinfo {author} {\bibfnamefont {B.}~\bibnamefont
  {Olmos}}, \bibinfo {author} {\bibfnamefont {I.}~\bibnamefont {Lesanovsky}}, \
  and\ \bibinfo {author} {\bibfnamefont {J.~P.}\ \bibnamefont {Garrahan}},\
  }\bibfield  {title} {\enquote {\bibinfo {title} {Facilitated spin models of
  dissipative quantum glasses},}\ }\href {\doibase
  https://doi.org/10.1103/PhysRevLett.109.020403} {\bibfield  {journal}
  {\bibinfo  {journal} {Phys. Rev. Lett.}\ }\textbf {\bibinfo {volume} {109}},\
  \bibinfo {pages} {020403} (\bibinfo {year} {2012})}\BibitemShut {NoStop}%
\bibitem [{\citenamefont {Macieszczak}\ \emph {et~al.}(2016)\citenamefont
  {Macieszczak}, \citenamefont {Gu{\c{t}}{\u{a}}}, \citenamefont {Lesanovsky},\
  and\ \citenamefont {Garrahan}}]{Macieszczak2016Towards}%
  \BibitemOpen
  \bibfield  {author} {\bibinfo {author} {\bibfnamefont {K.}~\bibnamefont
  {Macieszczak}}, \bibinfo {author} {\bibfnamefont {M.}~\bibnamefont
  {Gu{\c{t}}{\u{a}}}}, \bibinfo {author} {\bibfnamefont {I.}~\bibnamefont
  {Lesanovsky}}, \ and\ \bibinfo {author} {\bibfnamefont {J.~P.}\ \bibnamefont
  {Garrahan}},\ }\bibfield  {title} {\enquote {\bibinfo {title} {Towards a
  theory of metastability in open quantum dynamics},}\ }\href {\doibase
  https://doi.org/10.1103/PhysRevLett.116.240404} {\bibfield  {journal}
  {\bibinfo  {journal} {Phys. Rev. Lett.}\ }\textbf {\bibinfo {volume} {116}},\
  \bibinfo {pages} {240404} (\bibinfo {year} {2016})}\BibitemShut {NoStop}%
\bibitem [{\citenamefont {Maskara}\ \emph {et~al.}(2021)\citenamefont
  {Maskara}, \citenamefont {Michailidis}, \citenamefont {Ho}, \citenamefont
  {Bluvstein}, \citenamefont {Choi}, \citenamefont {Lukin},\ and\ \citenamefont
  {Serbyn}}]{Maskara2021discrete}%
  \BibitemOpen
  \bibfield  {author} {\bibinfo {author} {\bibfnamefont {N.}~\bibnamefont
  {Maskara}}, \bibinfo {author} {\bibfnamefont {A.~A.}\ \bibnamefont
  {Michailidis}}, \bibinfo {author} {\bibfnamefont {W.~W.}\ \bibnamefont {Ho}},
  \bibinfo {author} {\bibfnamefont {D.}~\bibnamefont {Bluvstein}}, \bibinfo
  {author} {\bibfnamefont {S.}~\bibnamefont {Choi}}, \bibinfo {author}
  {\bibfnamefont {M.~D.}\ \bibnamefont {Lukin}}, \ and\ \bibinfo {author}
  {\bibfnamefont {M.}~\bibnamefont {Serbyn}},\ }\bibfield  {title} {\enquote
  {\bibinfo {title} {Discrete time-crystalline order enabled by quantum
  many-body scars: Entanglement steering via periodic driving},}\ }\href
  {\doibase 10.1103/PhysRevLett.127.090602} {\bibfield  {journal} {\bibinfo
  {journal} {Phys. Rev. Lett.}\ }\textbf {\bibinfo {volume} {127}},\ \bibinfo
  {pages} {090602} (\bibinfo {year} {2021})}\BibitemShut {NoStop}%
\bibitem [{\citenamefont {Gong}\ \emph {et~al.}(2018)\citenamefont {Gong},
  \citenamefont {Hamazaki},\ and\ \citenamefont {Ueda}}]{gong2018discrete}%
  \BibitemOpen
  \bibfield  {author} {\bibinfo {author} {\bibfnamefont {Z.}~\bibnamefont
  {Gong}}, \bibinfo {author} {\bibfnamefont {R.}~\bibnamefont {Hamazaki}}, \
  and\ \bibinfo {author} {\bibfnamefont {M.}~\bibnamefont {Ueda}},\ }\bibfield
  {title} {\enquote {\bibinfo {title} {Discrete time-crystalline order in
  cavity and circuit qed systems},}\ }\href {\doibase
  https://doi.org/10.1103/PhysRevLett.120.040404} {\bibfield  {journal}
  {\bibinfo  {journal} {Phys. Rev. Lett.}\ }\textbf {\bibinfo {volume} {120}},\
  \bibinfo {pages} {040404} (\bibinfo {year} {2018})}\BibitemShut {NoStop}%
\bibitem [{\citenamefont {Ke{\ss}ler}\ \emph {et~al.}(2021)\citenamefont
  {Ke{\ss}ler}, \citenamefont {Kongkhambut}, \citenamefont {Georges},
  \citenamefont {Mathey}, \citenamefont {Cosme},\ and\ \citenamefont
  {Hemmerich}}]{kessler2021observation}%
  \BibitemOpen
  \bibfield  {author} {\bibinfo {author} {\bibfnamefont {H.}~\bibnamefont
  {Ke{\ss}ler}}, \bibinfo {author} {\bibfnamefont {P.}~\bibnamefont
  {Kongkhambut}}, \bibinfo {author} {\bibfnamefont {C.}~\bibnamefont
  {Georges}}, \bibinfo {author} {\bibfnamefont {L.}~\bibnamefont {Mathey}},
  \bibinfo {author} {\bibfnamefont {J.~G.}\ \bibnamefont {Cosme}}, \ and\
  \bibinfo {author} {\bibfnamefont {A.}~\bibnamefont {Hemmerich}},\ }\bibfield
  {title} {\enquote {\bibinfo {title} {Observation of a dissipative time
  crystal},}\ }\href {\doibase 10.1103/PhysRevLett.127.043602} {\bibfield
  {journal} {\bibinfo  {journal} {Phys. Rev. Lett.}\ }\textbf {\bibinfo
  {volume} {127}},\ \bibinfo {pages} {043602} (\bibinfo {year}
  {2021})}\BibitemShut {NoStop}%
\bibitem [{\citenamefont {Zanardi}\ \emph {et~al.}(2000)\citenamefont
  {Zanardi}, \citenamefont {Zalka},\ and\ \citenamefont
  {Faoro}}]{Zanardi2000Entangling}%
  \BibitemOpen
  \bibfield  {author} {\bibinfo {author} {\bibfnamefont {P.}~\bibnamefont
  {Zanardi}}, \bibinfo {author} {\bibfnamefont {C.}~\bibnamefont {Zalka}}, \
  and\ \bibinfo {author} {\bibfnamefont {L.}~\bibnamefont {Faoro}},\ }\bibfield
   {title} {\enquote {\bibinfo {title} {Entangling power of quantum
  evolutions},}\ }\href {\doibase 10.1103/PhysRevA.62.030301} {\bibfield
  {journal} {\bibinfo  {journal} {Phys. Rev. A}\ }\textbf {\bibinfo {volume}
  {62}},\ \bibinfo {pages} {030301} (\bibinfo {year} {2000})}\BibitemShut
  {NoStop}%
\bibitem [{\citenamefont {Zanardi}(2001)}]{Zanardi2001Entanglement}%
  \BibitemOpen
  \bibfield  {author} {\bibinfo {author} {\bibfnamefont {P.}~\bibnamefont
  {Zanardi}},\ }\bibfield  {title} {\enquote {\bibinfo {title} {Entanglement of
  quantum evolutions},}\ }\href {\doibase 10.1103/PhysRevA.63.040304}
  {\bibfield  {journal} {\bibinfo  {journal} {Phys. Rev. A}\ }\textbf {\bibinfo
  {volume} {63}},\ \bibinfo {pages} {040304} (\bibinfo {year}
  {2001})}\BibitemShut {NoStop}%
\bibitem [{\citenamefont {Prosen}\ and\ \citenamefont
  {Pi\ifmmode~\check{z}\else \v{z}\fi{}orn}(2007)}]{Prosen2007Operator}%
  \BibitemOpen
  \bibfield  {author} {\bibinfo {author} {\bibfnamefont {T.~c.~v.}\
  \bibnamefont {Prosen}}\ and\ \bibinfo {author} {\bibfnamefont
  {I.}~\bibnamefont {Pi\ifmmode~\check{z}\else \v{z}\fi{}orn}},\ }\bibfield
  {title} {\enquote {\bibinfo {title} {Operator space entanglement entropy in a
  transverse ising chain},}\ }\href {\doibase 10.1103/PhysRevA.76.032316}
  {\bibfield  {journal} {\bibinfo  {journal} {Phys. Rev. A}\ }\textbf {\bibinfo
  {volume} {76}},\ \bibinfo {pages} {032316} (\bibinfo {year}
  {2007})}\BibitemShut {NoStop}%
\bibitem [{\citenamefont {Pi\ifmmode~\check{z}\else \v{z}\fi{}orn}\ and\
  \citenamefont {Prosen}(2009)}]{Pizorn2009Operator}%
  \BibitemOpen
  \bibfield  {author} {\bibinfo {author} {\bibfnamefont {I.}~\bibnamefont
  {Pi\ifmmode~\check{z}\else \v{z}\fi{}orn}}\ and\ \bibinfo {author}
  {\bibfnamefont {T.~c.~v.}\ \bibnamefont {Prosen}},\ }\bibfield  {title}
  {\enquote {\bibinfo {title} {Operator space entanglement entropy in $xy$ spin
  chains},}\ }\href {\doibase 10.1103/PhysRevB.79.184416} {\bibfield  {journal}
  {\bibinfo  {journal} {Phys. Rev. B}\ }\textbf {\bibinfo {volume} {79}},\
  \bibinfo {pages} {184416} (\bibinfo {year} {2009})}\BibitemShut {NoStop}%
\bibitem [{\citenamefont {Zhou}\ and\ \citenamefont
  {Luitz}(2017)}]{Zhou2017Operator}%
  \BibitemOpen
  \bibfield  {author} {\bibinfo {author} {\bibfnamefont {T.}~\bibnamefont
  {Zhou}}\ and\ \bibinfo {author} {\bibfnamefont {D.~J.}\ \bibnamefont
  {Luitz}},\ }\bibfield  {title} {\enquote {\bibinfo {title} {Operator
  entanglement entropy of the time evolution operator in chaotic systems},}\
  }\href {\doibase 10.1103/PhysRevB.95.094206} {\bibfield  {journal} {\bibinfo
  {journal} {Phys. Rev. B}\ }\textbf {\bibinfo {volume} {95}},\ \bibinfo
  {pages} {094206} (\bibinfo {year} {2017})}\BibitemShut {NoStop}%
\bibitem [{\citenamefont {{Perez-Garcia}}\ \emph {et~al.}(2007)\citenamefont
  {{Perez-Garcia}}, \citenamefont {Verstraete}, \citenamefont {Wolf},\ and\
  \citenamefont {Cirac}}]{Perez-Garcia2007Matrix}%
  \BibitemOpen
  \bibfield  {author} {\bibinfo {author} {\bibfnamefont {D.}~\bibnamefont
  {{Perez-Garcia}}}, \bibinfo {author} {\bibfnamefont {F.}~\bibnamefont
  {Verstraete}}, \bibinfo {author} {\bibfnamefont {M.}~\bibnamefont {Wolf}}, \
  and\ \bibinfo {author} {\bibfnamefont {J.}~\bibnamefont {Cirac}},\ }\bibfield
   {title} {\enquote {\bibinfo {title} {Matrix product state
  representations},}\ }\href {http://dl.acm.org/citation.cfm?id=2011833}
  {\bibfield  {journal} {\bibinfo  {journal} {Quantum Inf. Comput.}\ }\textbf
  {\bibinfo {volume} {7}},\ \bibinfo {pages} {401} (\bibinfo {year}
  {2007})}\BibitemShut {NoStop}%
\bibitem [{\citenamefont {Shen}\ \emph {et~al.}(2023)\citenamefont {Shen},
  \citenamefont {Guo},\ and\ \citenamefont {Yang}}]{Shen2023Construction}%
  \BibitemOpen
  \bibfield  {author} {\bibinfo {author} {\bibfnamefont {R.}~\bibnamefont
  {Shen}}, \bibinfo {author} {\bibfnamefont {Y.}~\bibnamefont {Guo}}, \ and\
  \bibinfo {author} {\bibfnamefont {S.}~\bibnamefont {Yang}},\ }\bibfield
  {title} {\enquote {\bibinfo {title} {Construction of non-hermitian parent
  hamiltonian from matrix product states},}\ }\href {\doibase
  https://doi.org/10.1103/PhysRevLett.130.220401} {\bibfield  {journal}
  {\bibinfo  {journal} {Phys. Rev. Lett.}\ }\textbf {\bibinfo {volume} {130}},\
  \bibinfo {pages} {220401} (\bibinfo {year} {2023})}\BibitemShut {NoStop}%
\bibitem [{\citenamefont {\text{Google Quantum AI and
  Collaborators}}(2023)}]{Google2023Nonabelian}%
  \BibitemOpen
  \bibfield  {author} {\bibinfo {author} {\bibnamefont {\text{Google Quantum AI
  and Collaborators}}},\ }\bibfield  {title} {\enquote {\bibinfo {title}
  {Non-abelian braiding of graph vertices in a superconducting processor},}\
  }\href {\doibase https://doi.org/10.1038/s41586-023-05954-4} {\bibfield
  {journal} {\bibinfo  {journal} {Nature}\ }\textbf {\bibinfo {volume} {618}},\
  \bibinfo {pages} {264} (\bibinfo {year} {2023})}\BibitemShut {NoStop}%
\bibitem [{\citenamefont {M\"uller}\ \emph {et~al.}(2009)\citenamefont
  {M\"uller}, \citenamefont {Lesanovsky}, \citenamefont {Weimer}, \citenamefont
  {B\"uchler},\ and\ \citenamefont {Zoller}}]{Mesoscopic2009Muller}%
  \BibitemOpen
  \bibfield  {author} {\bibinfo {author} {\bibfnamefont {M.}~\bibnamefont
  {M\"uller}}, \bibinfo {author} {\bibfnamefont {I.}~\bibnamefont
  {Lesanovsky}}, \bibinfo {author} {\bibfnamefont {H.}~\bibnamefont {Weimer}},
  \bibinfo {author} {\bibfnamefont {H.~P.}\ \bibnamefont {B\"uchler}}, \ and\
  \bibinfo {author} {\bibfnamefont {P.}~\bibnamefont {Zoller}},\ }\bibfield
  {title} {\enquote {\bibinfo {title} {Mesoscopic rydberg gate based on
  electromagnetically induced transparency},}\ }\href {\doibase
  https://doi.org/10.1103/PhysRevLett.102.170502} {\bibfield  {journal}
  {\bibinfo  {journal} {Phys. Rev. Lett.}\ }\textbf {\bibinfo {volume} {102}},\
  \bibinfo {pages} {170502} (\bibinfo {year} {2009})}\BibitemShut {NoStop}%
\bibitem [{\citenamefont {Evered}\ \emph {et~al.}(2023)\citenamefont {Evered},
  \citenamefont {Bluvstein}, \citenamefont {Kalinowski}, \citenamefont {Ebadi},
  \citenamefont {Manovitz}, \citenamefont {Zhou}, \citenamefont {Li},
  \citenamefont {Geim}, \citenamefont {Wang}, \citenamefont {Maskara} \emph
  {et~al.}}]{Evered2023High}%
  \BibitemOpen
  \bibfield  {author} {\bibinfo {author} {\bibfnamefont {S.~J.}\ \bibnamefont
  {Evered}}, \bibinfo {author} {\bibfnamefont {D.}~\bibnamefont {Bluvstein}},
  \bibinfo {author} {\bibfnamefont {M.}~\bibnamefont {Kalinowski}}, \bibinfo
  {author} {\bibfnamefont {S.}~\bibnamefont {Ebadi}}, \bibinfo {author}
  {\bibfnamefont {T.}~\bibnamefont {Manovitz}}, \bibinfo {author}
  {\bibfnamefont {H.}~\bibnamefont {Zhou}}, \bibinfo {author} {\bibfnamefont
  {S.~H.}\ \bibnamefont {Li}}, \bibinfo {author} {\bibfnamefont {A.~A.}\
  \bibnamefont {Geim}}, \bibinfo {author} {\bibfnamefont {T.~T.}\ \bibnamefont
  {Wang}}, \bibinfo {author} {\bibfnamefont {N.}~\bibnamefont {Maskara}},
  \emph {et~al.},\ }\bibfield  {title} {\enquote {\bibinfo {title}
  {High-fidelity parallel entangling gates on a neutral atom quantum
  computer},}\ }\href {\doibase https://doi.org/10.1038/s41586-023-06481-y}
  {\bibfield  {journal} {\bibinfo  {journal} {Nature}\ }\textbf {\bibinfo
  {volume} {622}},\ \bibinfo {pages} {268} (\bibinfo {year}
  {2023})}\BibitemShut {NoStop}%
\bibitem [{\citenamefont {McEwen}\ \emph {et~al.}(2021)\citenamefont {McEwen},
  \citenamefont {Kafri}, \citenamefont {Chen}, \citenamefont {Atalaya},
  \citenamefont {Satzinger}, \citenamefont {Quintana}, \citenamefont {Klimov},
  \citenamefont {Sank}, \citenamefont {Gidney}, \citenamefont {Fowler} \emph
  {et~al.}}]{Mcewen2021Removing}%
  \BibitemOpen
  \bibfield  {author} {\bibinfo {author} {\bibfnamefont {M.}~\bibnamefont
  {McEwen}}, \bibinfo {author} {\bibfnamefont {D.}~\bibnamefont {Kafri}},
  \bibinfo {author} {\bibfnamefont {Z.}~\bibnamefont {Chen}}, \bibinfo {author}
  {\bibfnamefont {J.}~\bibnamefont {Atalaya}}, \bibinfo {author} {\bibfnamefont
  {K.}~\bibnamefont {Satzinger}}, \bibinfo {author} {\bibfnamefont
  {C.}~\bibnamefont {Quintana}}, \bibinfo {author} {\bibfnamefont {P.~V.}\
  \bibnamefont {Klimov}}, \bibinfo {author} {\bibfnamefont {D.}~\bibnamefont
  {Sank}}, \bibinfo {author} {\bibfnamefont {C.}~\bibnamefont {Gidney}},
  \bibinfo {author} {\bibfnamefont {A.}~\bibnamefont {Fowler}},  \emph
  {et~al.},\ }\bibfield  {title} {\enquote {\bibinfo {title} {Removing
  leakage-induced correlated errors in superconducting quantum error
  correction},}\ }\href {\doibase https://doi.org/10.1038/s41467-021-21982-y}
  {\bibfield  {journal} {\bibinfo  {journal} {Nat. Commun.}\ }\textbf {\bibinfo
  {volume} {12}},\ \bibinfo {pages} {1761} (\bibinfo {year}
  {2021})}\BibitemShut {NoStop}%
\bibitem [{\citenamefont {Miao}\ \emph {et~al.}(2022)\citenamefont {Miao},
  \citenamefont {McEwen}, \citenamefont {Atalaya}, \citenamefont {Kafri},
  \citenamefont {Pryadko}, \citenamefont {Bengtsson}, \citenamefont {Opremcak},
  \citenamefont {Satzinger}, \citenamefont {Chen}, \citenamefont {Klimov} \emph
  {et~al.}}]{Miao2022Overcoming}%
  \BibitemOpen
  \bibfield  {author} {\bibinfo {author} {\bibfnamefont {K.~C.}\ \bibnamefont
  {Miao}}, \bibinfo {author} {\bibfnamefont {M.}~\bibnamefont {McEwen}},
  \bibinfo {author} {\bibfnamefont {J.}~\bibnamefont {Atalaya}}, \bibinfo
  {author} {\bibfnamefont {D.}~\bibnamefont {Kafri}}, \bibinfo {author}
  {\bibfnamefont {L.~P.}\ \bibnamefont {Pryadko}}, \bibinfo {author}
  {\bibfnamefont {A.}~\bibnamefont {Bengtsson}}, \bibinfo {author}
  {\bibfnamefont {A.}~\bibnamefont {Opremcak}}, \bibinfo {author}
  {\bibfnamefont {K.~J.}\ \bibnamefont {Satzinger}}, \bibinfo {author}
  {\bibfnamefont {Z.}~\bibnamefont {Chen}}, \bibinfo {author} {\bibfnamefont
  {P.~V.}\ \bibnamefont {Klimov}},  \emph {et~al.},\ }\bibfield  {title}
  {\enquote {\bibinfo {title} {Overcoming leakage in scalable quantum error
  correction},}\ }\href {https://arxiv.org/abs/2211.04728} {\bibfield
  {journal} {\bibinfo  {journal} {arXiv:2211.04728}\ } (\bibinfo {year}
  {2022})}\BibitemShut {NoStop}%
\bibitem [{\citenamefont {Braunstein}\ and\ \citenamefont
  {Caves}(1994)}]{Braunstein1994Statistical}%
  \BibitemOpen
  \bibfield  {author} {\bibinfo {author} {\bibfnamefont {S.~L.}\ \bibnamefont
  {Braunstein}}\ and\ \bibinfo {author} {\bibfnamefont {C.~M.}\ \bibnamefont
  {Caves}},\ }\bibfield  {title} {\enquote {\bibinfo {title} {Statistical
  distance and the geometry of quantum states},}\ }\href {\doibase
  https://doi.org/10.1103/PhysRevLett.72.3439} {\bibfield  {journal} {\bibinfo
  {journal} {Phys. Rev. Lett.}\ }\textbf {\bibinfo {volume} {72}},\ \bibinfo
  {pages} {3439} (\bibinfo {year} {1994})}\BibitemShut {NoStop}%
\bibitem [{\citenamefont {Pezz\'e}\ and\ \citenamefont
  {Smerzi}(2009)}]{Pezze2009Entanglement}%
  \BibitemOpen
  \bibfield  {author} {\bibinfo {author} {\bibfnamefont {L.}~\bibnamefont
  {Pezz\'e}}\ and\ \bibinfo {author} {\bibfnamefont {A.}~\bibnamefont
  {Smerzi}},\ }\bibfield  {title} {\enquote {\bibinfo {title} {Entanglement,
  nonlinear dynamics, and the heisenberg limit},}\ }\href {\doibase
  https://doi.org/10.1103/PhysRevLett.102.100401} {\bibfield  {journal}
  {\bibinfo  {journal} {Phys. Rev. Lett.}\ }\textbf {\bibinfo {volume} {102}},\
  \bibinfo {pages} {100401} (\bibinfo {year} {2009})}\BibitemShut {NoStop}%
\bibitem [{\citenamefont {Hyllus}\ \emph {et~al.}(2012)\citenamefont {Hyllus},
  \citenamefont {Laskowski}, \citenamefont {Krischek}, \citenamefont
  {Schwemmer}, \citenamefont {Wieczorek}, \citenamefont {Weinfurter},
  \citenamefont {Pezz\'e},\ and\ \citenamefont {Smerzi}}]{Hyllus2012Fisher}%
  \BibitemOpen
  \bibfield  {author} {\bibinfo {author} {\bibfnamefont {P.}~\bibnamefont
  {Hyllus}}, \bibinfo {author} {\bibfnamefont {W.}~\bibnamefont {Laskowski}},
  \bibinfo {author} {\bibfnamefont {R.}~\bibnamefont {Krischek}}, \bibinfo
  {author} {\bibfnamefont {C.}~\bibnamefont {Schwemmer}}, \bibinfo {author}
  {\bibfnamefont {W.}~\bibnamefont {Wieczorek}}, \bibinfo {author}
  {\bibfnamefont {H.}~\bibnamefont {Weinfurter}}, \bibinfo {author}
  {\bibfnamefont {L.}~\bibnamefont {Pezz\'e}}, \ and\ \bibinfo {author}
  {\bibfnamefont {A.}~\bibnamefont {Smerzi}},\ }\bibfield  {title} {\enquote
  {\bibinfo {title} {Fisher information and multiparticle entanglement},}\
  }\href {\doibase https://doi.org/10.1103/PhysRevA.85.022321} {\bibfield
  {journal} {\bibinfo  {journal} {Phys. Rev. A}\ }\textbf {\bibinfo {volume}
  {85}},\ \bibinfo {pages} {022321} (\bibinfo {year} {2012})}\BibitemShut
  {NoStop}%
\bibitem [{\citenamefont {T\'oth}(2012)}]{Toth2012Multipartite}%
  \BibitemOpen
  \bibfield  {author} {\bibinfo {author} {\bibfnamefont {G.}~\bibnamefont
  {T\'oth}},\ }\bibfield  {title} {\enquote {\bibinfo {title} {Multipartite
  entanglement and high-precision metrology},}\ }\href {\doibase
  https://doi.org/10.1103/PhysRevA.85.022322} {\bibfield  {journal} {\bibinfo
  {journal} {Phys. Rev. A}\ }\textbf {\bibinfo {volume} {85}},\ \bibinfo
  {pages} {022322} (\bibinfo {year} {2012})}\BibitemShut {NoStop}%
\bibitem [{\citenamefont {Iadecola}\ \emph {et~al.}(2019)\citenamefont
  {Iadecola}, \citenamefont {Schecter},\ and\ \citenamefont
  {Xu}}]{Iadecola2019Quantum}%
  \BibitemOpen
  \bibfield  {author} {\bibinfo {author} {\bibfnamefont {T.}~\bibnamefont
  {Iadecola}}, \bibinfo {author} {\bibfnamefont {M.}~\bibnamefont {Schecter}},
  \ and\ \bibinfo {author} {\bibfnamefont {S.}~\bibnamefont {Xu}},\ }\bibfield
  {title} {\enquote {\bibinfo {title} {Quantum many-body scars from magnon
  condensation},}\ }\href {\doibase 10.1103/PhysRevB.100.184312} {\bibfield
  {journal} {\bibinfo  {journal} {Phys. Rev. B}\ }\textbf {\bibinfo {volume}
  {100}},\ \bibinfo {pages} {184312} (\bibinfo {year} {2019})}\BibitemShut
  {NoStop}%
\bibitem [{\citenamefont {Kitagawa}\ and\ \citenamefont
  {Ueda}(1993)}]{Kitagawa1993Squeezed}%
  \BibitemOpen
  \bibfield  {author} {\bibinfo {author} {\bibfnamefont {M.}~\bibnamefont
  {Kitagawa}}\ and\ \bibinfo {author} {\bibfnamefont {M.}~\bibnamefont
  {Ueda}},\ }\bibfield  {title} {\enquote {\bibinfo {title} {Squeezed spin
  states},}\ }\href {\doibase https://doi.org/10.1103/PhysRevA.47.5138}
  {\bibfield  {journal} {\bibinfo  {journal} {Phys. Rev. A}\ }\textbf {\bibinfo
  {volume} {47}},\ \bibinfo {pages} {5138} (\bibinfo {year}
  {1993})}\BibitemShut {NoStop}%
\end{thebibliography}%

\clearpage

\onecolumngrid
\makeatletter

\setcounter{MaxMatrixCols}{10}

\setcounter{figure}{0}
\renewcommand{\thefigure}{S\@arabic\c@figure}
\setcounter{equation}{0} \makeatletter
\renewcommand \theequation{S\@arabic\c@equation}
\renewcommand \thetable{S\@arabic\c@table}

\begin{center} 
	{\large \bf Supplementary Materials for: Embedding Quantum Many-Body Scars into Decoherence-Free Subspaces}
\end{center}

\section{Obtaining local projectors via the compressed matrix product state}

In this section, we present the compressed matrix product state (MPS) technique for finding local projectors annihilating the whole scar subspace.  For typical models hosting many-body scar towers, there exists one reference state $\ket{S_0}$ in the scar subspace, which admits a simple entanglement structure. Other scarred eigenstates are generated by acting certain ladder operator $Q^\dagger$ on $\ket{S_0}$ repeatedly (up to some normalization constants):
\begin{equation}
    \ket{S_n}=(Q^\dagger)^n\ket{S_0}.
\end{equation}
In order to make the non-decaying eigenmodes of the constructed Liouvillians equally spaced on the imaginary axis (which lead to the persistent periodic oscillations), we require that the original scarred Hamiltonians exhibit the restricted spectrum generating algebra in the scar subspace $W$, i.e., $([H,Q^\dagger]-\omega Q^\dagger)W=0$. 

Before proceeding on, we clarify the connections between the spectrum generating algebra generated by $Q^\dagger$, and the Shiraishi-Mori embedding formalism. 
In general, for quantum many-body scarred Hamiltonians, there is no direct relation between these two concepts. On the one hand, for a scarred Hamiltonian following the Shiraishi-Mori formalism, the embedded scar subspace is not required to host the spectrum generating algebra (e.g., the two original examples in the Shiraishi-Mori paper, Ref.~\cite{Shiraishi2017Systematic}). In these cases, the energy spacings between scar states are not guaranteed to be equal. On the other hand, scarred Hamiltonians hosting the spectrum generating algebra can go beyond the formalism of Shiraishi-Mori embedding (e.g., the last two of our examples, the spin-$1$ AKLT model~\cite{Moudgalya2018Exact} and the domain-wall preserving model~\cite{Iadecola2020Quantum}). However, in certain scarred models (e.g., the first two of our examples, the toy model~\cite{Choi2019emergent} and the spin-$1$ $XY$ model~\cite{Schecter2019Weak}), the scarred Hamiltonians can simultaneously follow the Shiraishi-Mori embedding formalism and host the spectrum generating algebra. 

Now we introduce the compressed MPS technique to obtain local projectors annihilating the whole scar tower (especially useful for scarred models beyond the Shiraishi-Mori formalism).  The method is to compress the scar tower into the following single state with a parameter $\beta$ \cite{shibata2020onsager,Chattopadhyay2020quantum,mark2020unified}:
\begin{equation}
    \ket{S(\beta)}=\exp(\beta Q^\dagger)\ket{S_0}=\sum_{n=0}\frac{\beta^n}{n!}\ket{S_n}.
\end{equation}
Typically, the reference state $\ket{S_0}$ can be represented as the MPS form:
\begin{equation}\label{eqn:MPS}
\ket{S_0}=\sum_{\mu_1,\mu_2,\cdots,\mu_L}\textnormal{Tr}\left[A^{(\mu_1)}_1A^{(\mu_2)}_2\cdots A^{(\mu_L)}_L\right]\ket{\mu_1,\mu_2,\cdots,\mu_L},
\end{equation}
where $\{\mu_j\}$ label the local bases on site $j$. The dimension $\chi$ of matrices $\{A^{(\mu_j)}_j\}$ (bond dimension) is usually of order $\mathcal{O}(1)$ for $\ket{S_0}$. If $\exp(\beta Q^\dagger)$ further admits a matrix product operator (MPO) expression with bond dimension $\chi_O$, the bond dimension of the compressed state $\ket{S(\beta)}$ will be no larger than $\chi \times \chi_O$. Finding local projectors annihilating $\ket{S(\beta)}$ for \textit{any} $\beta$ is necessary and sufficient for obtaining local projectors to annihilate the whole scar tower. The $\mathcal{O}(1)$ bond dimension of $\ket{S(\beta)}$ allows us to apply the standard linear algebra techniques (see the next subsection) to construct local projectors annihilating the local tensors of $\ket{S(\beta)}$ for any $\beta$ (could be two-local, three-local, ...).
In contrast, the bond dimensions of the MPS representations for the scar-tower states $\{\ket{S_n}\}$ increase linearly with the label $n$, such that $\{\ket{S_n}\}$ near the middle of the spectrum typically possess logarithmic entanglement entropy. It would be very challenging to directly find the common local projectors annihilating all the $\{\ket{S_n}\}$ through the standard approach. This demonstrates the necessity of applying the compressed MPS technique as a general and systematic method to construct the local projectors in the dissipators.

\subsection{Finding projectors for local MPSs}
\label{subsec:local_proj_method}

Before proceeding to specific models, we give a brief review on how to construct local projectors for a given MPS \cite{Perez-Garcia2007Matrix}. The following method is sufficient in the sense that the $k$-local projectors annihilate the $k$-local MPS, thus annihilating the whole MPS. We denote the $k$-local MPS as
\begin{equation}
    \ket{M(l,r)}=\sum_{\mu_1,\cdots,\mu_k}(A^{(\mu_1)}_1 \cdots A^{(\mu_k)}_k )_{l,r} \ket{\mu_1,\cdots,\mu_k},
\end{equation}
where $l,r$ are the left and right uncontracted indices (dangling bonds) of the local MPS. 
Therefore, there exist $\chi^2$ such states within the $d^k$-dimensional Hilbert space ($d$ is the physical dimension of $\{\mu_j\}$). 
The $k$-local projectors are expected to annihilate all the $\{\ket{M(l,r)}\}_{l,r=1}^\chi$, so we need to find a set of orthonormal bases $\{\ket{v_p}\}$ for the linear subspace spanned by $\{\ket{M(l,r)}\}_{l,r=1}^\chi$. The local projector annihilating all the $\{\ket{M(l,r)}\}_{l,r=1}^\chi$ can be constructed as $P = 1 -\sum_{p} \ket{v_p}\bra{v_p}$. 
Specifically, one can first figure out the maximal linearly independent set of $\{\ket{M(l,r)}\}_{l,r=1}^\chi$, and then compute their Gram matrix to decide the orthonormal bases $\{\ket{v_p}\}$~\cite{Shen2023Construction}.
In this paper we would like to keep the designed dissipators as local as possible, thus only considering the case of $k=2$ or $3$, which could readily be implemented with existing quantum simulation technologies.

\subsection{Toy model hosting Dicke states as scars}

The first model we consider is a one dimensional spin-$1/2$ chain that hosts all the $x$-direction Dicke states as quantum many-body scars, namely the $L+1$ components of the largest spin representation of $S=L/2$, $\{\ket{S=L/2,S_x = m}\}$, $S_x = \sum_j \sigma_j^x /2$, $m = -L/2, -L/2+1 \cdots, L/2 $  \cite{Choi2019emergent}
\begin{equation}
    H_{\text{toy}} = \frac{\Omega}{2} \sum_j \sigma_j^x + \sum_j P_j h_j P_j.
\end{equation}
$P_j = (1 -\vec{\sigma}_j\cdot\vec{\sigma}_{j+1}) / 4$ and $h_j=\sum_{\mu\nu}J_{\mu\nu}\sigma_{j-1}^\mu\sigma_{j+2}^\nu$ is a generic two-spin operator. 
Since all the Dicke states locally are spin triplets, the two-local projectors annihilating them exactly corresponds to $\{P_j\}$, which project onto neighboring pairs of singlets.
We will alternatively show how to obtain these projectors through the compressed MPS technique.

Within the scar subspace we choose the $x$-direction ferromagnetic state with all spins pointing to $\ket{-}=(\ket{\uparrow} + \ket{\downarrow})/\sqrt{2}$ as the reference state
\begin{equation}
    \ket{S_0}=|-,\cdots,-\rangle.
\end{equation}

The ladder operator corresponds to the summation of all the raising operators in the $x$-direction
\begin{equation}
    Q^\dagger=\sum_j (\sigma_j^y+i\sigma_j^z)/2.
\end{equation}
The ladder operator increases the $x$-direction magnetization by one, so $(Q^\dagger)^{L+1}=0$. 
The compressed MPS turns out to be 
\begin{equation}
    \ket{S(\beta)}=\exp(\beta Q^\dagger)\ket{S_0}=\bigotimes_j (\ket{-}_j+\beta\ket{+}_j),
\end{equation}
with bond dimension $\chi=1$. 
All the two-local components in $\ket{S(\beta)}$ have the forms of $\ket{--}, \ket{++}$ and $\ket{+-}+\ket{-+}$, which are indeed spin-$1$ triplet states of two spin-$1/2$. Therefore, the two-local projectors annihilating the whole scar tower should be taken as the projectors onto the spin-$0$ singlet state as expected
\begin{equation}
    P_j=\frac{1}{4}(1-\vec{\sigma}_j\cdot\vec{\sigma}_{j+1}).
\end{equation}

\subsection{Spin-$1$ $XY$ model}
The second example we present is the spin-$1$ $XY$ model \cite{Schecter2019Weak}
\begin{equation}
    H_{XY} = \sum_{j} [ S^x_j S^x_{j+1} + S^y_j S^y_{j+1} + h S^z_j + D (S^z_j)^2 ].
\end{equation}
The reference state and compressed MPS of this model also manifest as product states:
\begin{align}\label{eqn:XY}
    \ket{S_0}&=\ket{-1,-1,\cdots,-1},\quad Q^\dagger=\sum_j (-1)^j(S_j^{+})^2,\nonumber\\
    \ket{S(\beta)}&=\bigotimes_{j=1}^{L/2} (\ket{-1}_{2j-1}-\beta\ket{1}_{2j-1})(\ket{-1}_{2j}+\beta\ket{1}_{2j}).
\end{align}
Here $\ket{S(\beta)}$ is two-site translational invariant. The two-local Hilbert space of the spin-$1$ model is spanned by nine orthogonal bases. We adopt the same notation $\ket{T^{S,m}}_{j,j+1}$ [$ T_{j,j+1}^{S,m} = (\ket{T^{S,m}}\bra{T^{S,m}})_{j,j+1} $] as in the main text to label them as the representation of SU(2) symmetry group, where $S=0,1,2$ is the total spin, and $m=-S,-S+1,\cdots,S$ is the total spin-$z$ polarization. 
We expand the nine two-local bases on the computational bases as follows~\cite{mark2020unified}:
\begin{align}
    &\ket{T^{S=2,m=-2}}=\ket{-1,-1},&& \ket{T^{S=2,m=-1}}=\frac{1}{\sqrt{2}}(\ket{0,-1}+\ket{-1,0}),\nonumber\\
    &\ket{T^{S=2,m=0}}=\frac{1}{\sqrt{6}}(\ket{1,-1}+2\ket{0,0}+\ket{-1,1}), &
    &\ket{T^{S=2,m=1}}=\frac{1}{\sqrt{2}}(\ket{0,1}+\ket{1,0}),\nonumber\\ &\ket{T^{S=2,m=2}}=\ket{1,1},&&\ket{T^{S=1,m=-1}}=\frac{1}{\sqrt{2}}(\ket{0,-1}-\ket{-1,0}),\nonumber\\ &\ket{T^{S=1,m=0}}=\frac{1}{\sqrt{2}}(\ket{1,-1}-\ket{-1,1}),&&\ket{T^{S=1,m=1}}=\frac{1}{\sqrt{2}}(\ket{1,0}-\ket{0,1}),\nonumber\\
    &\ket{T^{S=0,m=0}}=\frac{1}{\sqrt{3}}(\ket{1,-1}-\ket{0,0}+\ket{-1,1}).
\end{align}

The two-local components in $\ket{S(\beta)}$ include three of them: $\ket{T^{S=2,m=-2}}_{j,j+1},\ket{T^{S=2,m=2}}_{j,j+1}$ and $\ket{T^{S=1,m=0}}_{j,j+1}$, such that the two-local projectors annihilating the whole scar tower correspond to the summation of projectors onto the other six bases. Interestingly, the null space of the two-local $XY$ interaction terms $\{S^x_j S^x_{j+1} + S^y_j S^y_{j+1}\}$ coincides with the subspace spanned by the six two-local bases \cite{Schecter2019Weak,mark2020unified}, so in the main text we design the jump operators in the simple form as $L_j = S_j^x (S^x_j S^x_{j+1} + S^y_j S^y_{j+1}) $. 

\subsection{AKLT model}
\label{subsec:AKLT_scar}
The Hamiltonian of the spin-$1$ AKLT model reads
\begin{equation}
    H_{\text{AKLT}}=\sum_j T_{j,j+1}^{S=2}=\sum_j \left(\frac{1}{3} + \frac{1}{2} \vec{S}_j\cdot \vec{S}_{j+1} + \frac{1}{6} ( \vec{S}_j\cdot \vec{S}_{j+1} )^2 \right).
\end{equation}
Below we all consider even system sizes $L$. 
The reference state $\ket{S_0}$ is the ground state $\ket{G}$ of $H_{\text{AKLT}}$, which admits a $\chi=2$ MPS representation as Eq.~\eqref{eqn:MPS}:
\begin{equation}
    A^{(\pm 1)}=\mp\sqrt{\frac{2}{3}}\sigma^\mp, \quad  A^{(0)}=-\frac{1}{\sqrt{3}}\sigma^z
\end{equation}
The ladder operator $ Q^\dagger=\sum_j (-1)^j(S_j^{+})^2 $ coincides with that of the $XY$ model. 
$\ket{S_{L/2}}=(Q^\dagger)^{L/2}\ket{S_0}$ is the fully polarized ferromagnetic state $\ket{1,\cdots,1}$ if $L/2$ is even, and $(Q^\dagger)^{L/2}\ket{S_0}=0$ if $L/2$ is odd. There exist at most $L/2 + 1$ scarred eigenstates within the tower.
 
The compressed state exhibits a $\chi=2$ two-site translational invariant MPS representation
\begin{equation}
    \ket{S(\beta)}=\exp(\beta Q^\dagger)\ket{S_0}
    =\sum_{\mu_1,\cdots, \mu_L} \text{Tr}\left[\prod_{j=1}^{L/2}(A^{(\mu_{2j-1})}-\beta B^{(\mu_{2j-1})})(A^{(\mu_{2j})}+\beta B^{(\mu_{2j})})\right]|\mu_1,\cdots, \mu_L\rangle,
\end{equation}
where 
\begin{equation}
    B^{(+1)}=A^{(-1)}=\sqrt{\frac{2}{3}}\sigma^+, \quad B^{(0)}=B^{(-1)}=0.
\end{equation}

The two-local projectors annihilating the whole  scar tower should annihilate the two-local tensors $AA$ and $AB-BA$ ($BB=0$), which are found as 
\begin{equation} P_j=T_{j,j+1}^{S=2,m=-2}+T_{j,j+1}^{S=2,m=-1}+T_{j,j+1}^{S=2,m=0}.
\end{equation}

In terms of these projectors, we could rewrite the AKLT Hamiltonian as
\begin{equation}
    H_{\text{AKLT}}=\sum_j (P_j+T_{j,j+1}^{S=2,m=2}+T_{j,j+1}^{S=2,m=1})=(\sum_j P_j) + H'.
\end{equation}

Since $[P_j, H']\neq 0$ \cite{mark2020unified}, the AKLT model goes beyond the Shiraishi-Mori embedding formalism.
The common null space of all the $\{P_j\}$ is larger than the scar subspace. 
One particular tricky point for the AKLT model (different from the domain-wall preserving model) is that, even after utilizing the Hamiltonian part $H_{\text{AKLT}}$, there still exist several undesired eigenstates with irrational eigenvalues within the decoherence-free subspace, corresponding to the eigenstates of $H_{\text{NH}}=H_{\text{AKLT}}-i\gamma\sum_j P_j$ with real irrational eigenvalues [Fig. 2(c) of the main text].
We leave the discussions about these undesired states in Sec.~\ref{sec:special}.

A systematic method to refine the decoherence-free subspace and expel the undesired states is to add the three-local projector that annihilates the three-local tensors in $\ket{S(\beta)}$, which include $AAA$, $BAB$, and $BAA-ABA+ AAB$. 
After carefully working out the maximal linearly independent set for these three-local MPS tensors (the method mentioned in Subsec.~\ref{subsec:local_proj_method}), we find the following simple state 
\begin{equation}
    \ket{T'}=\frac{1}{\sqrt{2}}(\ket{0,1,1}+\ket{1,1,0}).
\end{equation}

One can readily verify that $A^{(0)}A^{(1)}A^{(1)} + A^{(1)}A^{(1)}A^{(0)} = 0$ and the same for $BAB$ and $BAA-ABA+ AAB$. $\ket{T'}$ is therefore orthogonal to these three-local MPS tensors.
The corresponding three-local projectors $\{T'_{j-1,j,j+1}=(\ket{T'}\bra{T'})_{j-1,j,j+1}\}$ annihilate the whole scar tower and can effectively kill the unwanted states within the decoherence-free subspace [see derivations in Sec.~\ref{sec:special} and the numerical verification in Fig. 2(c) of the main text], even though $\ket{T'}$ is just one of many three-local states that are orthogonal to the three-local MPS tensors.

\subsection{Domain-wall preserving model}

In the domain-wall preserving model \cite{Iadecola2020Quantum}
\begin{equation}
    H_{\text{DW}} = \sum_j (\sigma_j^x - \sigma_{j-1}^z \sigma_j^x \sigma_{j+1}^z) + \Delta \sum_j\sigma_j^z + J\sum_j \sigma_j^z\sigma_{j+1}^z,
\end{equation}
the ladder operator 
\begin{equation}    
Q^\dagger = \sum_j (-1)^j P^0_{j-1} \sigma_j^+ P^0_{j+1}, \quad P^0_j = \frac{1-\sigma_j^z}{2}, \quad P^1_j = \frac{1+\sigma_j^z}{2},
\end{equation}
generates the scar tower from the reference state $\ket{S_0}=\ket{\downarrow\downarrow\cdots \downarrow}$. 

The compressed state 
\begin{equation}
    \ket{S(\beta)}=\exp(\beta Q^\dagger)\ket{S_0} \propto \prod_j ( 1 - P_j^1 P_{j+1}^1)
    \prod_j \left[1+(-1)^j \beta \sigma^+_j \right] \ket{S_0}
\end{equation}
admits a $\chi=2$ two-site translational invariant MPS representation \cite{Iadecola2020Quantum}
\begin{equation}
A^{(\downarrow)}_j =
\left( \begin{array}{cc}
0 & 0  \\
-1 & 1 
\end{array} \right),\qquad A^{(\uparrow)}_j =(-1)^j\frac{\beta}{2}
\left( \begin{array}{cc}
-1 & -1  \\
1 & 1 
\end{array} \right).
\end{equation}

The two-local components in $\ket{S(\beta)}$ for any $\beta$ includes $\ket{\downarrow\uparrow},\ket{\uparrow\downarrow}$ and $\ket{\downarrow\downarrow}$, such that the only two-local projectors annihilating the whole scar tower correspond to the Rydberg-blockade constraints $\{(\ket{\uparrow\uparrow}\bra{\uparrow\uparrow})_{j,j+1}\}$, namely two neighboring spins are forbidden to be in the up states simultaneously.

We once more mention that the null space $W'$ of the two-local projectors is \textit{exponentially} large with respect to $L$, while the dimension of the scar subspace $W$ is only $L/2+2$ (the extra state comes from another N\'eel cat state $\ket{\uparrow\downarrow\cdots \uparrow\downarrow} - (-1)^{L/2}\ket{\downarrow\uparrow\cdots \downarrow\uparrow}$).
In order to create a scar-state-only decoherence-free subspace, we crucially rely on the Hamiltonian part of the Liouvillian, $H_{\text{DW}}$, to drive unwanted
states out of $W'$ and make them decay away.

\section{Special States in the AKLT model}\label{sec:special}
Ref.~\cite{Moudgalya2018Exact} identified several classes of exact excited eigenstates in the AKLT model. Besides the scar-tower states discussed in Subsec.~\ref{subsec:AKLT_scar}, there exist 
some other eigenstates of $H'=\sum_j (T_{j,j+1}^{S=2,m=2}+T_{j,j+1}^{S=2,m=1})$ within the null space of the two-local projectors $\{P_j=T_{j,j+1}^{S=2,m=-2}+T_{j,j+1}^{S=2,m=-1}+T_{j,j+1}^{S=2,m=0}\}$, which therefore enter the decoherence-free subspace of the corresponding Liouvillian. Below we consider even system sizes $L$ and the periodic boundary condition.

A particular class of these special eigenstates corresponds to the single-magnon excitations upon the ferromagnetic state $\ket{F}=\ket{1,1,\cdots,1}$.  $\ket{F}$ apparently is annihilated by all the $\{P_j\}$ and belongs to the eigenstate of $H'$ with energy $L$ (actually it is the $n=L/2$ scar state if $L/2$ is even). As we demonstrate below, these single-magnon excitation states have irrational eigenvalues for generic $L$, which will ruin the periodic oscillations within the decoherence-free subspace.

After adding the three-local projectors $\{T'_{j-1,j,j+1}=(\ket{T'}\bra{T'})_{j-1,j,j+1}\}$, most of these single-magnon excitation states are expelled from the decoherence-free subspace. There still remain a few special states with eigenvalues $L-1$ or $L-2$ for the non-Hermitian Hamiltonian $H_{\text{NH}}=H_{\text{AKLT}}-i\gamma\sum_j (P_j + T'_{j-1,j,j+1})$ [Fig. 2(c) of the main text], for which we present their analytical expressions.

\subsection{Single-magnon excitation states}
\label{subsec:single_magnon_excit}

Consider the single-magnon excitations (flipping a spin from $\ket{1}$ to $\ket{0}$) upon $\ket{F}$ with the momentum $k$, up to normalization constants:
\begin{equation}
    \ket{0_k} = \sum_j e^{i k j} S_j^- \ket{F},
\end{equation}
where $k=2\pi l/ L, \  l= 0,1,2,\cdots,L-1$. 

These states are annihilated by the two-local projectors $\{P_j\}$: Every computational basis in $\ket{0_k}$ only contains one $\ket{0}$ and $L-1$ $\ket{1}$, while $\ket{T^{S=2,m=-2,-1,0}}_{j,j+1}$ contain $\ket{-1}$ or adjacent two $\ket{0}$'s. 

On the other hand, the action of $T_{j,j+1}^{S=2,m=1} + T_{j,j+1}^{S=2,m=2}$ onto $\ket{0_k}$ is given by
\begin{equation}
(T_{j,j+1}^{S=2,m=1} + T_{j,j+1}^{S=2,m=2}) \ket{0_k} = \frac{1}{2}\left[ e^{i k j} + e^{i k (j+1)} \right](S_j^- + S_{j+1}^-)\ket{F} + \sum_{n \neq j, j+1} e^{i k n} S_n^- \ket{F}.
\end{equation}
Summing over the site index $j$, we obtain 
\begin{equation}
   \sum_j (T_{j,j+1}^{S=2,m=1} + T_{j,j+1}^{S=2,m=2}) \ket{0_k} = \left[L -1 + \cos(k) \right] \ket{0_k},
\end{equation}
such that $\ket{0_k}$ is an eigenstate of $H'$ with energy $E=L-1+\cos(k)$. For generic $L$, for example $L = 8$, $\ket{0_k}$ acquire irrational eigenvalues $ 7 \pm \sqrt{2}/2 $ for the momentum $k=\pi/4$ and $3\pi/4$ [Fig. 2(c) of the main text], and for larger $L$ more irrational energy levels come in. 

The action of the three-local projectors onto $\ket{0_k}$ goes as
\begin{equation}
    T'_{j-1,j,j+1}\ket{0_k}=e^{ikj}\cos(k)(S_{j-1}^-+S_{j+1}^-)\ket{F}.
\end{equation}
The right hand side vanishes if and only if $\cos(k)=0$. Therefore, after adding the three-local dissipators, among these single-magnon excitation states only $\ket{0_k}$ with $k=\pi/2$ and $k=3\pi/2$ survive in the decoherence-free subspace, with energy $L-1$. Note that such momentum $k$ only exists if $L/2$ is even. 

In Fig. 2(c) of the main text we compute the spectrum of the non-Hermitian spin-$1$ AKLT Hamiltonian instead of the whole Liouvillian, in order to access to larger system sizes ($L\ge 8$) to observe the irrational real eigenvalues.

\subsection{One remaining special state at $E=L-2$}
After adding the three-local projectors, we carefully analyze the states remaining in the decoherence-free subspace obtained by numerical diagonalization. Besides the scar-tower states and the $\ket{0_k}$ states with $k=\pi/2$ and $k=3\pi/2$ for $L/2$ is even, we find one more special state with eigenvalue $E=L-2$ for all the even $L$. We denote
\begin{align}
    &\ket{S_1'}=\sum_{j=1}^L (-1)^j (S_j^-)^2\ket{F} = \sum_{j=1}^L (-1)^j\ket{(-1)_j},\nonumber\\
    &\ket{S_2'}=\sum_{j=1}^L \sum_{j'=1}^{L/2-1} (-1)^{j+j'} S_j^{-} S_{j+2j'}^{-} \ket{F} = \sum_{j=1}^L (-1)^j \sum_{j'=1}^{L/2-1}(-1)^{j'}\ket{0_j 0_{j+2j'}}.
\end{align}

$\ket{S'_1}$ can be interpreted as a single bi-magnon (flipping a spin from $\ket{1}$ to $\ket{-1}$) moving with the momentum $\pi$ on the reference state $\ket{F}$. $\ket{S'_2}$ corresponds to the case that two magnons spaced by even-site distance move with the center-of-mass momentum of $\pi$. Since the bi-magnon $\ket{(-1)_j}$ in $\ket{S'_1}$ has the momentum of $\pi$,
and $\ket{S'_2}$ does not contain $\ket{-1}$ or adjacent two $\ket{0}$'s, these two states are annihilated by the tow-local projectors $\{P_j=T_{j,j+1}^{S=2,m=-2}+T_{j,j+1}^{S=2,m=-1}+T_{j,j+1}^{S=2,m=0}\}$. 
Below we further prove that $\ket{S'_1}$ and $\ket{S'_2}$ are annihilated by the three-local projectors $ \{T'_{j-1,j,j+1} =(\ket{T'}\bra{T'})_{j-1,j,j+1} \}$, and are eigenstates of $H'=\sum_j (T_{j,j+1}^{S=2,m=2}+T_{j,j+1}^{S=2,m=1})$ with $E=L-2$. We then reveal their relationship with the scar-tower state of $n=L/2-1$, $\ket{S_{L/2-1}}$, which possesses the same eigenenergy $E=L-2$.

Since $\ket{S'_1}$ does not contain $\ket{0}$, $T'_{j-1,j,j+1} \ket{S'_1} = 0$. The action of the three-local projector $T'_{j-1,j,j+1}$ on $\ket{S'_2}$ is given by
\begin{align}
    T'_{j-1,j,j+1}\ket{S'_2}=
    &\frac{1}{2}[\sum_{j'=2}^{L/2-1} (-1)^{j-1+j'}(\ket{0_{j-1} 0_{j+2j'-1}}+\ket{0_{j+1} 0_{j+2j'-1}})\nonumber\\
    &+\sum_{j'=1}^{L/2-2} (-1)^{j+1+j'}(\ket{0_{j-1} 0_{j+2j'+1}}+\ket{0_{j+1} 0_{j+2j'+1}})\nonumber\\
    &+\sum_{j'=1}^{L/2-2} (-1)^{j-1-j'} (\ket{0_{j-2j'-1}0_{j-1}}+\ket{0_{j-2j'-1}0_{j+1}})\nonumber\\
    &+\sum_{j'=2}^{L/2-1} (-1)^{j+1-j'} (\ket{0_{j-2j'+1}0_{j-1}}+\ket{0_{j-2j'+1}0_{j+1}})]\nonumber\\
    &=0.
\end{align}

As for the action of $H'$ on $\ket{S'_1}$, $T_{j,j+1}^{S=2,m=1}$ annihilates $\ket{S'_1}$ locally, and the other term $T_{j,j+1}^{S=2,m=2}$ acts like
\begin{equation}
\sum_{j=1}^L T_{j,j+1}^{S=2,m=2}\ket{S'_1} = \sum_{j=1}^L (-1)^j (\sum_{j'\neq j-1,j} T_{j',j'+1}^{S=2,m=2})\ket{(-1)_j} = (L-2)\ket{S'_1}.
\end{equation}
The eigenenergy of $\ket{S'_1}$ for the non-Hermitian AKLT Hamiltonian equals $L-2$.

Next, for $\ket{S'_2}$ we obtain
\begin{align}
    \sum_{j=1}^L T_{j,j+1}^{S=2,m=1}\ket{S'_2}
    =&\sum_{j=1}^L (-1)^j \sum_{j'=1}^{L/2-1}(-1)^{j'}(T_{j-1,j}^{S=2,m=1}+T_{j,j+1}^{S=2,m=1}+T_{j+2j'
    ,j+2j'+1}^{S=2,m=1}+T_{j+2j'-1
    ,j+2j'}^{S=2,m=1})\ket{0_j 0_{j+2j'}}\nonumber\\
    =&\sum_{j=1}^L (-1)^j \sum_{j'=1}^{L/2-1}(-1)^{j'}\frac{1}{2}(\ket{0_{j-1}0_{j+2j'}}+\ket{0_{j+1}0_{j+2j'}}+2\ket{0_{j}0_{j+2j'}})\nonumber\\
    &+\sum_{j=1}^L (-1)^j\sum_{j'=1}^{L/2-1}(-1)^{j'}\frac{1}{2}(\ket{0_{j}0_{j+2j'-1}}+\ket{0_{j}0_{j+2j'+1}}+2\ket{0_{j}0_{j+2j'}})\nonumber\\
    =&2\ket{S_2'}.
\end{align}
By counting the number of $\ket{11}$ strings in $\ket{S'_2}$ we have $\sum_j T_{j,j+1}^{S=2,m=2}\ket{S'_2}=(L-4)\ket{S'_2}$. Combining these two terms together, we deduce that the eigenenergy of $\ket{S'_2}$ is also $L-2$. 

To sum up, both $\ket{S'_1}$ and $\ket{S'_2}$ are annihilated by the two-local projectors $\{P_j\}$ and three-local projectors $\{T'_{j-1,j,j+1}\}$, and are eigenstates of $H'$ with the eigenenergy $E=L-2$. When $L/2$ is even, the scar-tower state $\ket{S_{L/2-1}}$ is the linear superposition of these two states. $\ket{S'_1}$ and $\ket{S'_2}$ can combine to generate another superposition state orthogonal to $\ket{S_{L/2-1}}$. When $L/2$ is odd, the two-magnon excitation state $\ket{S'_2}$ is ill-defined and self-vanishing as
\begin{equation}
    \ket{S'_2}=\sum_{j=1}^L (-1)^j\sum_{j'=1}^{L/2-1}(-1)^{j'}\ket{0_j 0_{j+2j'}}
    =\sum_{j=1}^L (-1)^j \sum_{j'=1}^{L/2-1}(-1)^{j'} \ket{0_j 0_{j+2j'+L}}=(-1)^{L/2}\ket{S'_2}.
\end{equation}
The scar-tower state of $n=L/2-1$ now becomes $\ket{S_{L/2-1}}=\sum_j \ket{(-1)_j}$, which corresponds to the single bi-magnon excitation state with zero momentum (revealed by numerical calculations), which is orthogonal to $\ket{S'_1}$. For both cases we observe the two-fold degeneracy at $E=L-2$ in the spectrum of the non-Hermitian AKLT Hamiltonian with two-local and three-local projectors.


\section{More numerical results}
\begin{figure}
\hspace*{-0.7\textwidth}
\includegraphics[width=0.7\linewidth]{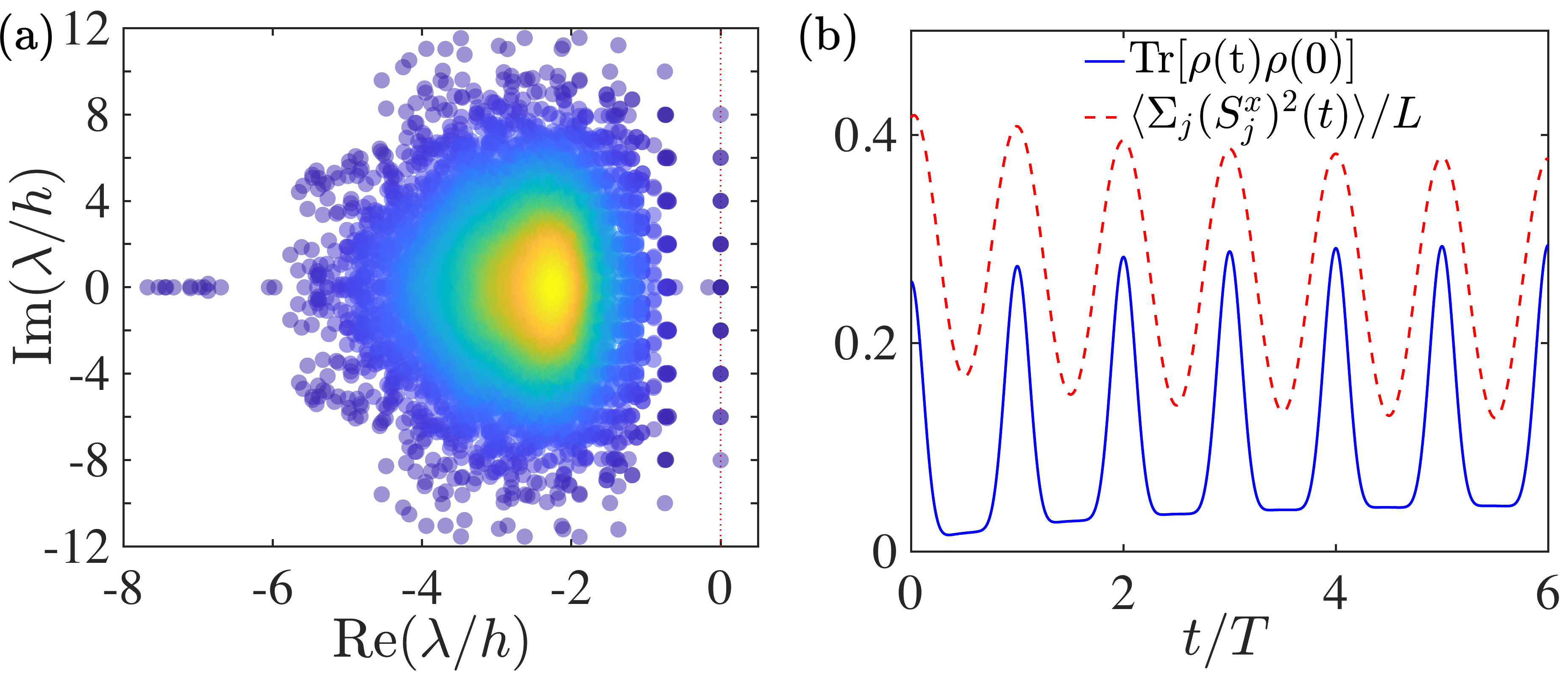} 
\caption{Numerical results for the spin-$1$ $XY$ model.  (a) Liouvillian spectrum of the spin-$1$ $XY$ model. $L=4, h=1, D=1, \gamma=1, V_{j,j+1}=S_j^x$. (b) Liouvillian dynamics of the Loschmidt echo and $\sum_j(S_j^x)^2/L$ for the spin-$1$ $XY$ model, starting from $(\ket{\psi_0}\bra{\psi_0} + I/2^L)/2$, $\ket{\psi_0} = \bigotimes_j \left[(\ket{1}_j -(-1)^j \ket{-1}_j)/\sqrt{2}\right]$~\cite{Schecter2019Weak}.}
\label{fig:XY_supp}
\end{figure}

\begin{figure}
\hspace*{-0.85\textwidth}
\includegraphics[width=0.85\linewidth]{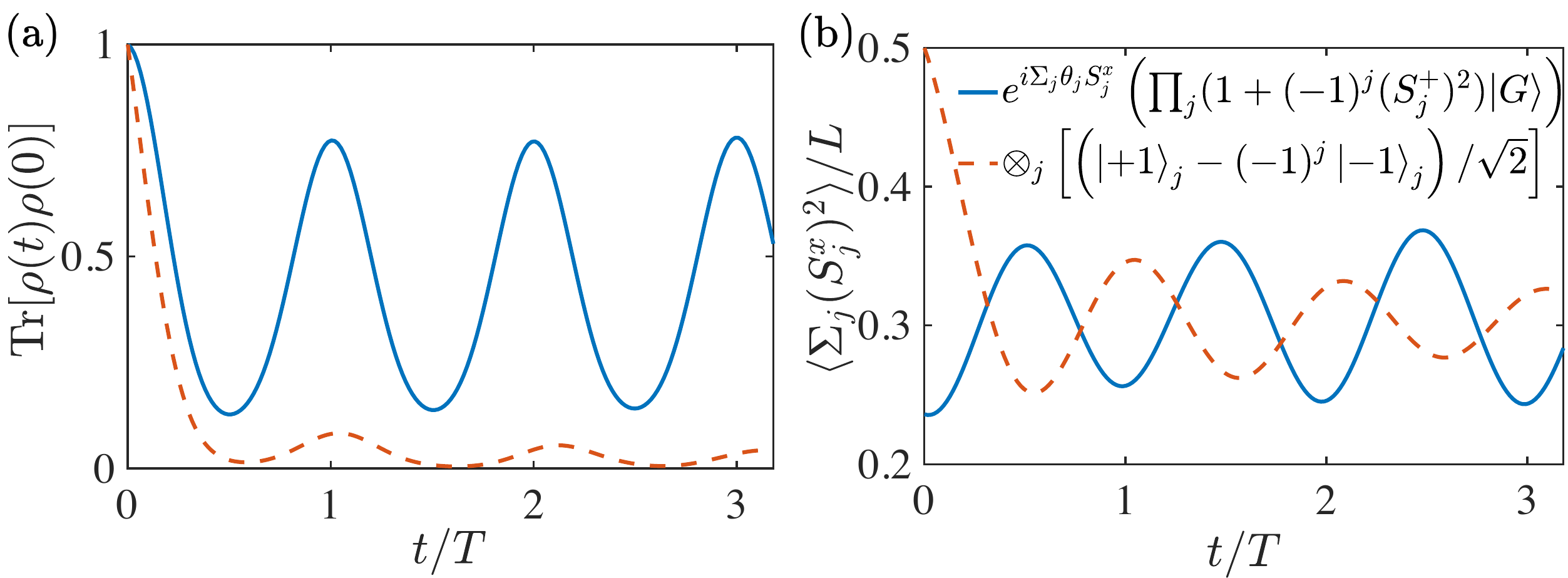} 
\caption{Liouvillian dynamics of the Loschmidt echo (a) and the observable average $\langle \sum_j (S_j^x)^2 \rangle / L$ (b) for the AKLT model, starting from the imperfectly prepared compressed MPS ($\beta = 1$, $\chi = 2$) and the product state same as that in the spin-$1$ $XY$ model \cite{Schecter2019Weak}. $L=6, \gamma =1,  V_{j,j+1}= (\ket{1}\bra{-1} + \ket{0}\bra{0} + \ket{-1}\bra{1})_j, V'_{j-1,j,j+1}= (\ket{1}\bra{-1} + \ket{0}\bra{0} + \ket{-1}\bra{1})_j , \theta_j \in [0,0.2\pi]$.}
\label{fig:AKLT_dynamics}
\end{figure}

\begin{figure}
\hspace*{-0.45\textwidth}
\includegraphics[width=0.45\linewidth]{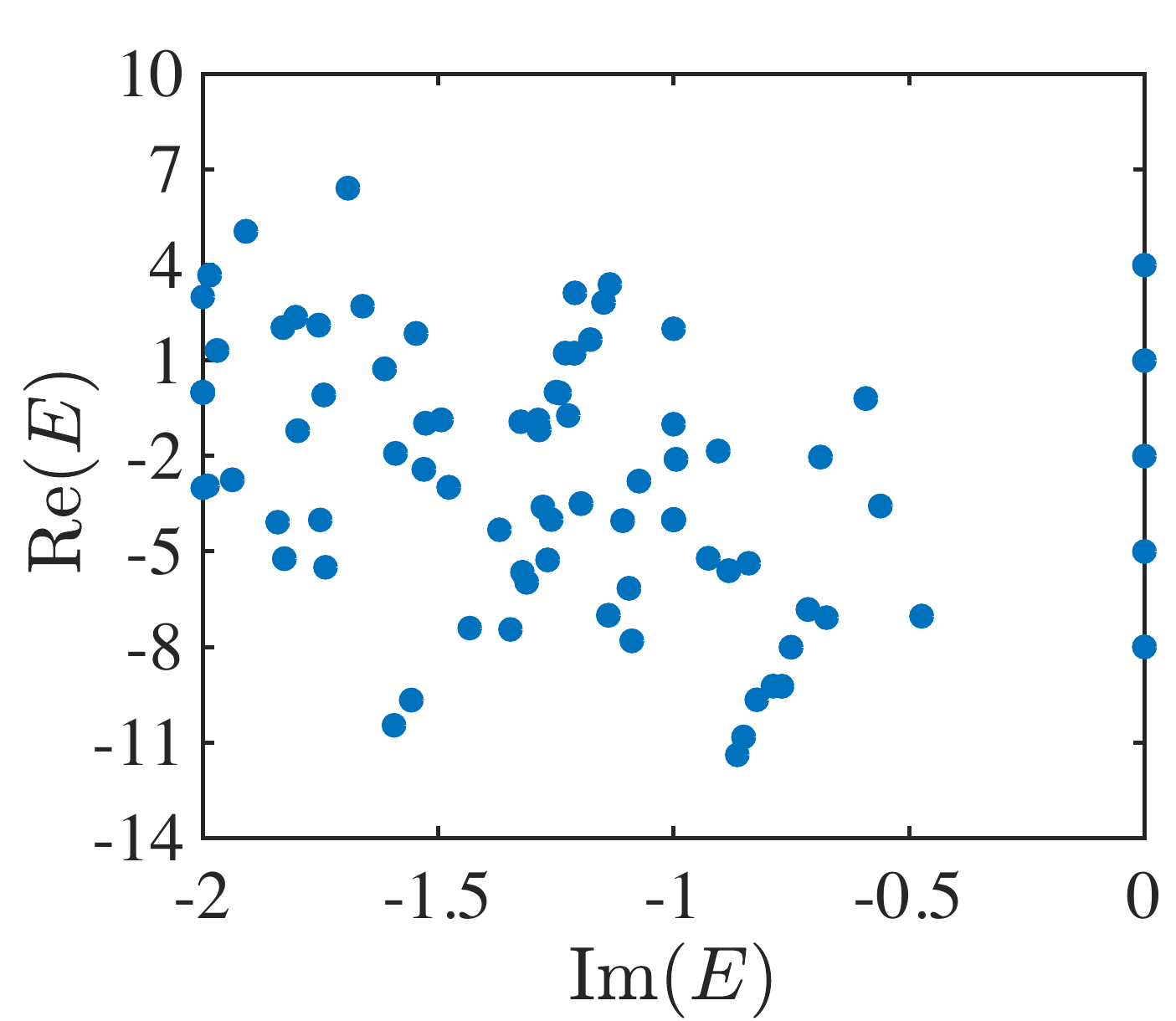} 
\caption{Spectrum of the non-Hermitian Hamiltonian for the domain-wall preserving model. $L=8,\gamma=1,\Delta=0.5, J=1$.}
\label{fig:DW_spectrum}
\end{figure}

In this section, as the supplement to Fig. 2 of the main text, we present more numerical results about the constructed Liouvillians in Table. I of the main text.
In Fig.~\ref{fig:XY_supp}, we display the Liouvillian spectrum and coherent revivals for the spin-1 $XY$ model, which follows the Shiraishi-Mori embedding formalism~\cite{Schecter2019Weak,mark2020unified}. As shown in Fig.~\ref{fig:AKLT_dynamics} for the the Liouvillian dynamics of the AKLT model, the Loschmidt echo $\textnormal{Tr}[\rho(t)\rho(0)]$ and the observable average $\langle \sum_j (S_j^x)^2 \rangle / L$ both exhibit persistent oscillations starting from two different initial states that are easy to prepare. Note that since $L=6$, $L/2$ is odd, the special states with $E=L-1$ (see Subsec.~\ref{subsec:single_magnon_excit}) do not exist. The energy spacing $\omega =2 $ and the oscillation period $T=2\pi/\omega = \pi$. 
In Fig.~\ref{fig:DW_spectrum} we compute the spectrum of the non-Hermitian domain-wall preserving Hamiltonian $H_{\text{NH}}=H_{\text{DW}}-i\gamma\sum_j P_j$, where $P_j = (\ket{\uparrow\uparrow}\bra{\uparrow\uparrow})_{j,j+1} $. All the $L/2+1$ scar-tower states are equidistantly distributed on the real axis with energy spacing $\omega = 2\Delta - 4J = -3$. There exists one additional state degenerate with the highest scar-tower state of $n=L/2$, $\ket{S_{L/2}} = \ket{\uparrow\downarrow\cdots \uparrow\downarrow} + (-1)^{L/2}\ket{\downarrow\uparrow\cdots \downarrow\uparrow}$ (here at $E=-8$), which is the other N\'eel cat state $\ket{\uparrow\downarrow\cdots \uparrow\downarrow} - (-1)^{L/2}\ket{\downarrow\uparrow\cdots \downarrow\uparrow}$.

\section{More discussions about the experimental realizations}
\label{sec:exp}

In this section we rigorously prove the equivalence between the dynamical process described by the experimental scheme in the main text and the corresponding Lindblad master equation. Suppose at evolution time $t$ the system qubits are decoupled with the ancilla qubits in the form of
\begin{equation}
\ket{\Psi(t)}=\ket{\psi(t)}\otimes\ket{\downarrow\downarrow\cdots\downarrow}.
\end{equation}

After the unitary operator $\exp(-i H_{\text{DW}} \delta t)$ on $\ket{\psi(t)}$ and the coupling local unitary gates $\prod_j \exp(-i H_\text{coup}^j \sqrt{\delta t} )$, where $ H_\text{coup}^j = \sqrt{2\gamma} (L_j \tau_j^+ + L_j^\dagger \tau_j^-) $, the system and ancilla qubits get entangled with each others. We decompose the final state on the computational bases of ancilla qubits as 
\begin{align}\label{eq:trajectory}
    &\left(\prod_j\exp(-iH_{\text{coup}}^j\sqrt{\delta t})\right) \exp(-iH_{\text{DW}}\delta t) \left(\ket{\psi(t)}\otimes\ket{\downarrow\downarrow\cdots\downarrow} \right) \nonumber\\
    =&(1-iH_{\text{DW}}\delta t-\gamma\delta t\sum_j L_j^\dagger L_j)\ket{\psi(t)}\otimes\ket{\downarrow\downarrow\cdots\downarrow} - i\sqrt{2\gamma\delta t} \sum_j (L_j\ket{\psi(t)}) \otimes (\tau_j^+\ket{\downarrow\downarrow\cdots\downarrow})\nonumber\\
    -&\gamma\delta t\sum_{j<k}(L_jL_k\ket{\psi(t)})\otimes (\tau_j^+\tau_k^+\ket{\downarrow\downarrow\cdots\downarrow})+O(\delta t^{3/2}).
\end{align}
The three terms in Eq.~\eqref{eq:trajectory} correspond to the cases where quantum jumps induced by $\{L_j\}$ do not occur, occur once and occur twice
respectively.

Next we reset all the ancilla qubits back to $\ket{\downarrow}$ via measurements or optical pumping~\cite{Shankar2013Autonomously,Han2021Experimental,Cai2021High,Google2023Stable,Barreiro2011Open,Lin2013Dissipative,Schindler2013quantum,Weimer2010Rydberg}. The process of resetting an ancilla qubit can be described by the following Kraus map from an arbitrary initial reduced density matrix $\rho_\text{init}$
\begin{equation}
    \rho_\text{init} \to \ket{\downarrow}\bra{\downarrow}\rho_\text{init}\ket{\downarrow}\bra{\downarrow}+\ket{\downarrow}\bra{\uparrow}\rho_\text{init}\ket{\uparrow}\bra{\downarrow}.
\end{equation}

After the resetting the system qubits decouple from the ancilla again, and become the following mixed state
\begin{align}
    \rho(t+\delta t)=&\ket{\psi(t)}\bra{\psi(t)}+\delta t\left[(-iH_{\text{DW}}-\gamma\sum_j L_j^\dagger L_j)\ket{\psi(t)}\bra{\psi(t)}+\ket{\psi(t)}\bra{\psi(t)}(iH_{\text{DW}}-\gamma\sum_j L_j^\dagger L_j)\right] \nonumber\\
    &+2\gamma\delta t\sum_j L_j\ket{\psi(t)}\bra{\psi(t)}L_j^\dagger+O(\delta t^2).
\end{align}

According to the statistical interpretation of density matrices, we deduce that the above joint dynamical evolution faithfully reproduces the many-body Liouvillian dynamics, up to the error of order $O(\delta t^2)$. 
For the quantum trajectory simulations \cite{Daley2014Quantum} in Fig. 3(b) of the main text, instead of manipulating the density matrices, after each resetting operation we pick up one particular pure state of the system qubits according to its jump probability. We repeat the dynamical evolution for many times and calculate the trajectory ensemble average.

As a supplement for the experimental scheme mentioned in the main text, here we further provide concrete procedures to implement each step on current quantum devices: First, in our dissipative protocol, we couple the system qubits to another array of ancilla qubits in order to make the engineered dissipators as local as possible. The arrangement of qubit positions [Fig. 3(a) of the main text] naturally fits the two-dimensional architecture of the superconducting chip~\cite{Google2023Stable} and the programmable lattice geometry of two-dimensional Rydberg-atom arrays~\cite{Bluvstein2021Controlling}. 
Second, the implementation of two-body jump operators requires three-qubit unitary gates, which can be decomposed into finite-depth single- and two-qubit elementary gates (see a demonstration in Ref.~\cite{Google2023Nonabelian}). Ref.~\cite{Weimer2010Rydberg} also proposed the design of $n$-qubit jump operators in Rydberg-atom arrays, which are based on the electromagnetically induced transparency~\cite{Mesoscopic2009Muller}. 
The simultaneously operated two-qubit gate fidelities of superconducting qubits and Rydberg atoms have exceeded 99.5\% nowadays~\cite{Google2023Stable,Evered2023High}, which are high enough to support the large circuit depths for realizing the desired dissipative scarred dynamics. 
The Hamiltonian evolution part can be either realized by analog simulation (e.g., the global Rabi oscillation term in the toy model), or can be Trotterized into local elementary gates through standard approaches.
Finally, all the ancilla qubits can be reset back to the spin down state via measurements~\cite{Han2021Experimental,Cai2021High}, optical pumping in trapped ions or Rydberg atoms~\cite{Barreiro2011Open,Lin2013Dissipative,Schindler2013quantum,Weimer2010Rydberg}, or recent fast resetting protocols~\cite{Mcewen2021Removing,Miao2022Overcoming,Google2023Stable}.

Besides, we stress that any environmental decoherence satisfying the same forms of the dissipators $L_j = V_j P_j$ ($V_j$ could be arbitrary local operators) will not affect the open scarred dynamics. Since we are consecutively resetting all the ancilla qubits, our dissipative protocol intrinsically tolerates instantaneous perturbations and gate errors that drive the system out of the scar-state-only decoherence-free subspaces. The dissipative scarred dynamics or dissipatively prepared scar states can be preserved for a time scale longer than the coherence time of the physical qubits~\cite{Google2023Stable}.

\section{Dissipative preparation of scar states with quantum metrology applications}

In this section, we demonstrate that our dissipative Liouvillian dynamics towards the scar-state-only decoherence-free subspaces can be utilized to prepare scar states with quantum metrology applications.
Quantum many-body scars stemming from the $su(2)$ spectrum generating algebra have been shown to possess extensive multipartite entanglement, manifested by their extensive \textit{quantum Fisher information} density~\cite{Dooley2021Robust,Desaules2022Extensive,Dooley2023Entanglement}.
This property makes the scar-tower states bear significant potential as resources in the quantum enhanced metrology.
However, in closed systems these scar-tower states are highly excited eigenstates of many-body Hamiltonians, which lack effective and systematic methods to prepare them on experimental quantum platforms. 
Below we show through the concrete example of the Dicke state scars that: based on the constructed Liouvillians in the main text, if we impose the corresponding strong symmetries to the designed Liouvillians, we can prepare any desired scar-tower states by starting the dissipative evolution from an easily-prepared initial state in the same symmetry sector.

Before proceeding on, we give a brief introduction to the central concept of quantum Fisher information in quantum metrology, which sets the ultimate bounds on the precision of estimating an unknown parameter (see~\cite{Pezze2018Quantum} for a review).
Suppose we prepare a quantum state $\rho$ and want to use it to measure certain unknown field strength $\lambda$ that linearly couples to an operator $O$.
We evolve the quantum state as $\rho_{\lambda}=e^{-i\lambda O}\rho e^{i\lambda O}$, and perform quantum measurements on $\rho_{\lambda}$ to estimate $\lambda$ according to the measurement outcomes. It has been proved that the variance or precision of $\lambda$ is lower bounded by the quantum Cram\'er-Rao bound:
\begin{equation}
    (\Delta\lambda)^2 \ge \frac{1}{M \mathcal{F}_Q(O, \rho)},
\end{equation}
where $M$ is the number of independent measurements. The quantum Fisher information $\mathcal{F}_Q(O, \rho)$ has the expression as~\cite{Braunstein1994Statistical}:
\begin{equation}\label{eq:QFI}
    \mathcal{F}_Q(O, \rho) = 2\sum_{n,m} \frac{(p_n - p_m)^2 }{p_n + p_m } |\langle \psi_n| O | \psi_m \rangle |^2 \leq 4\langle \Delta O^2 \rangle, 
\end{equation} 
where $\{ p_n, \ket{\psi_n} \}$ is the spectrum decomposition of the density matrix $\rho=\sum_n p_n |\psi_n \rangle\langle\psi_n|$. The equality is saturated for the pure state $\rho = \ket{\psi}\bra{\psi}$ and $\langle \Delta O^2 \rangle = \mathrm{Tr}(\rho O^2) - \mathrm{Tr}(\rho O)^2$. 

One important physical application of the quantum Fisher information is to characterize the genuine multipartite entanglement~\cite{Pezze2009Entanglement,Hyllus2012Fisher,Toth2012Multipartite}. 
If the quantum Fisher information density for an extensive summation of local operators $O = \sum_{j=1}^L o_j$ satisfies $f_Q(O, \rho) = \mathcal{F}_Q(O, \rho) / L > m$, there exist at least $(m+1)$ parties in the system forming a genuinely entangled block.
For short-range entangled states, $\mathcal{F}_Q$ scales as $\mathcal{O}(L)$, thus $\Delta \lambda \ge 1 / \sqrt{M L} $, which corresponds to the standard quantum limit.
In contrast, for long-range entangled states, $\mathcal{F}_Q$ can scale super-extensively as $\mathcal{O}(L^2)$, thus $\Delta \lambda \ge 1 / \sqrt{M} L $, which is known as the Heisenberg limit.
Scar-tower states stemming from the $su(2)$ spectrum generating algebra have been shown to exhibit extensive multipartite entanglement and extensive quantum Fisher information density~\cite{Dooley2021Robust,Desaules2022Extensive,Dooley2023Entanglement}, because they possess off-diagonal long-range order with respect to the ladder operator $Q^\dagger$ generating the $su(2)$ algebra~\cite{Iadecola2019Quantum}.

In the main text, we mainly focus on the persistent periodic oscillations within the scar-state-only decoherence-free subspaces, where the non-stationary steady states are always mixed states containing the coherence terms like $\ket{S_n}\bra{S_m}$. In contrast, below we illustrate the principle for how to prepare each single scar state through the Liouvillian dynamics. 
Consider the case when the Liouvillians respect the following strong symmetry conditions~\cite{Buca2012Note} for the operator $A$ lifting the energy degeneracy of the scar-tower states (i.e., the scar-tower states carry different good quantum numbers $A \ket{S_n} = a_n\ket{S_n}$), such as the $\sum_{j} \sigma^x_j$, $\sum_j (S^z_j)^2$ terms in the scarred Hamiltonians:
\begin{equation}
    [H,A]=0,\quad [L_j,A]=[L_j^\dagger,A]=0,\forall j.
\end{equation}
During the dissipative evolution, the strong symmetry implies the conservation of the good quantum number $a_n$. In other words, if we denote the projector onto the symmetry sector (Hilbert subspace) with the quantum number $a_n$ by $\Pi_n$, we will straightforwardly deduce that
$\mathcal{L} ( \Pi_n \rho \Pi_m ) = \Pi_n \mathcal{L}( \rho ) \Pi_m $.
Hence, once we choose an easily-prepared initial state within the $a_n$ symmetry sector (e.g., a product state or low-bond-dimension MPS), the steady state is guaranteed to become the desired single scar state $\ket{S_n}$.

\begin{figure}
\hspace*{-0.45\textwidth}
\includegraphics[width=0.45\linewidth]{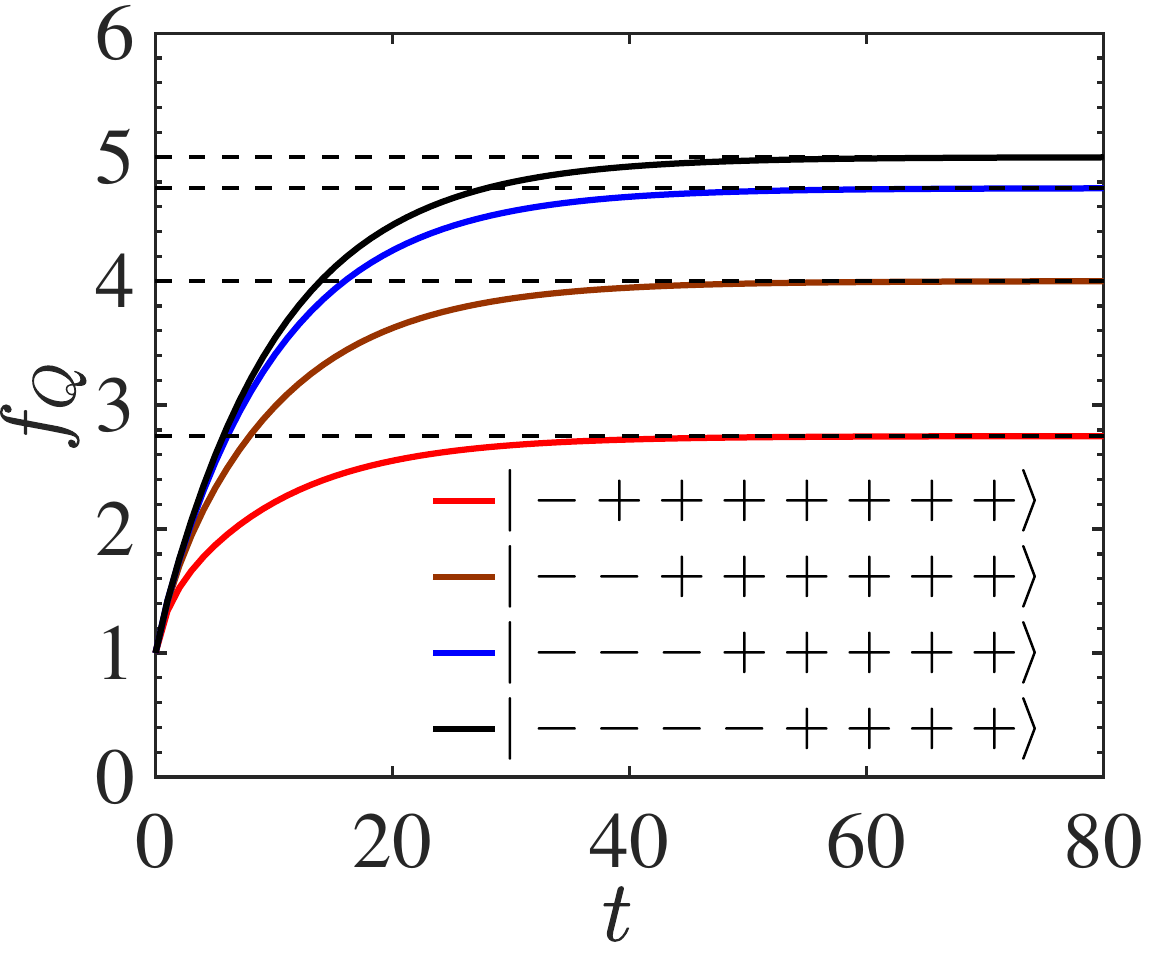} 
\caption{The Liouvillian dynamics of the quantum Fisher information density $f_Q(\sum_{j=1}^L \sigma_j^z, \rho)$ for the toy model hosting $x$-direction Dicke states as scars, starting from different $x$-direction spin product states.
$L = 8$, $\gamma = 1$. $\ket{\pm}$ denotes the $\pm 1$ eigenstate of $\sigma^x$. The horizontal dashed lines indicate the exact quantum Fisher information density $f_Q = 2(-n^2 / L + n) + 1$ for the $n$-th Dicke state.}
\label{fig:QFI_dynamics}
\end{figure}

Specifically, we demonstrate the aforementioned principle by the toy model hosting Dicke states as scars~\cite{Choi2019emergent}. The quantum Fisher information density of the $x$-direction Dicke states with respect to the operator $\sum_{j=1}^L \sigma^z_j$ can be calculated as
\begin{equation}
    f_Q(\sum_{j=1}^L \sigma^z_j, \ket{S_n}) = 2(- \frac{n^2}{L} + n) + 1,
\label{eqn:QFI_Dicke}
\end{equation}
where $\ket{S_n}=\ket{S=L/2,S_x=n-L/2}$ are the $n$-th $x$-direction Dicke state ($n=0,1,\cdots,L$). The quantum Fisher information density obtains the maximum value 
$L/2 + 1$ in the middle of the scar tower $n=L/2$ (known as the twin-Fock state~\cite{Pezze2018Quantum}), which scales extensively with the system size and implies the extensive multipartite entanglement of the Dicke state.

Based on the constructed Hamiltonians and jump operators of the toy model shown in Table. I of the main text, we further require that the Liouvillian superoperator respects the strong symmetry $S_x = \sum_{j=1}^L \sigma^x_j/2 $, $[H_{\text{toy}}, S_x]=[ L_j, S_x ]= 0, \forall j$. After imposing the strong symmetry condition onto the constructed Liouvillians, we obtain $ H_{\text{toy}}=\Omega \sum_j\sigma_j^x / 2 + \sum_j P_j h_j P_j$, where $P_j = (1-\vec{\sigma}_j\cdot\vec{\sigma}_{j+1}) / 4$ and 
\begin{equation}
h_j=J_1(\sigma_{j-1}^y\sigma_{j+2}^y+\sigma_{j-1}^z\sigma_{j+2}^z)+J_2(\sigma_{j-1}^y\sigma_{j+2}^z-\sigma_{j-1}^z\sigma_{j+2}^y)+J_3\sigma_{j-1}^x\sigma_{j+2}^x.
\end{equation}
Meanwhile, the two-local dissipators take the form of
\begin{equation}
L_j=\sigma_j^x (1 - \vec{\sigma}_j\cdot\vec{\sigma}_{j+1} ).
\end{equation} 
By utilizing the Liouvillians above, we can dissipatively prepare any desired $x$-direction Dicke state starting from an $x$-direction product state in the same symmetry sector. For the sake of simplicity in experimental implementations, we could set $H_{\text{toy}}=0$ and only exploit the two-local dissipators.
As displayed in Fig.~\ref{fig:QFI_dynamics}, we numerically calculate the Liouvillian dynamics of the quantum Fisher information density starting from different $x$-direction spin product states. The final values of $f_Q$ all converge to the exact analytical values of the corresponding Dicke states in Eq.~\eqref{eqn:QFI_Dicke}.

Several remarks come in order. First, we emphasize that due to their extensive quantum Fisher information density, the Dicke states are valuable resources for quantum enhanced metrology~\cite{Pezze2018Quantum}. 
In contrast to the conventionally used all-to-all interactions (e.g., the one-axis twisting term~\cite{Kitagawa1993Squeezed}), our dissipative protocol utilizes two-local short-range dissipators to effectively prepare arbitrary Dicke states. Second, following the procedures discussed in Sec.~\ref{sec:exp}, the dissipative preparation of scar states can be readily implemented on current quantum devices, especially for superconducting qubits and Rydberg-atom arrays. 
As illustrated previously, through consecutively resetting the ancilla qubits, our protocol naturally tolerates instantaneous perturbations and certain gate errors, and can preserve the prepared target states for time scales longer than the coherence time of the physical qubits. 
All the above merits demonstrate that our constructions of scar-state-only decoherence-free subspaces are not just some exact mathematical forms. Their practical applications and experimental feasibility are of equal importance to their theoretical innovations and mathematical elegance.

\end{document}